Ain Shams University
Faculty of Engineering
Electrical Power and Machines Engineering Department

# Coupling Wind Farm with Nuclear Power Plant

## A Thesis

Submitted in partial fulfillment of the requirements of the degree of
Master of Science in Electrical Engineering

Submitted by

Mohamed Kareem Abdel Rahman Al-Ashery
B.Sc. of Electrical Engineering
Electrical Power and Machines Engineering
Ain Shams University, 2008

Supervised by:
Prof. Dr. Mohamad Abd Al Rahim Badr
Dr. Walid El-Khattam
Dr. Moustafa Saleh El Koliel

Ain Shams University
Faculty of Engineering
Electrical Power and Machines Engineering Department

# Coupling Wind Farm with Nuclear Power Plant

A Master of Science thesis Submitted by

Mohamed Kareem Abdel Rahman Al-Ashery

## EXAMINERS' COMMITTEE

| Name | Signature |
|---|---|
| Prof. Essam El-Din Mohamed Abou-El-Zahab<br>Cairo University, Faculty of Engineering,<br>Electrical Power and Machines Engineering Dept. | ………… |
| Prof. Almoataz Youssef Abdelaziz Mohammed<br>Ain Shams University, Faculty of Engineering,<br>Electrical Power and Machines Engineering Dept. | ………… |
| Prof. Mohamad Abd Al Rahim Badr<br>Future University,<br>Dean, Faculty of Engineering and Technology | ………… |
| Dr. Walid Ali Seif El-Eslam El-Khattam<br>Ain Shams University, Faculty of Engineering,<br>Electrical Power and Machines Engineering Dept. | ………… |

Date: / / 2014

# Researcher Data

| | |
|---|---|
| Name: | Mohamed Kareem Abdel Rahman Al-Ashery |
| Date of Birth: | 25/6/1986 |
| Place of Birth: | Saudi Arabia |
| Academic Degree: | B.Sc. in Electrical Engineering |
| Field of specialization: | Electrical Power and Machines Engineering |
| University issued the degree: | Ain Shams University |
| Date of issued degree: | 2008 |
| Current job: | Research assistant, Egyptian Atomic Energy Authority |

# Statement

This dissertation is submitted as partial fulfillment of Master of Science in Electrical Engineering, Faculty of Engineering, Ain Shams University.

The author carried out the work included in this thesis, and no part of it has been submitted for a degree or qualification at any other scientific entity.

Student Name: Mohamed Kareem Abdel Rahman Al-Ashery
Signature:
Date: 22 / 7 / 2014

# Acknowledgments

الحمد لله رب العالمين

All praise is due to ALLAH, the Lord of the Worlds, Most Gracious, Most Merciful; Who taught man that which he knew not.

I would like to thank all the people who contributed in some way to the work described in this thesis. First and foremost, I wish to express my gratitude to my supervisors, Prof. Dr. Moh Abd El Rehim Badr, Dr. Walid El-Khattam, and Dr. Moustafa El Koliel for their exceptional guidance, encouragement, insightful thoughts and useful discussions. I would like to thank Prof. Badr for accepting to be my supervisor in this project. My sincere appreciation goes to Dr. Walid for all I have learned from him and for his continuous help and support in all stages of this thesis. I would also like to thank him for being an open person to ideas, and for encouraging and helping me to shape my interest and ideas. I would like to thank Dr. Moustafa for his valuable advices that help me not only in my thesis, but also in my life.

I would like to thank my colleagues and friends for their support and help. My sincere thanks go to my colleagues at the Egyptian Atomic Energy Authority. Special thanks to Ahmed Sallam, my colleague at Egypt Era, for useful discussions during my thesis. I would like to acknowledge my friends at Nuclear Power Plant Authority, Mohamed Essa, Mohamed Ramadan, and Mostafa Fouad for their valuable technical discussions.

Last but not least, I would like to thank my parents. Their unlimited care, their patience and love have always guided my through my whole life. I wish ALLAH help me reward them only part of what they did for me.

Moh Kareem Al-Ashery,
Cairo, Egypt,
2014

# Abstract

Mohamed Kareem Abdel-Rahman Al-Ashery, Coupling Wind farm with Nuclear Power Plant, Master of Science dissertation, Ain Shams University, 2014.

Climate change has been identified as one of the greatest challenges facing nations, governments, businesses and citizens of the globe. The threats of climate change demand an increase in the share of renewable energy from the total of energy generation. Meanwhile, there are tremendous efforts to decrease the reliance on fossil fuel energies which opens the venue for increasing the usage of alternative resources such as nuclear energy. Many countries (e.g. Egypt) are planning to meet increasing electricity demands by increasing both renewable (especially wind energy) and nuclear energies contributions in electricity generation.

In the planning phase of siting both new Wind Farms (WFs) and Nuclear Power Plants (NPPs), many benefits and challenges exist. An important aspect taken into consideration during the NPP siting is the existence of ultimate heat sink which is sea water in most cases. That is why most NPPs are sited on sea coasts. On the other hand, during WF siting, the main influential aspect is the existence of good wind resources. Many coastal areas around the world fulfill this requirement for WF siting. Coupling both NPPs and WFs in one site or nearby has many benefits and obstacles as well. In this thesis, based on international experience and literature reviews, the benefits and obstacles of this coupling/adjacency are studied and evaluated. Various case studies are carried out to verify the coupling/adjacency concept.

# Summary

This thesis studies and evaluates the benefits and obstacles of coupling of Wind Farms (WFs) with Nuclear Power Plants (NPPs). The dissertation is divided into five chapters organized as follows:

Chapter One: It is an introduction to the idea of coupling of WF with NPP. Motivation of the idea and the thesis outline are discussed.

Chapter Two: This chapter presents the literature survey, and the characteristics of the site that can accommodate both NPP and WF are illustrated. It also evaluates different sites in Egypt that can accommodate coupling. Finally, advantages and disadvantages of coupling of WFs with NPPs are explained.

Chapter Three: It illustrates the benefits of connecting WF and NPP to the same point of the grid. Two case studies are conducted to verify two main benefits of this adjacency, which are the impact of high short circuit power level on the voltage quality aspects of WF, and increasing reliability and availability of NPP Emergency Power Systems (EPSs) by the on-site WF.

Chapter Four: The benefit of helping in geographical distribution of WFs in the grid is discussed in details. A case study is conducted to illustrate the smoothing effect of WFs geographical distribution in the Egyptian grid. After that, wind energy capacity credit assessment is done considering the case in Egypt. Finally, Strategic plan for the coupling in Egypt is illustrated, considering the coordination between WFs and NPPs new installations.

Finally, the thesis ends by extracting conclusions and stating future work that might be done based on this work.

# Contents

# List of Figures

# List of Tables

# List of Symbols

| | |
|---|---|
| $\%P_{WF}\ (V_t)$ | The power generation from WF in percentage of installed capacity |
| $\underline{A}$ | Transition rate matrix |
| C | Scale parameter of the 2-parameter Weibull distribution |
| $c(\Psi_k, Va)$ | Flicker coefficient |
| dso | Switching operations voltage change of the grid at PCC (normalised to nominal voltage) |
| dss | Steady state voltage change of the grid at PCC (normalised to nominal voltage); |
| F | Set of failure states of the system |
| f(v) | Probability density function of the 2-parameter Weibull distribution of wind speed (v) |
| K | Shape parameter of the 2-parameter Weibull distribution |
| $k_f(\Psi_k)$ | Flicker step factor; |
| $k_u(\Psi_k)$ | Voltage change factor |
| N | Number of states |
| n | Reading number; |
| $N_{120}$ | Number of switchings within a 2 hours period |
| $N_r$ | Total number of readings for each case (24*12=288) |
| $N_{WT}$ | Number of WTs in the WF |
| Ø | Phase angle between voltage and current |
| $Ø_{06}$ | Probability that wind speed is in the region of $S_0$ ($V_i < V < V_r$) and decreases to be in the region of state $S_6$ (V< Vi) |
| $P(V_t)$ | Power output of the WT at wind speed ($V_t$) |
| $P_0(t)$ | Probability function in time of state $S_0$ |
| $P_E$ | Equipment installed capacity (MW) |
| $P_i(t)$ | Probability of system to be in state (i) function in time |
| $P_{lt}$ | Flicker distortion |
| $P_n$ | Output power (% installed capacity) according to reading number (n) |
| $P_r$ | Rated power output of the WT |
| $P_r\ (V_i<V<V_r)$ | Probability of wind speed to be above (Vi) and below (Vr) |
| $P_{WF}(V_t)$ | Electrical power generated by WF (MW) |
| $Q_A$ | Unreliability or unavailability of system A |
| $Q_p$ | Unreliability or unavailability of the parallel system |
| $R_A$ | Reliability or availability of the system A |
| $R_L$ | Short circuit resistance of the grid |
| $R_p$ | Reliability or availability of the parallel system |
| $R_{SC}$ | Short Circuit Ratio |
| S | Set of success states of the system |
| $S_{60}$ | Apparent power at the 1-min. active power peak |
| $S_i$ | State number (i) |
| $S_n$ | Apparent power of the WT at rated power |
| Ssc | Short circuit power level of the grid at PCC |
| t | Year number |
| T | WF's Life Time (Year)+ WF's Construction Time (Year) |
| $U_S$ | Sending-end voltage of the constant voltage source |
| $U_t$ | Terminal voltage |
| V | Wind speed in m/s at H height in m |
| Va | Annual average wind speed at hub height |
| $V_i$ | Cut-in wind speed of WT |
| $V_o$ | Cut-out wind speed of WT |
| $V_r$ | Rated wind speed of WT |
| $V_T$ | Wind speed at the WT hub height |
| $V_t$ | Wind speed at time (t) |
| $X_L$ | Short circuit reactance of the grid |
| $Z_L \angle \theta_L$ | Line impedance |
| $\alpha_{10}$ | Transition rate from $S_1$ to $S_0$ |
| $\Delta P_n$ | Variation in output power (% installed capacity)from one reading to the next |
| $\Delta U$ | Voltage drop |
| $\lambda$ | Failure rate of one unit of DG |
| $\lambda_{ccf}$ | Common cause failure rate of DGs set |
| $\mu$ | Repair rate of one unit of DG |
| $\Psi_k$ | Grid impedance angle at PCC |
| $\boldsymbol{\alpha}$ | Correction factor |

# List of Abbreviations

| | |
|---|---|
| a.g.l. | above ground level |
| CF | Capacity Factor |
| CPUC | California Public Utilities Commission |
| DC | Direct Current |
| DG | Diesel Generator |
| EEHC | Egyptian Electricity Holding Company |
| ELCC | Effective Load Carrying Capability |
| EPSs | Emergency Power systems |
| EPZ | Emergency Planning Zone |
| ESCE | Egyptian Supreme Council of Energy |
| LCOE | Levelized Cost of Energy |
| LOOP | Loss of Off-site Power |
| LPZ | Low Population Zone |
| LVRT | Low Voltage Ride Through |
| NPP | Nuclear Power Plant |
| NYISO | New York Independent System Operator |
| OGU | Ordinary Generation Unit |
| OLS | Obstacle Limitation Surface |
| PCC | Point of Common Coupling |
| PDF | Probability Distribution Function |
| PJM | Pennsylvania-New Jersey-Maryland |
| SBO | Station Black Out |
| SPP | Southwest Power Pool |
| TSO | Transmission System Operator |
| WF | Wind Farm |
| WT | Wind Turbine |

# Chapter 1
# Introduction

## 1.1 Electricity in Egypt: current and future

The regular annual reports issued by the Egyptian Electricity Holding Company (EEHC) indicate that the installed generating power capacity in 2012 is about 29,074 MW based mainly on oil and natural gas sources and including only 2% of wind energy and no nuclear energy. The country has limited fossil fuel energy resources and almost fully utilized hydro energy. The continuous increase in demand of energy (7% annually) raises a significant stress on the governmental policy to look for energy sources alternatives as a vital national priority. The ambitious plan of Egypt is to increase the renewable energy share to 20% of the total demand by the year 2020. Meanwhile, nuclear energy is planned to enter the country energy mix by the year 2019 [1].

Among others, the promising renewable energies are the sun and wind energy sources. However, the wind energy production is much more cost effective than other alternatives. Therefore, in April 2007, the Egyptian Supreme Council of Energy (ESCE) elaborated on a plan to increase the share of wind energy such that it would represent 12% of total electricity demand by 2020. This would require wind capacity to reach 7500 MW by 2020 [2]. The national database (Atlas) of wind energy confirms that Egypt has suitable wind energy potential areas around the country especially some areas along the Mediterranean and Red sea coasts, and the west and east of the Nile valley.

As stated above, nuclear energy is planned to enter the country energy mix by the year 2019. A main requirement during Nuclear Power Plant (NPP) site selection is the existence of ultimate heat sink, which is sea water in most cases. Therefore NPP site in Egypt must be a coastal area. First NPP is to be built in El Dabaa site on the Mediterranean Sea coast. The site can accommodate four units each of 1000-1600 MW installed capacity. The first unit is planned to enter service in 2019 and the last in 2025 [1].

## 1.2 Thesis Motivation

From the previous section it can be realized that in Egypt coastal areas are preferred for both NPPs and Wind Farms (WFs). The country energy plan is to increase the installed capacity of both nuclear and wind energies in the grid. Therefore, in the future, it is possible to find a site where WF is erected near NPP site or vice versa. The thesis motivation is to propose siting of WF around NPP and study the whole advantages and disadvantages of this coupling. This coupling will be feasible if it has technical and economical benefits for all entities (NPP, WF, and grid) from various aspects.

The coupling of NPPs and WFs exists in Illinois (USA) where Grand Ridge WF of capacity 98MW is located adjacent to LaSalle NPP of capacity 2309MW [3]. Other examples in Ontario (Canada): 1) The Ripley Wind Power Project of capacity 76MW is located near Bruce Power NPP of capacity 4820MW, with some kilometers between the two sites [4]; 2) A single Wind Turbine (WT) of capacity 1.8 MW is erected in the site of Pickering Nuclear Generating Station of capacity 3100 MW for research purposes. In Finland, Olkiluoto WF of capacity 1 MW is located near Olkiluoto NPP of capacity 1760 MW.

## 1.3 Thesis outline

The main objective of this thesis is to illustrate and quantify the technical and economical advantages and disadvantages of the idea of coupling of WFs with NPPs. This will help decision makers to choose the right decision during planning of the generation expansion siting. This may differ from case to case, coupling may be useful in some cases, and it may not be so in others. The thesis consists of five chapters in addition to a list of references as detailed below.

In Chapter 2, first, the previous work done in this point is mentioned briefly. After that, the study starts by pointing out the properties of the site that can accommodate both NPP and WF. Many siting requirements for both NPP and WF are the same, such as requirement for existence of grid connection, transport infrastructure to the site, etc. Main requirement during the WF siting is the existence of potential wind resource. Main requirement during the NPP siting is to decrease the risks posed to the community and environment. A case study is conducted to find sites in Egypt that can accommodate coupling. There are sites in Egypt that passes the initial site selection requirements for both NPP and WF. Among these sites is El Dabaa site which is proposed for the first NPP in Egypt and has a potential wind resource according to [5]. Next, the benefits of NPP and WF coupling are mentioned and illustrated. Finally, some disadvantages of coupling and the proper solution for each disadvantage are illustrated.

Chapter 3 starts by detailed illustration of the benefits of connecting the WF and NPP to the same point of the grid. Also, a case study is conducted to illustrate the impact of the connection point short circuit power level on the voltage quality aspects of WF. The higher the short circuit power level, the lower the impacts of connecting WF on the voltage quality aspects of Point of Common Coupling (PCC). Finally, the chapter ended by detailed illustration of the benefit of increasing reliability and availability of NPP EPSs by the on-site WF. A case study using MARKOV process is used for verification and quantification of this benefit.

In chapter 4, the benefit of helping in geographical distribution of WFs in the grid is discussed in details. A case study is conducted to illustrate the effect of WFs geographical distribution on the aggregated wind power production in the Egyptian grid. The geographical distribution of WFs should lead to smoothing the wind power production in the grid. Next, wind energy capacity credit assessment is done considering the case in Egypt. Also, a strategic plan for the coupling in Egypt is illustrated, considering the coordination between WFs and NPPs new installations. Finally, the impact of coupling on Levelized Cost Of Energy from WF is illustrated and discussed using a case study.

In chapter 5, the conclusion and future work are discussed.

# Chapter 2

# Literature Survey

This chapter presents the previous work carried out in this point of research. Then, it gives overview over the important considerations during NPP site selection and WF site selection. A site selection procedure for coupling is then presented, followed by evaluating sites in Egypt that can accommodate that coupling. Finally the advantages and disadvantages of coupling WF and NPP are discussed.

## 2.1 Previous Work

Coupling or adjacency of NPP and WF is a new trend with economic evaluation examples of existing projects rather than research or mathematical modeling. In [3] the authors mentioned some benefits of existence of both NPP and WF in the same site. WTs on a standby operational mode are net importers of power for their control and yaw mechanisms. They also require the vicinity of a power grid with excess capacity to export their generated power. Therefore, WFs can be constructed in the immediate vicinity of low population density zones around NPPs [3]. As an example, the Grand Ridge WF surrounds the site of LaSalle NPP near Versailles, Illinois (USA). The NPP consists of two Boiling Water Reactor units. The plant started operation in 1982 and generates about 2,309 MW. The Grand Ridge WF's construction was completed in 2008 with 98 MW of installed capacity. The WF covers an area of 6,000 acres around the LaSalle NPP in the Brookfield, Allen and Grand Rapids townships. An expansion to Grand Ridge with a capacity of 111 MW is slated with an operational start in 2009 [3].

In [4], the study is mainly concerned with the economical benefit of adding hydrogen storage facility to the wind-nuclear system. The study considers a system consisting of NPP and WF selling electricity to the Ontario electricity market. This is a realistic scenario for southwestern Ontario with a nuclear power producer, i.e. Bruce Power, and a number of large-scale WFs located nearby. The proposed system includes hydrogen storage and distribution facilities. The nuclear–wind system is assumed to be located in Bruce County in Ontario, Canada, because Bruce Power's NPP operates in the region and the Ripley WF Project being built there. The hydrogen storage system is assumed to be located in the greater Toronto area, which is Ontario's major load center, and assumes that this region has a well-developed hydrogen, heat and oxygen markets. The Ripley Wind Power Project consists of 38 Enercon E82 2MW WTs with total capacity of 76MW. Bruce Power is Canada's first privately owned NPP, with installed capacity of 4820 MW [4].

Finally, in [6, 7, 8], the coupling of WF and NPP is proposed. Each study proposes a different motivation for coupling and different configuration for the system.

It is important to mention that, no one study the advantages and disadvantages of the coupling of WF and NPP considering different technical and economical aspects illustrated.

# 2.2 Site selection for coupling of WF with NPP

## 2.2.1 NPP site selection

The use of nuclear energy must be safe; it shall not cause injury to people, or damage to the environment or property. The main objective in site evaluation for nuclear installations is to protect the public and the environment from the radiological consequences of radioactive releases due to accidents or normal operation [9].

In NPP site selection, there are two main objectives [10]:

1. Ensuring the technical and economical feasibility of the plant; and
2. Minimising potential adverse impacts on the community and environment.

To account for these objects, two sets of criteria have been developed. The primary criteria are concerned with the technical and economic feasibility of NPPs. The secondary criteria relate to the risks that NPPs pose to the community and environment [10]. A balance must be achieved between these two criteria.

## A. Primary criteria (technical and economic feasibility of NPPs)

A NPP is basically a large electrical generating facility, and it shares a number of siting factors or requirements with large fossil-fueled plants. There are four primary criteria for NPP site selection [10]:

1. *Proximity to appropriate existing electricity infrastructure;*
2. *Proximity to major load centers (i.e. large centers of demand);*
3. *Proximity to transport infrastructure to facilitate the movement of nuclear fuel, waste and other relevant materials;* and
4. *Access to large quantities of water for cooling.*

## B. Secondary criteria (Hazards and risks to the community and environment)

It is divided into two main sections which are:

1. *Specific hazards from external events* [9, 11]*:*
   - a) Earthquakes and Surface faulting;
   - b) Meteorological and climatological characteristics of site region;
   - c) Flooding potential due to natural causes;
   - d) Geotechnical hazards;
   - e) External human induced events such as aircraft crashes, chemical explosions, and other important human induced events; and
   - f) Other important considerations such as volcanism, sand storms, severe precipitation, snow, ice, hail, and subsurface freezing of subcooled water (frazil).

2. *Site characteristics and the potential effects of the NPP in the region [9, 11]:*
   - a) Atmospheric dispersion of radioactive material;
   - b) Dispersion of radioactive material through surface water;
   - c) Dispersion of radioactive material through ground water;
   - d) Distance from densely populated areas is necessary to minimize community opposition and security risks and to reduce the complexity associated with emergency planning [10]; and
   - e) Uses of land and water in the region.

## 2.2.2 WF Site Selection

The first phase in any wind generation development is the initial site selection. For many developers the starting point of this process involves looking at a chosen area in order to identify one or more sites which may be suitable for development [12]. The selection of an appropriate site for a WF involves examining and balancing a number of technical, environmental and planning issues.

The factors affecting site selection can be divided to two main topics: technical considerations and environmental considerations.

### A. Initial technical considerations

The site selection process will largely involve desk-based studies to determine whether sites satisfy five crucial technical criteria for successful development [12, 13]:

1. *Potential wind resource, promising values are average wind speeds above 6 m/s at 50 m above ground level (a.g.l.) [12];*
2. *Potential size of site, to make the development commercially viable;*
3. *Cost effective electrical connection access;*
4. *Suitable landownership - current, previous and future usage;* and
5. *Construction issues (ease of construction) such as site access constraints.*

### B. Initial environmental considerations

At the same time as carrying out technical analyses, proponents will consider potential environmental impacts on potential sites [12, 13]:

1. *Landscape values of the site and its surrounds;*
2. *Proximity to dwellings to avoid shadow flicker and visual impacts;*
3. *Ecology of the site;*
4. *Cultural heritage such as the existence of items and places of cultural significance to the Aboriginal and non-Aboriginal community;*
5. *Conservation and recreational uses such as national parks and conservation reserves, as well as sites of international significance;*
6. *Electromagnetic interference;*
7. *Aircraft safety;* and

*8. Restricted areas such as military installations and telecommunications installations.*

### 2.2.3 Site selection for coupling

The site selection factors for each of NPP and WF are illustrated above. The site selection factors for coupling can be divided to three main groups as follows:

The first group consists of similar site selection requirements for both NPP and WF, which can be summarized as:

1. *Proximity to appropriate existing electricity infrastructure;*
2. *Proximity to transport infrastructure:*

   It is important for NPP to facilitate the movement of nuclear fuel, waste and other relevant materials [10]. Also for WF to transport construction components including long ones such as blades and towers;
3. *Sufficient distance from permanent human activities:*

   Sufficient distance from WTs must be kept to minimize impacts from noise, shadow flicker and visual impacts generated by WTs [12, 13]. Also, NPPs should be located in sparsely populated areas that are distant from large population centers to minimize community opposition and security risks and to reduce the complexity associated with emergency planning [9]; and
4. *Siting away from ecological areas, and cultural heritage areas; and did not impact the landscape values of the site and its surrounds.*

The second group consists of the site selection requirements important for NPP alone, which can be summarized as:

1. *Requirement of ultimate heat sink for cooling in NPP steam cycle [10];* and
2. *Siting away from external hazards such as earthquakes, surface faulting, flooding, geotechnical hazards, and other hazards from external human induced events [9].*

The third group consists of the site selection requirements important for WF alone, which can be summarized as:

1. *Requirement for potential wind resource in the site; promising values are average wind speeds above 6 m/s [12];*

2. *Avoidance of electromagnetic interference of WTs with microwave, television, radar or radio;* and
3. *Avoidance of WTs physical obstruction for aviation operations in the site [12, 13].*

Therefore, simple site selection procedure will be as follow:

1. *Prepare a list of NPPs and large thermal power plants sites in the country, already in operation or to be built in the near future;*
2. *Study the wind source potentials in each site and the feasibility of erecting WF, considering the whole technical and economical advantages and disadvantages of the coupling;* then
3. *Select the best sites for coupling which will provide technical and economical benefits than other sites.*

## 2.2.4 Proposed sites for coupling in Egypt

In Egypt, coastal areas are proposed for NPP siting due to the requirement of ultimate heat sink for cooling, which is needed in the steam power cycle. This ultimate heat sink in most cases is the sea water. According to [5] the wind energy potential along the coast of Mediterranean Sea in north Egypt is quite promising and the wind energy potential along the east coast of Red Sea in Egypt is high. Therefore, the coastal areas in Egypt are appealing for both NPP and WF, and the coupling between them is applicable.

According to [14], the selected site for the first NPP in Egypt is El Dabaa site. Besides this site, there are other five sites under evaluation, and they are:

1. *El Negeila East (north-western Mediterranean coast near West Mersa Matrouh)*
2. *El Negeila West (West of the El Negeila East site area)*
3. *Hammam Pharoaun (on the east bank of the Suez Gulf)*
4. *South Safaga (along the Red Sea coast)*
5. *South Mersa Alam (along the Red Sea coast)*

The geographical locations of these sites are shown in Figure 2.1.

Now, the wind resource potentials of each site will be evaluated by using the wind atlas of Egypt shown in Figure 2.2 that can generate Table 2.1.

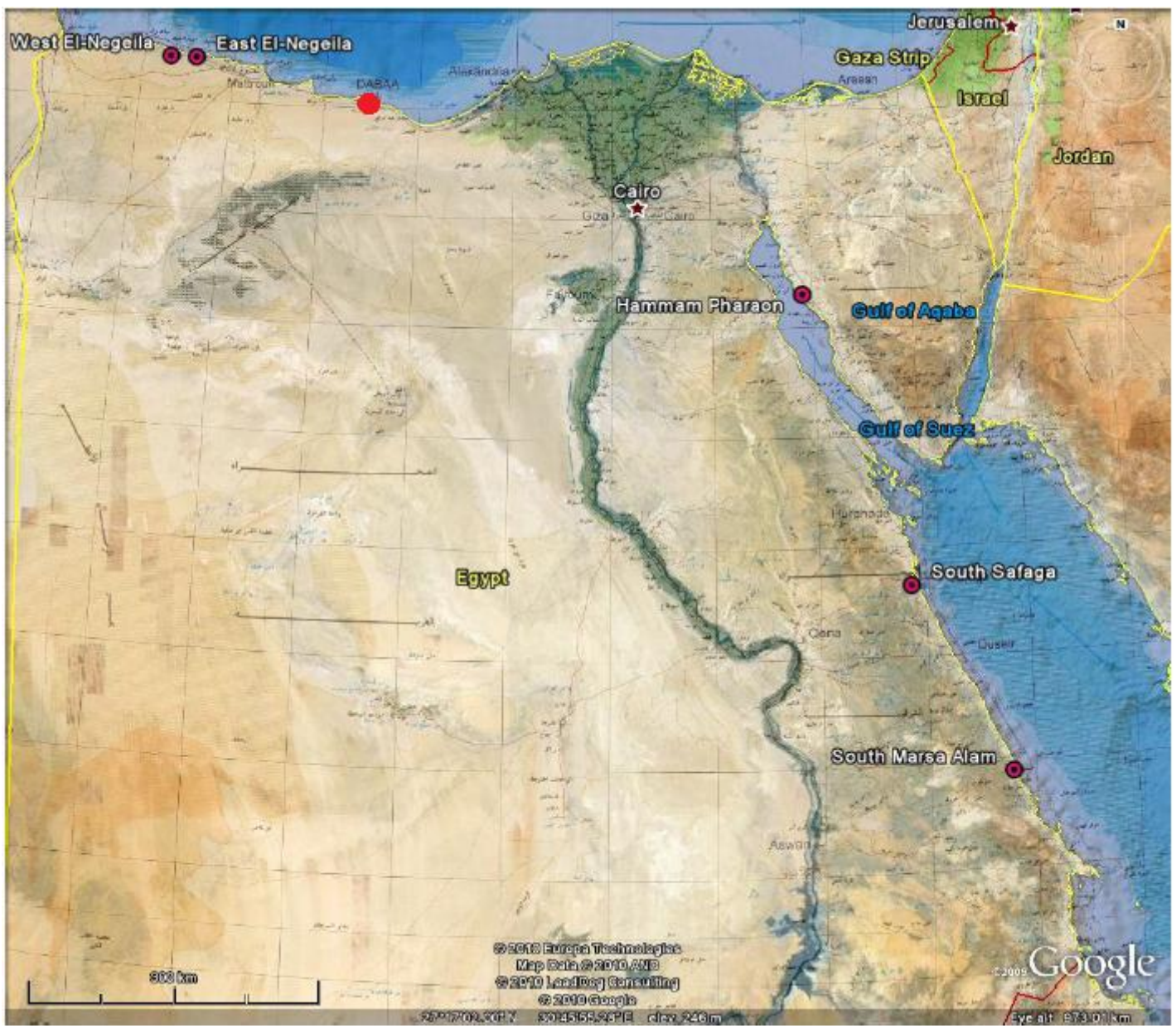

Figure 2.1: Geographical locations for sites under evaluation to accommodate NPP in Egypt [14]

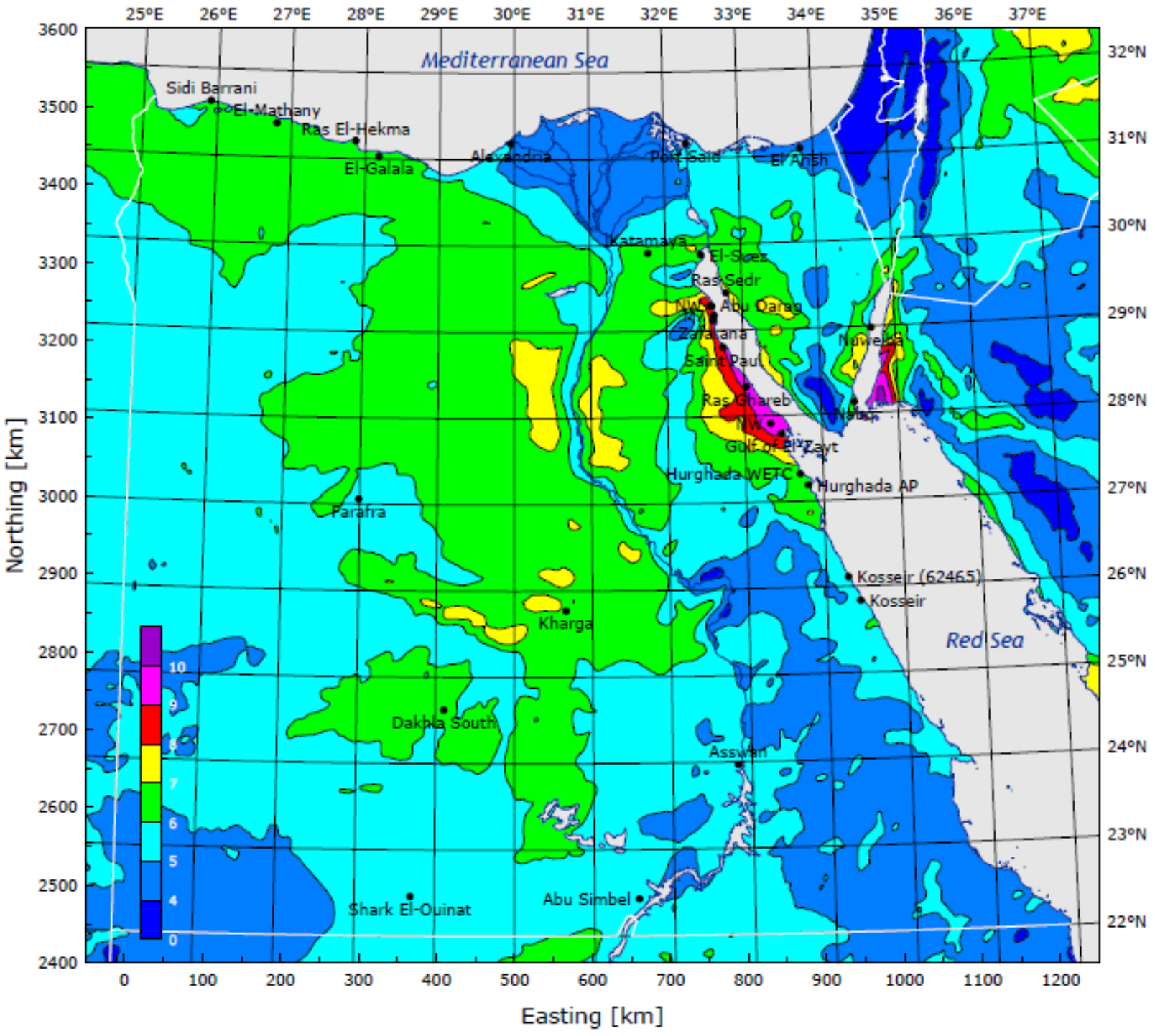

Figure 2.2: Wind Atlas of Egypt [15]

Table 2.1: Average wind speeds in the selected sites at 50 m/s a.g.l.

| Site | Average wind speed m/s |
|---|---|
| ElDabaa | 6-7 m/s |
| El Negeila East | 6-7 m/s |
| El Negeila West | 6-7 m/s |
| Hammam Pharoaun | 6-8 m/s |
| South Safaga | 6-7 m/s |
| South Mersa Alam | 5-6 m/s |

As mentioned before, wind energy can be produced economically only in areas that have average annual wind speeds above 6 m/s at 50-m height [12,16]. So, most of these sites fulfill the main requirements for WF siting.

Among these sites, El Dabaa site will be considered for more illustration. This is due to the available data, and as it is proposed for the first NPP in the country. El Dabaa site extends from (28° 21’ 33’’ E) to (28° 35’ 11’’ E) and from (30° 58’ 50’’ N) to (31° 5’ 22’’ N). According to the Wind Atlas of Egypt [15], the average wind speed is about 6.5 m/sec at the height of 50 m *a.g.l.*. Also, according to [5], the average wind speed is 5.4 m/sec at a height of 10 m *a.g.l.*. Also the author in [5] studied the electricity generation for WF in El Dabaa and concluded that [5]:

1. *The power density obtained from the wind, is ranging from 340 to 425 W/m$^2$ and 150 to 555 W/m$^2$ at the heights of 70–100 m, respectively. This power density is equally as high as the inland potential close to Vinde by (Denmark) and is similar to the power density in European countries;*
2. *The maximum yearly energy gain from a 2 MW WT in El Dabaa is found to be 7975 MWh; and*
3. *The electricity production cost was found to be less than 2 € cent/kWh.*

This study encourages the construction of large WTs with a power level of 2 MW in El Dabaa, where the expected cost of electricity generation was found to be very competitive with the cost of 1 kWh produced by the Egyptian Electricity Authority as shown at the actual tariff system in Egypt [5]. Therefore, one of the best proposed sites for coupling is El Dabaa, which is proposed to accommodate NPP in the near future and has good wind characteristics.

## 2.3 Advantages of Coupling of WF with NPP

### 2.3.1 Good use of the low population density areas around NPPs

Multiple planning areas or zones are required around a NPP. This is important to minimize health and safety risks of the NPPs.

The first zone, called exclusion area (the site of the NPP), extends to approximately one kilometer from the facility. Within this area, permanent settlement is prohibited and the operator of the facility should have authority over all activities carried out in the area [10]. The exclusion distances for most US reactors fall in the range 0.5–1.6 km [17].

The second zone, called Low Population Zone (LPZ), extends to approximately five kilometers from the facility. Development is restricted in this zone to exclude sensitive activities (for example, hospitals) and high density settlements and prevent unsuitable growth in the number of permanent residents [10].

The third zone, called Emergency Planning Zone (EPZ), consists of two zones, a plume exposure pathway EPZ and an ingestion pathway EPZ. They have radii that range from around 16 km in relation to the plume exposure pathway EPZ to 80 km for the ingestion pathway EPZ. Plans are required to be prepared for this area to ensure the evacuation of people in an emergency [10].

These areas around NPP are very good to WF siting if other siting requirement for WF is fulfilled. Erecting WF in these areas will make use of them and ensuring the requirements of these areas such as low population densities and avoidance of sensitive activities.

## 2.3.2 Existence of connection to grid

WFs have to be installed in the immediate proximity of the resource (wind). The best conditions for an installation of wind power can usually be found in remote, open areas with low population densities [16, 18]. Due to low population densities and so low energy consumption after wind energy is harvested; it must be transported/ transmitted to population centers where most energy is used [16].

The power transmission capacity of the electricity supply system usually decreases with falling population density. Areas for WFs are generally located in regions with low population density and with low power transmission capacity [19]. Therefore, WFs are usually located in areas having weak transmission infrastructure, meaning that the transmission lines operate at lower voltages and with higher impedances than stronger parts of the grid. Such lines are poorly suited to accommodate wind power [20]. Therefore, WFs are built in remote areas where the grid reinforcements are more urgent and more expensive than in areas close to industrial loads, where conventional generation is usually situated. Owing to the low utilization rate of the WTs, the energy produced per megawatt of new transmission is low [18]. Also, for WF, there may be significant environmental impact associated with the electrical connection (e.g., the construction of a substation and new circuit). Although this may be dealt with formally as a separate planning application it still needs to be considered [21].

By integrating WF near NPP we avoid all above problems. In NPP site there are at least two substations with different voltage levels (ex: 500KV & 220 KV) and a strong connection to grid due to the large generation capacity of the NPP. So, this solution will not only decrease costs of connecting WF to grid but also have an environmental benefit by avoiding the construction of new electrical connection.

Also, this solution provides the possibility of integrating WF with large installed capacity. Determining the size of a WF which may be connected to a particular point in the electrical network requires a series of calculations based on the specific project data [16, 19, 21]. However, Table 2.2 gives some indication of the maximum capacities which experience has indicated may be connected [16]. From Table 2.2, it can be concluded that WF located around NPP will be connected to transmission lines which will allow large penetration of wind energy in this part of grid.

Table 2.2: Guide to the Amount of Power Carried Through Power Networks [16]

| Electricity Network | Voltage | Current, amps | Power |
|---|---|---|---|
| Transmission | 110-750 kV | ~500 - 1000 | 60 – 1300 MW |
| Primary Distribution | 11-69 kV | ~500 | 10 – 60 MW |
| Secondary Distribution | 400 V -11 kV | 5 - 200 | 0.8 kW – 3.6 MW |

The cost saving of grid connection will be great. The costs for grid connection for WF can be split up in two; the costs for the local electrical installation and the costs for connecting the WF to the electrical grid. The local electrical installation comprises the medium voltage grid in the WF up to a common point and the necessary medium voltage switch gear at that point. Cited total costs for this item ranges from 3 to 10 % of the total costs of the complete WF. The cost for connection to the electrical grid ranges from almost 0% for a small WF connected to an adjacent medium voltage line and upwards. For a 150 MW on-shore WF a figure of 7.5% has been given for this item [19]. A typical breakdown of costs for a 10 MW WF in UK gives 6% of the total cost for the connection to the electrical grid [21]. In case of integrating WF near NPP the cost for connection to the electrical grid is almost zero due to the already existed grid connection.

### 2.3.3 Existence of main facilities required for power plant construction, operation and maintenance

NPP site contains many facilities that will be useful for WF during different periods of construction and operation. There is a strong road network which is used for transporting very heavy weight equipment of NPP. Also, usually there is sea port in the site for transporting any large equipment by ships if required. Large workshop is built for helping in NPP maintenance and operation activities; this will be helpful for WF. Camp for NPP workers housing near the site is also existed.

During WF site selection, in addition to assessing the wind resource it is also necessary to confirm that road access is available, or can be developed at reasonable cost, for transporting the turbines and other equipment. Blades of large WTs can be more than 40 m in length and so clearly can pose difficulties for transport on minor roads. For a large WF, the heaviest piece of equipment is likely to be the main transformer if a substation is located at the site [21].

All above mentioned available facilities and others will be useful for the WF and decrease effectively the capital and running cost, making the site of NPP appealing for WF siting if other required factors are fulfilled.

### 2.3.4 Other benefits of coupling

Remaining benefits of coupling are illustrated in the following chapters, and they can be summarized as follows:

1. Benefits of coupling on the grid integration issues of WF;
2. Increasing reliability and availability of NPP EPSs by the on-site WF; and
3. Geographical distribution of WFs in the grid.

## 2.4 Disadvantages of Coupling of WFs with NPPs

Some of these disadvantages are technical problems which can be overcome by proper solutions, and others are related to public safety as this coupling may increase the hazards and the probability of accidents occurrence in the site.

### 2.4.1 Technical problems which can be overcome by proper solutions

#### 2.4.1.1 High short circuit power level of PCC and WF's protection switchgear

Short circuit level of a point in the grid is used to assess the capability of equipment to make, break and pass the severe currents experienced under short circuit conditions. If the short circuit level is excessive, switchgear may not be able to deal safely with the currents involved [22]. Connecting a WF to the substation where NPP is connected (high short circuit power level) will require the short-circuit duty of the WF switchgear to be high. Excessive short-circuit currents may be encountered and thus require heavier mechanical bracing in the switchgear as well as high-capacity interrupting circuit breakers which supply the system. This high-capacity equipment may be expensive or not commercially available.

#### 2.4.1.2 Noise from WTs

Noise is a form of pollution generated by WTs. However, the impact of sound is limited to a few hundred meters from the base of a WT. Noise is generated in a WT from two primary sources, aerodynamic interactions between the blades and wind and mechanical noise from different parts of the WT [16]. This induced noise from WF adjacent to or around NPP site may have a bad impact on the human activities related to NPP operation.

Strategies to lower the induced noise level include [16]:

1. Use of smaller length blades when installed in noise-sensitive areas;
2. Setting of WT control to lower rotor’s revolutions per minute; and
3. Use of direct-drive WTs, which are quieter because of the absence of gearbox.

#### 2.4.1.3 Shadow flicker and blade glint

Shadow flicker occurs when the sun passes behind the WT and casts a shadow. As the rotor blades rotate, shadows pass over the same point causing an effect termed shadow flicker. Shadow flicker may become a problem when residences are located near, or have a specific orientation to the WF [23]. This effect generally lasts no more than 30 minutes and only appears in very specific situations. More details about this problem are illustrated in [24].

This shadow flicker from WF adjacent to or around NPP site may have a bad impact on the human activities related to NPP operation. Prevention and control measures to address shadow flicker impacts include the following:

1. Site and orient WTs so as to avoid residences located within the narrow bands [25].
2. Turning off particular WTs at certain times [23].

Similar to shadow flicker, blade or tower glint occurs when the sun strikes a rotor blade or the tower at a particular orientation. This can impact a community, as the reflection of sunlight off the rotor blade may be angled toward nearby residences. Blade glint is a temporary phenomenon for new WTs only, and typically disappears when blades have been soiled after a few months of operation. Also, tower glint can be mitigated by painting the WT tower with non-reflective coating to avoid reflections from towers [23].

#### 2.4.1.4 Interference of WTs with the communication systems of NPP

In NPP site many telecommunication systems are used. Locating WF around the site may have bad impact on these communication systems, unless control measures are used. WTs could potentially cause electromagnetic interference with telecommunication systems (e.g. microwave, and radio). The nature of the potential impacts depends primarily on the location of the WT relative to the transmitter and receiver, characteristics of the rotor blades, signal frequency receiver, characteristics, and radio wave propagation characteristics in the local atmosphere [23].

Although, WTs have the potential to interfere with electromagnetic signals that form part of a wide range of modern communication systems, this is normally only a local effect and can usually be remedied at modest cost [21]. Prevention and control measures to address impacts to telecommunication systems include the following [23]:

1. Modify placement of WTs to avoid direct physical interference of point-to-point communication systems.
2. Install a directional antenna.
3. Modify the existing aerial.
4. Install an amplifier to boost the signal.

#### 2.4.1.5 Wind energy power quality aspects and its impact on NPP

Wind energy power quality aspects are described briefly in section 3.3.1. Grid connected WTs may affect power quality. The power quality depends on the interaction between the grid and the WT. Voltage variations, flicker, harmonics, and transients can be produced by WT at PCC. Although integrating WF near NPP (stiff point in the grid) decreases the power quality problems of WF, this may have a side effect on the NPP auxiliaries and therefore on the NPP operation. This disadvantage can be solved by proper technical solutions, which will be selected according to technical studies and simulations for case by case of each problem.

#### 2.4.1.6 Lower energy production from WF

Energy plan in Egypt is to site WF in sites of highest wind speeds in the country so as for highest energy generation for the installed capacity. On the other hand, idea of coupling mainly depends on siting WF around or near NPP or large fossil-fueled power plant, and this site in most cases has lower wind speeds compared to other sites of best wind speeds in the country. This means lower energy production for the same installed WF capacity.

This disadvantage can be compensated by other technical and economic benefits of the coupling which are illustrated in this thesis. So, a detailed techno-economical study considering different technical and economic factors of the two choices should be conducted to get the right choice. Also, using taller blades and towers for WTs can increase the total energy production of the WF and so overcome the problem.

### 2.4.2 Accident hazards from WF

WF accidents may have no major impact in site containing WF alone. In case of coupling, part of WTs are located near NPP building and human activities in the plant. Therefore, the occurrence of accident may lead to major consequences on the humans and / or the operation of the NPP. So, before the construction of WF, these hazards must be considered carefully and proper measures must be taken to minimize the possibility of occurrence of these accidents.

The wind energy industry enjoys an outstanding health & safety record [25]. However, it involves mechanical hazards characteristic of rotating machinery, as well as electrical hazards typical of electrical production equipment. To minimize such risks an offset from human normal activities is adopted according to different local, national and international norms, regulations and laws [26].

#### 2.4.2.1 Fire hazards

The risk of fire at WFs, or the risk of fire damage to WT generators, is very low. WTs manufactured today incorporate the highest quality and safety standards, but the potential for a fire always exists when electronics and flammable oils and hydraulic fluids exist in the same enclosure. With normal maintenance and servicing practices in place, a WF will not impose an increased fire hazard to the host community. Fires due to machine failure in modern WTs are extremely rare. Cases of fire damage to land neighbouring WFs are practically nonexistent [25].

#### 2.4.2.2 Unique industrial risks

The blade system in a WT has incorporated in it mechanisms designed to avoid excessive rotational speeds, including blade feathering and hydraulic and friction braking. Blade failure could still occur ejecting the blade away for a distance from the machine. Stress and vibration could cause tower failure. Failure mode analysis and safety zones around WT must be used at the design and site selection stage [26].

Several potential major types of accidents can be considered for WTs:

1. Turbine tower failure, but a properly constructed WT poses negligible risk of this type [27].
2. Blade failure and ejection, which may affect public safety, although overall risk of blade throw is extremely low [23, 25]. There are some blade throw management strategies illustrated in [23] & [26].

#### 2.4.2.3 Aviation operations hazards (Aircraft safety)

Near NPP site, usually there is an airport for servicing the NPP operating staff long transportations and required urgent maintenance jobs. There are basically two ways in which the construction of a WT or WF may impact upon aviation operations [28]:

1. The physical obstruction caused by a tall structure:

   In addition to the hazard posed to aircraft in approaching or departing from an airfield, WTs can also pose a potential danger to aircraft flying at low level for any other reason [26, 28]. The Obstacle Limitation Surface (OLS) is a series of surfaces that set the height limits of objects around an airport. The height of each WT will need to be individually reviewed to check if any part of the machine will protrude through the OLS surface [25].

2. The effects on communications, navigation and surveillance systems:

   WTs could potentially cause electromagnetic interference with aviation radar and telecommunication systems (e.g. microwave, television, and radio) [23]. Prevention and control measures to address this impact are illustrated in [23] and [25].

# Chapter 3

# Connecting WF and NPP to the Same Point in the Grid

In this chapter, benefits of coupling on the grid integration issues of WFs are presented. A case study is used to verify the benefit of coupling on voltage quality at PCC of WF. After that, the benefit of increasing reliability and availability of NPP EPSs by the on-site WF is discussed. A study using MARKOV process is conducted for verification and quantification of this benefit.

## 3.1 Benefits of coupling on the grid integration issues of WF

Based on a number of parameters, such as grid stiffness, transmission voltage levels, WT topology, Transmission System Operators (TSOs) issued a number of requirements for WTs to fulfill in order to get grid connection agreement. These requirements were called grid codes, and they became more demanding as wind penetration levels grew [29]. Although, the grid codes regarding wind power are country specific, the grid codes of most countries generally aim to achieve the same thing. Electricity networks are constructed and operated to serve a huge and diverse customer demographic [30].

### 3.1.1 Voltage quality at the coupling point of WF

Grid integration of WF affects the power quality. The power quality depends on the interaction between the grid and the WT. Figure 3.1 shows a classification of different power quality phenomena [18]. The scope here will be on the voltage quality aspects. In grid codes a set of requirements are imposed to cope with problems arising due to rapid voltage changes/jumps, voltage flicker and harmonics.

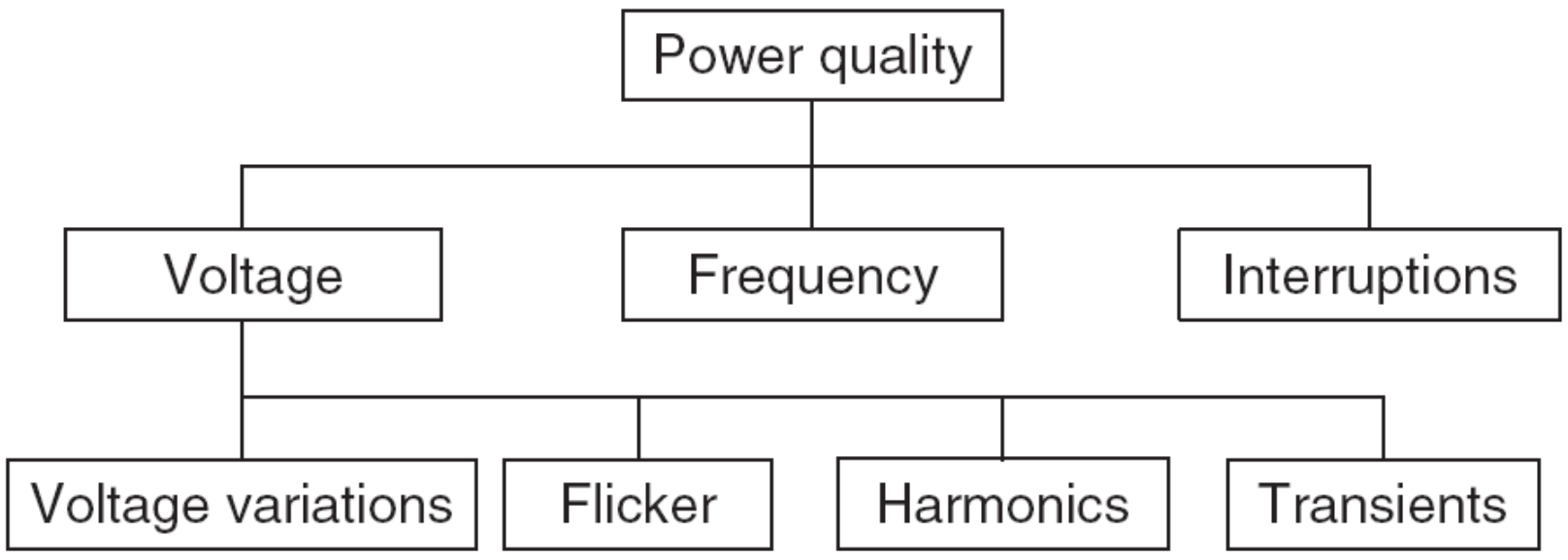

Figure 3.1: Classification of different power quality phenomena

The voltage quality aspects are [18]:

1. Voltage variations: which are changes in the RMS value of the voltage during a short period of time, mostly few minutes. If wind power is introduced, voltage variations also emanate from the power produced by the WT.
2. Flicker: it is the traditional way of quantifying voltage fluctuations. The method is based on measurements of variations in the voltage amplitude (i.e. the duration and magnitude of the variations). Flicker from WTs occurs in two different modes of operation: continuous operation (due to power fluctuations) and switching operations.
3. Voltage harmonics: they are virtually always present in the utility grid. Nonlinear loads, power electronic loads, and rectifiers and inverters in motor drives are some sources that produce harmonics. Fixed-speed WTs are not expected to cause significant harmonics and inter-harmonics. Variable-speed WTs equipped with a converter produce harmonic currents during continuous operation.
4. Transients: they seem to occur mainly during the startup and shutdown of fixed speed WTs. The transients may reach a value of twice the rated WT current and may substantially affect the voltage of the low-voltage grid. The voltage transient can disturb sensitive equipment connected to the same part of the grid.

#### 3.1.1.1 Main requirement for WF grid integration

Short circuit capacity ($S_{SC}$) is the product of the pre-fault voltage (or rated voltage) and the current which would flow if a three-phase symmetrical fault were to occur (Eq. 3.1). $S_{SC}$ is a useful parameter which gives an immediate understanding of the capacity of the circuit to deliver fault current and resist voltage variations [21, 31]. The short circuit capacity in a given point in the electrical network is a measure of its strength that has a heavy influence [19, 21, 31]. The ability of the grid to absorb disturbances is directly related to the short circuit capacity [19].

$S_{sc}$ (MVA) = √3 × line voltage (KV) × short circuit current (KA) (3.1)

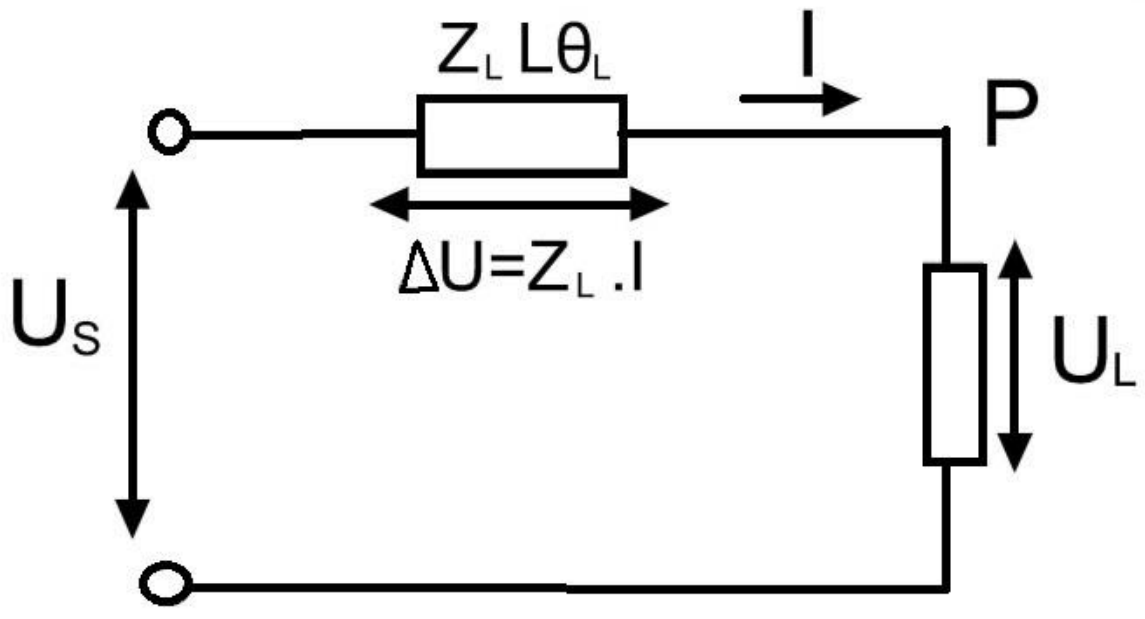

Figure 3.2: Equivalent circuit [19]

Any point (P) in the network can be modeled as an equivalent circuit as shown in Figure 3.2. Far away from the point the voltage can be taken as constant ($U_S$) i.e. infinite bus. The short circuit capacity $S_{SC}$ in MVA can be found using Eq. 3.1 as $U_S^2$ / $Z_L$ where $Z_L$ is the magnitude of the line impedance. Variations in the load (or production) in (P) causes current variations in the line and these in turn a varying voltage drop (ΔU) over the line impedance $Z_L \angle \theta_L$. It is obvious from Figure 3.2, that if the impedance $Z_L$ is small then the voltage variations at (P) will be small (the grid is strong) and consequently, if $Z_L$ is large, then the voltage variations will be large (the grid is weak) [18, 19].

Stiff grid and weak grid are relative terms with respect to installation [19], so it is important to compare the size of equipment to the strength of the power system. A simple comparison is to divide the system strength by the equipment capacity, which is named the Short Circuit Ratio ($R_{SC}$) [19, 31]. A high $R_{SC}$ means good performance and a low $R_{SC}$ means trouble [31].

$$R_{SC} = \frac{S_{SC}}{P_E} \qquad (3.2)$$

Where:

$P_E$: Equipment installed capacity (MW)

Another characteristic grid measure is the grid impedance angle at PCC (Ψk) [32]:

$$\Psi_K = Arctan\left(\frac{X_L}{R_L}\right) \qquad (3.3)$$

Where:

in Figure 3.2: $Z_L \llcorner \theta_L = R_L + j X_L$

Ψk: Grid impedance angle at PCC;

$R_L$: Short circuit resistance of the grid

$X_L$: Short circuit reactance of the grid

The main requirement for grid integration of WF is the limitation of voltage deviation caused by the WTs at the PCC, for which 2% of nominal voltage is a commonly established limit. The short circuit power level at PCC is the crucial value for the permissible installed power ratings [33]. The grid is strong with respect to the WF if $R_{SC}$ is above 20, so the variations in voltage are minor and predictable. The grid is weak with respect to the WF if $R_{SC}$ is below 10 [16, 19].

#### 3.1.1.2 Fulfillment of the main requirement for WF grid integration using Coupling

High short circuit capacity at the PCC allows integration of large WF without negative effect on the voltage quality. The NPP PCC with the grid has very high short circuit level. The short-circuit level at PCC decreases, as the distance of PCC from the substation increases [16]. So integrating WF near NPP will have a positive impact on voltage quality instead of integrating it in remote area with low short circuit level.

The Egyptian Energy plan is to locate WFs in areas with the highest wind speeds in the country to maximize the energy yield of the installed capacity. This leads to locating WFs in the same area on the coast of Suez Bay, which may has lower short circuit capacity compared to the PCC of NPP with the grid. In this thesis, it is proposed to locate part of the future planned WFs near NPP if wind resource is promising. This coupling will allow integration of large capacity WF without negative effect on voltage quality due to the high short circuit capacity existed at the PCC.

#### 3.1.1.3 Case study illustrating the effect of short circuit capacity on the voltage quality aspects of WF

This study is conducted to illustrate the positive impact on voltage quality aspects when integrating WFs near NPPs or large thermal power plants, if the wind resource is good. PCC voltage profile of two choices will be compared, these choices are as follows:

1. Locating WFs as it is in the Egyptian Energy plan, which is based on maximizing energy yield of installed capacity. This means locating them on the shore of Suez Bay where short circuit power level may be small compared to other points in the grid;
2. Locating part of the future planned WFs, through coupling, near NPP, if wind resource is promising in areas such as El Dabaa.

The assessment is performed according to the methods given in the IEC 61400-21 [34], which can be summarized as follows:

*Steady-State voltage change:*

The steady-state voltage change for one WT is given by Equation (3.4), and for the whole WF the steady state voltage change can be estimated by Equation (3.5). $S_{60}$ and Ø can be calculated from $P_{60}$ and $Q_{60}$ [19, 34].

$$d_{SS} = \frac{S_{60}}{S_{SC}} \times \{\cos(\Psi_K + \emptyset)\} \quad \text{Only valid for } \cos(\Psi_k+\emptyset) > 0.1 \tag{3.4}$$

$$d_{SS\Sigma} = N_{WT} \times \frac{S_{60}}{S_{SC}} \times \{\cos(\Psi_K + \emptyset)\} \quad \text{Only valid for } \cos(\Psi_k+\emptyset) > 0.1 \tag{3.5}$$

Where:

dss: Steady state voltage change of the grid at PCC due to one WT (normalised to nominal voltage);

$dss_{\Sigma}$: Steady state voltage change of the grid at PCC due to the WF (normalised to nominal voltage);

$S_{60}$: Apparent power at the 1-min. active power peak;

Ssc: Short circuit power level of the grid at PCC;

$\Psi_k$: Grid impedance angle at PCC;
*Ø:* Phase angle between voltage and current; and
$N_{WT:}$ Number of WTs in the WF.

*Flicker distortion:*

The flicker distortion for continuous operation of one WT can be calculated from Equation (3.6), and for the whole WF the flicker emission can be estimated by Equation (3.7) [19, 34].

$$P_{lt} = C(\Psi_k, V_a) \times \frac{S_n}{S_{SC}} \tag{3.6}$$

$$P_{lt\sum} = \sqrt{N_{WT}} \times C(\Psi_k, V_a) \times \frac{S_n}{S_{SC}} \tag{3.7}$$

Where:
$P_{lt}$: Flicker distortion (emission) of one WT;
$P_{lt\sum}$: Flicker distortion (emission) of the whole WF;
c($\Psi_k$, Va): Flicker coefficient;
Va: Annual average wind speed at hub height; and
$S_n$: Apparent power of the WT at rated power.

*Switching operations:*

For switching operations two criterions must be checked: the voltage change due to the inrush current of a switching and the flicker effect of the switching. The control system of a WF ensures that two or more WTs in a WF are not switched on simultaneously. Therefore, only one WT has to be taken into account for the calculation of the voltage change. The worst case of switchings concerning the voltage change is the cut-in of the WT at rated wind speed, and can be calculated using Equation (3.8) [19, 34].

$$d_{SO} = K_u(\Psi_k) \times \frac{S_n}{S_{SC}} \tag{3.8}$$

Where:

dso: Switching operations voltage change of the grid at PCC (normalised to nominal voltage); and

$k_u(\Psi_k)$: Voltage change factor.

On the other hand, the flicker effect has to be calculated for both types of switching: for the cut-in at cut-in wind speed and for the cut-in at rated wind speed. The flicker emission due to switching operations of a single WT can be estimated from Equation (3.9), and Equation (3.10) can be used for the whole WF. The flicker effect has to be calculated for both types of switching: for the cut-in at cut-in wind speed and for the cut-in at rated wind speed [19, 34].

$$P_{lt} = 8 \times (N_{120})^{0.31} \times K_f(\Psi_k) \times \frac{S_n}{S_{SC}} \tag{3.9}$$

$$P_{lt\Sigma} = N_{WT} \times 8 \times (N_{120})^{0.31} \times K_f(\Psi_k) \times \frac{S_n}{S_{SC}} \tag{3.10}$$

Where:

$P_{lt}$: Flicker distortion (emission) of one WT;

$P_{lt\Sigma}$: Flicker distortion (emission) of the whole WF;

$k_f(\Psi_k)$: Flicker step factor; and

$N_{120}$: Number of switchings within a 120-minute period.

Definitions for different factors and coefficients mentioned above can be found in details in [34]. Also, methods to calculate their values and some examples are provided in [34].

Two types of WTs are selected, Type A and Type D. Description of different WT's Types and the configuration of each type can be found in [18]. WTs are selected as typical representatives of solutions and developmental directions in wind technology. Table 3.1 shows the data for the selected sites. Tables 3.2 and 3.3 show the data for the used WTs. Four scenarios will be compared, with the details illustrated in Table 3.4. Appling Equations (3.4) to (3.10), to all scenarios, results given in Table 3.5 can be obtained. Short circuit power level of the grid at PCC has a significant effect on the calculated parameters for both types of WTs used (Type A and D). As, the short circuit power level of the grid at PCC increases, all these parameters decrease, and so better voltage quality at PCC.

Table 3.1: Data of the proposed sites

| | El Dabaa | Zafarana |
|---|---|---|
| Annual average wind speed (m/s) | 7.5 | 8.5 |
| Nominal voltage of the grid (KV) | 500 | 220 |
| Short circuit power level of the grid (MW), Ssc | 1000 | 600 |
| Grid impendance angle (°), ($\Psi_k$) | 85° | 50° |

Table 3.2: Data of the used Type A WT [19]

| Wind Turbine Type | Type A |
|---|---|
| Hub height (m) | 50 |
| Rated power (KW), $P_n$ | 600 |
| Rated apparent power (KVA), $S_n$ | 607 |
| Rated voltage (V), $U_n$ | 690 |
| Rated current (A), $I_n$ | 508 |
| Max. power (KW), $P_{60}$ | 645 |
| Max. Reactive power (KVAR), $Q_{60}$ | 114 |

| Flicker: | | | | |
|---|---|---|---|---|
| Grid impedence angle $\Psi_k$ (°): | 30° | 50° | 70° | 85° |
| Annual average wind speed $V_a$ (m/s): | Flicker coefficient, c($\Psi_k$, Va): | | | |
| 6.0 | 7.1 | 5.9 | 5.1 | 6.4 |
| 7.5 | 7.4 | 6.0 | 5.2 | 6.6 |
| 8.5 | 7.8 | 6.5 | 5.6 | 7.2 |
| 10.0 | 7.9 | 6.6 | 5.7 | 7.3 |

| Switching operations: | | | | | | | | |
|---|---|---|---|---|---|---|---|---|
| Case of switching operation: | Cut-in at cut in wind speed | | | | Cut-in at rated wind speed | | | |
| Max number of switchings $N_{10}$ | 3 | | | | 1 | | | |
| Max number of switchings $N_{120}$ | 30 | | | | 8 | | | |
| Grid impedence angle $\Psi_k$ (°) | 30° | 50° | 70° | 85° | 30° | 50° | 70° | 85° |
| Flicker step factor $k_f(\Psi_k)$ | 0.35 | 0.34 | 0.38 | 0.43 | 0.35 | 0.34 | 0.38 | 0.43 |
| Voltage change factor $k_u(\Psi_k)$ | 0.7 | 0.7 | 0.8 | 0.9 | 1.30 | 0.85 | 1.05 | 1.60 |

Table 3.3: Data of the used Type D WT [35]

| Wind Turbine Type | Type D |
|---|---|
| Hub height (m) | 85 |
| Rated power (KW), $P_n$ | 2000 |
| Rated apparent power (KVA), $S_n$ | 2000 |
| Rated voltage (V), $U_n$ | 690 |
| Rated current (A), $I_n$ | 1660 |
| Max. power (KW), $P_{60}$ | 2011 |
| Max. Reactive power (KVAR), $Q_{60}$ | 14.3 |

| Flicker: | | | | |
|---|---|---|---|---|
| Grid impedence angle $\Psi_k$ (°): | 30° | 50° | 70° | 85° |
| Annual average wind speed $V_a$ (m/s): | Flicker coefficient, c($\Psi_k$, Va): | | | |
| 3.0 | 3 | 3 | 2 | 2 |
| 6.0 | 3 | 3 | 2 | 2 |
| 9.0 | 3 | 3 | 2 | 2 |
| 12.0 | 3 | 3 | 2 | 2 |
| 15.0 | 3 | 3 | 2 | 2 |

| Switching operations: | | | | | | | | |
|---|---|---|---|---|---|---|---|---|
| Case of switching operation: | Cut-in at cut in wind speed | | | | Cut-in at rated wind speed | | | |
| Max number of switchings $N_{10}$ | 10 | | | | 1 | | | |
| Max number of switchings $N_{120}$ | 120 | | | | 12 | | | |
| Grid impedence angle $\Psi_k$ (°) | 30° | 50° | 70° | 85° | 30° | 50° | 70° | 85° |
| Flicker step factor $k_f(\Psi_k)$ | 0.1 | 0.1 | 0.1 | 0.1 | 0.1 | 0.1 | 0.1 | 0.1 |
| Voltage change factor $k_u(\Psi_k)$ | 0.2 | 0.1 | 0.1 | 0.0 | 0.9 | 0.7 | 0.3 | 0.1 |

Table 3.4: Scenarios under study

| Site | El Dabaa | | Zafarana | |
|---|---|---|---|---|
| WT Type | Type A | Type D | Type A | Type D |
| Installed Capacity | 333*0.6 MW<br>= 199.8 MW | 100*2 MW<br>= 200 MW | 333*0.6 MW<br>= 199.8 MW | 100*2 MW<br>= 200 MW |
| Scenario | Scenario 1 | Scenario 3 | Scenario 2 | Scenario 4 |

Table 3.5: Calculation results for voltage quality aspects for all scenarios

| Site | El Dabaa | Zafarana | El Dabaa | Zafarana |
|---|---|---|---|---|
| WT Type | Type A | Type A | Type D | Type D |
| Scenario | Scenario 1 | Scenario 2 | Scenario 3 | Scenario 4 |
| $V_a$ (m/s) | 7.5 | 8.5 | 7.5 | 8.5 |
| $S_{60}$ (MVA) | 0.655 | 0.655 | 2.01105 | 2.01105 |
| $S_{n,i}$ (MVA) | 0.607 | 0.607 | 2 | 2 |
| $S_{SC}$ (MVA) | 1000 | 600 | 1000 | 600 |
| Ø (°) | 10 | 10 | 0.41 | 0.41 |
| $\Psi_k$ (°) | 70 | 50 | 70 | 50 |
| Number of WTs | 333 | 333 | 100 | 100 |
| Steady-state voltage change ($d_{ss}$) | | | | |
| dss (%) (One WT) | 0.00011 | 0.00055 | 0.00067 | 0.00214 |
| $dss_{\sum}$ (%) (100 WT) | 0.03788 | 0.18176 | 0.06743 | 0.21360 |
| Flicker distortion ($P_{lt}$) (%) | | | | |
| $C(\Psi_k, V_a)$ | 5.2 | 6.5 | 2 | 3 |
| $P_{lt}$ (%) | 0.003 | 0.007 | 0.004 | 0.010 |
| $P_{lt\sum}$ (%) | 0.058 | 0.120 | 0.040 | 0.100 |
| Switching Operations: | | | | |
| Voltage change due to cut-in at rated wind speed ($d_{so}$) | | | | |
| $k_u(\Psi_k)$ | 1.05 | 0.85 | 0.3 | 0.7 |
| $d_{SO}$ (%) | 0.00063735 | 0.0008599 | 0.0006 | 0.00233333 |
| Flicker distortion due to cut-in at cut-in wind speed ($P_{lt}$ ) (%) | | | | |
| $N_{120}$ | 30 | 30 | 120 | 120 |
| $k_f(\Psi_k)$ | 0.38 | 0.34 | 0.1 | 0.1 |
| $P_{lt}$ (%) | 0.005 | 0.008 | 0.007 | 0.012 |
| $P_{lt\sum}$ (%) | 1.764 | 2.630 | 0.706 | 1.176 |
| Flicker distortion due to cut-in at rated wind speed ($P_{lt}$ ) (%) | | | | |
| $N_{120}$ | 8 | 8 | 12 | 12 |
| $k_f(\Psi_k)$ | 0.38 | 0.34 | 0.1 | 0.1 |
| $P_{lt}$ (%) | 0.004 | 0.005 | 0.003 | 0.006 |
| $P_{lt\sum}$ (%) | 1.171 | 1.746 | 0.346 | 0.576 |

## 3.1.2 Active power / frequency control capability

To provide frequency control, the WTs must operate below the potential output level for the current wind conditions. For All types of WTs, load-frequency control can be obtained by slightly increasing the nominal blade pitch angle, deloading the WT by a corresponding amount. Thus, the WT output can be adjusted in sympathy with frequency variations, akin to governor control on a conventional generator. Such an approach is feasible with conventional pitch control. For variable-speed WTs, power output regulation can also be achieved by varying the rotor speed – a small decrease in rotational speed away from the optimum tip-speed ratio will cause a reduction in electrical output [22]. So, the WT pitch angle could be adjusted for partial output, while maintaining a reserve margin of, say, 1–3 % of rated output (delta control). There are other control strategies that can be implemented which are illustrated in [22].

### 3.1.2.1 Benefits of coupling on the active power / frequency control capability of the WF

A grid employing large amounts of wind energy requires the balance of the system to be highly flexible to respond to the increased variability of the net load. Therefore, when wind energy penetration in the power system becomes high, the use of storage units becomes interesting. At the same time, storage units may ensure power delivery from the WF, even when the wind speed is low. Consequently, storage units can provide large WFs with power regulation capabilities, which can enable WFs to replace, and not just to supplement, other power plants. Besides a capacity reduction of other power plants, such regulation capabilities may further reduce the requirement for spinning reserve and thus reduce the total fuel costs in the power system [32].

On the other hand, NPPs generally provide constant, base load power and are most economic when operated at constant power levels. Operating NPPs in load-following modes decreases the plants' annual energy output and increases the levelized cost of energy. By coupling energy storage facility to NPPs, the reactor remains at nearly constant output, while cycling the energy storage facility in response to the variability of the net load. Meanwhile, the added energy storage facility can work as onsite power source to supply safety related loads during Loss of Off-site Power (LOOP) and Station

Black Out (SBO) accidents, which increases the capacity and availability of the onsite power supply. Finally, the energy storage facility can store the energy spilled through the NPP turbine by-bass valves.

Therefore, the addition of energy storage facility to the system will have technical and economical benefits for both WF and NPP. Different energy storage technologies are illustrated in [33]. Some research works were done in the coupling of hydrogen storage facility to wind-NPP system, as in [4, 36, 37]. Hydrogen can be generated from part of the generated heat in the reactor core of NPP and from electrolyzers which are supplied from WF. Also, NPP already has in its electrical system a large sets of Lead-Acid battery banks. These banks can be integrated to the energy storage facility.

### 3.1.3 Reactive power / voltage control capability

WFs that are large enough to be connected to the transmission system tend to be erected in remote areas or offshore because of their dimension and impact on the scenery. Given that the node voltage is a local quantity, it can be difficult to control the voltage at these distant places by use of conventional power stations elsewhere in the grid. Therefore, WTs have to have reactive power / voltage control capabilities. The reactive power / voltage control capabilities of WTs erected in large-scale WFs and connected to the transmission network become increasingly important [18]. The exchange rate and level of reactive power with the grid is specified by the TSO [29]. Utilities around the world have started requiring WFs to maintain a power factor of 0.95 or higher [16]. WF reactive power requirements stated by US Federal Energy Regulatory Commission are illustrated in [16]. Danish grid code requirements for WF reactive power are given in [29]. TSO in Germany gives requirements that must be fulfilled by any power plant to be connected to grid regardless of its type. These requirements are presented in Figure 3.3. This figure shows one of the three possible variants of TSO demands. TSO selects one of the versions on the basis of relevant network characteristics [30].

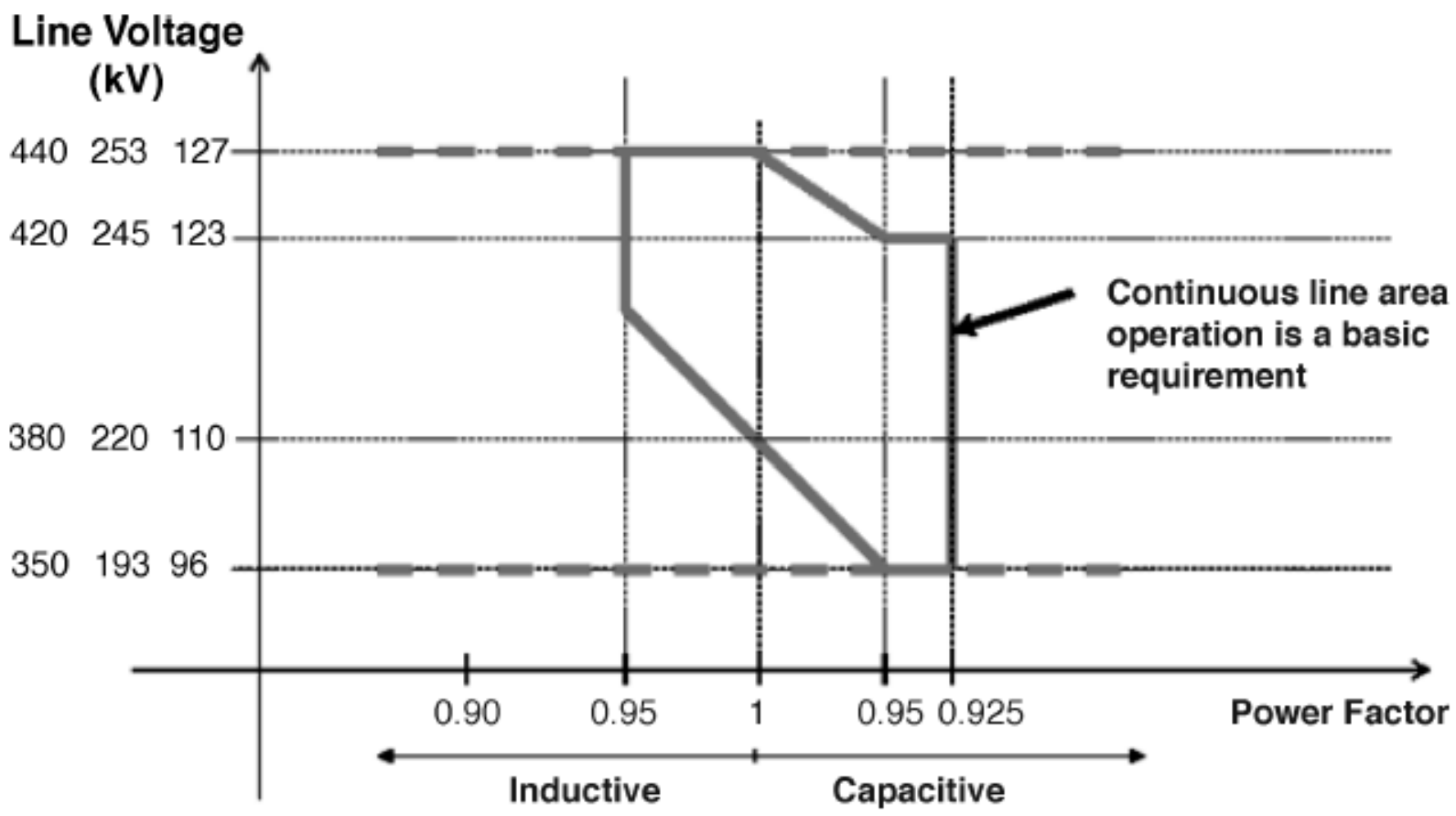

Figure 3.3: Reactive power control requirement as a function of grid voltage (EON Netz GmbH) [30]

#### 3.1.3.1 Benefits of coupling on reactive power / voltage control capability of WF

There are many solutions for supplying reactive power within WF. Some of them are: increasing converter size for Types C and D WTs [18], provision of centralized versus distributed reactive compensating mechanically switched technologies, and rotating synchronous compensators. Vestas turbines have estimated that WTs with a reactive power capability corresponding to a power factor of 0.9 at rated active power output incur the costs shown in Table 3.6 over WTs constructed for unity power factor [22].

Table 3.6: Cost implications of WT power factor capability of 0.9 [22]

| Method | Cost increase (%) |
|---|---|
| SVC at MV | 100 |
| Converter resizing | 41-53 |
| Passive elements at LV | 47 |

By coupling of WF with the adjacent NPP, the NPP can carry the task of reactive power/ voltage support perfectly. This is due to the large capacity synchronous generator used in NPPs which has a capability to supply large amounts of reactive power in cheap costs and with high reliability and availability. For WF utilizing Type A or B WTs, NPP will supply reactive power for the WF collecting system, WTs themselves, and voltage support to PCC. This will achieve the grid requirements of reactive power / voltage control by WF. Also, avoiding the use of switched capacitors at each WT and so transients associated with them. For WF utilizing Type C or D WTs, each WT will work at unity power factor at the low voltage side of its step up transformer, which connects it to the collecting system. NPP will supply reactive power for the WF collecting system, and carry the task of voltage support at PCC. Therefore, grid requirements of reactive power / voltage control by WF is achieved without increasing the converters rating, and so lower cost.

## 3.2 Increasing Reliability and Availability of NPP EPSs by the on-site WF

The safe and economic operation of a NPP requires the plant to be connected to an electrical grid system that has adequate capacity for exporting the power from the NPP, and for providing a reliable electrical supply to the NPP for safe startup, operation and normal or emergency shutdown of the plant. An important characteristic of all NPPs is that after a nuclear reactor is shut down, it continues to produce a significant amount of heat for an extended period. Hence the reactor cooling systems must continue to operate for several days after a reactor shuts down, to prevent overheating and damage to the reactor core. Therefore, reliable cooling arrangements must be provided, and this requires robust and diverse sources of reliable electrical supply [38]. This reliable electrical supply comes either from the grid (off-site power), or from on-site emergency back-up power, such as batteries, diesel generators or gas turbines [39].

The transmission system is the source of power to the offsite power system, and it has higher availability and reliability than the on-site emergency power system because of the diverse and multiple generators connected to the transmission system. Hence NPPs generally consider offsite power as the primary source (preferred source) of power for cooling down the reactor during normal and emergency shutdowns [38]. The reliability of off-site power is usually assured by two or more physically independent transmission circuits to the NPP to minimize the likelihood of their simultaneous failure. Similarly, the reliability of on-site power is enhanced by sufficient independence, redundancy and testability of batteries, diesel generators, gas turbines and the on-site electric distribution systems to perform safety and other functions even if a single failure occurs [39].

The full electrical load of the auxiliaries of a NPP is typically 5–8% of the NPP rated load. Hence the electrical connection to the NPP must be able to supply this load during reactor startup, and immediately after reactor shutdown, whether from a planned shut down or an unplanned reactor trip that may occur at any time [38].

### 3.2.1 NPP electrical system

The typical structure of NPP electrical systems is illustrated in Figure 3.4. The onsite power system is the name of the electrical components and switchgears in the system, which is supporting the process system components with electric power. All the station auxiliaries are divided into two main groups. First is the non class 1E loads, and they should have two independent power sources with automatic switchover provisions. They allow power interruption during automatic switchover and may not require power after reactor scram. The second is the EPSs, with a simplified single line diagram shown in Figure 3.5. The purpose of the EPSs is to provide the plant with the necessary power in all relevant conditions within the design basis so that the plant can be maintained in a safe state after postulated initiating events, in particular during the LOOP accident [38]. The EPSs can be powered from the normal power supply or (optionally) from the alternative on-site power supplies as shown in Figure 3.5 [40].

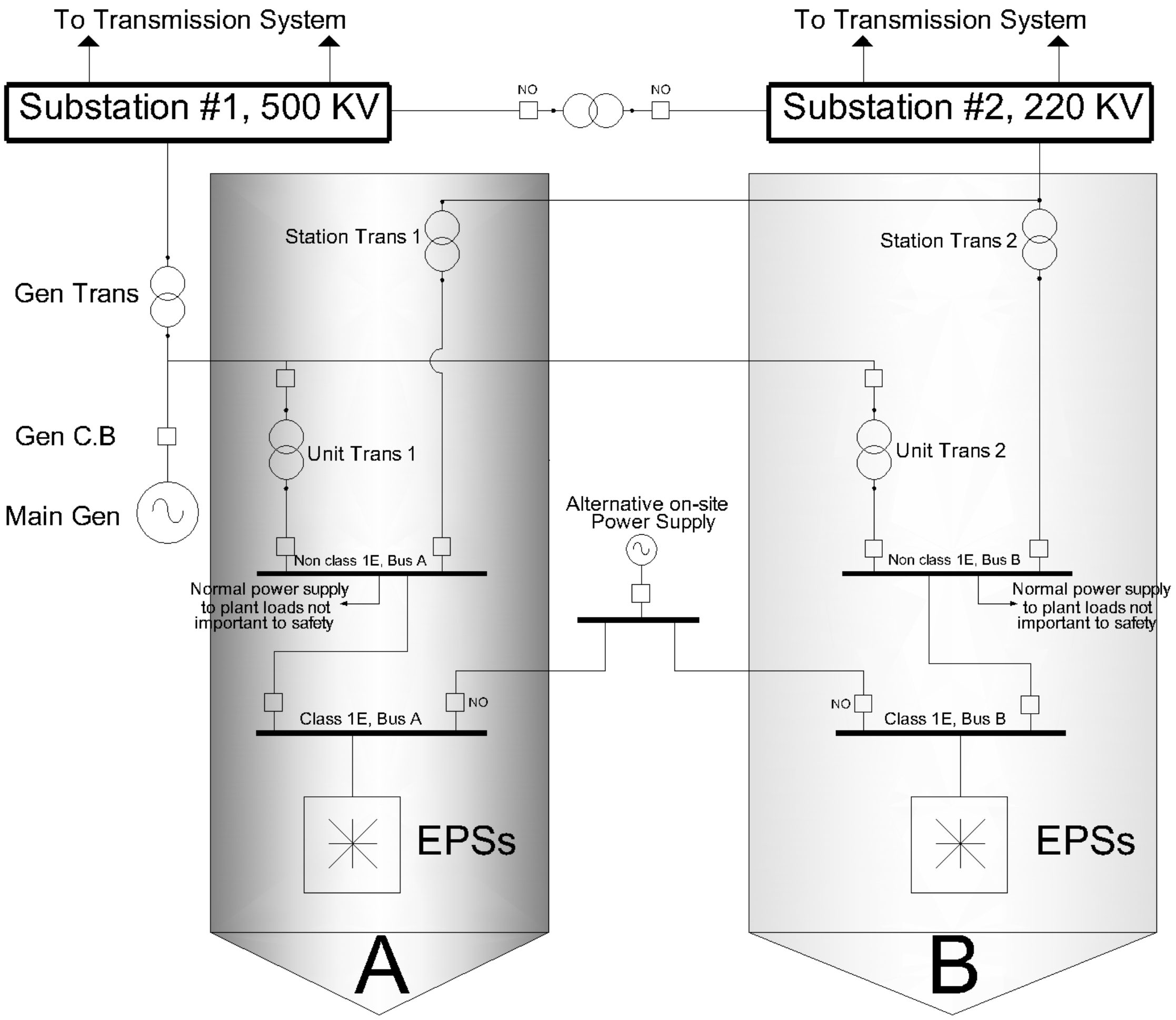

Figure 3.4: Simplified single line diagram of the NPP electrical system [40]

EPSs are generally divided into three systems as follows [40]:

1. An AC power system: the assigned AC loads allow a certain interruption of the power supply.
2. A DC power system: that supplies DC loads without interruption from a battery. This DC system includes a battery charger that is connected to the AC system of the EPSs.
3. A non-interruptible AC power system: that is supplied by the DC power system of the EPSs by means of inverters and is also connectable to the AC power system of the EPSs. The most important components of the safety systems belong to this group.

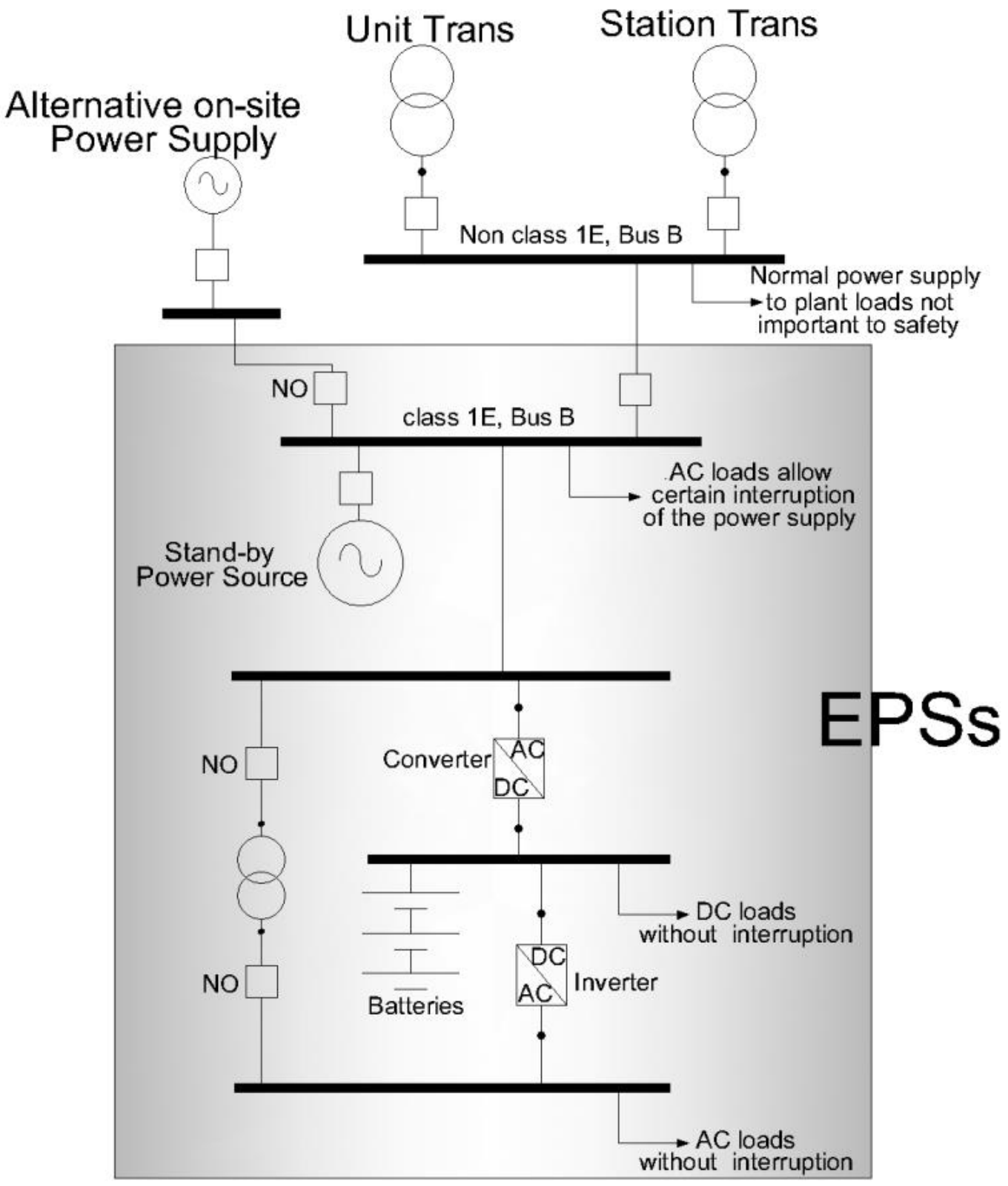

Figure 3.5: A simplified single line diagram of EPSs block [40]

### 3.2.2 The benefit of adjacency of WF & NPP on the NPP electrical system

To decrease the probability of SBO accident, some NPPs chose to add additional emergency ac power sources, typically diesel generators or gas turbine generators [40]. In case of coupling, there is new onsite power source which is the WF. By integrating WF in the NPP electrical system using proper measures, the reliability of onsite power is enhanced by sufficient independency and redundancy of batteries, diesel generators, gas turbines and the new power source (WF).

This integration of onsite WF in the EPSs has some benefits:

1. Increasing the reliability and availability of power source in the EPSs by adding another type of generation (WF). Formerly, tens and hundreds of WTs have been distributed over an area, offer the advantages of wind diversity and distributed reliability [22].

2. Increasing the time of availability of power source in the EPSs in case of LOOP accident. This is due to the lower fuel consumption of diesel generators.
3. Allowance to supply more loads than safety related loads due to the increase on the onsite power source capacity.

However, this integration requires proper technical measures to work probably. The integration of WF in the NPP electrical system can be achieved by providing a reliable connection from WF output to the bus bar of alternative onsite power supply in the NPP electrical system (Figure 3.5). In case of LOOP accident, WF or part of it will be allowed to work in parallel to the standby diesel generators as wind-diesel system. In wind-diesel system, as the wind power or load changes, the diesel power must change since the diesel is required to balance the difference between the required load and the wind generation. In case of diesel generators fails due to any reason, energy storage facility may be used to smooth out fluctuations in the load and wind power [18]. In this case, the WF has to be connected to the battery bus in the EPSs (Figure 3.5). This is achieved by the use of rectifiers for the WF output or the power electronic devices impeded in WTs themselves. The power generated from the WF will charge the batteries and supplies power to both DC loads and uninterruptable AC loads. Supplying power to AC loads which allow certain interruption of power will require an inverter. Therefore, all EPSs loads can be supplied even if diesel generators fail to supply power.

### 3.2.3 Using MARKOV Process to illustrate EPS's Reliability and Availability enhancement using on-site WF

Availability is defined as the probability that the system is operating properly when it is requested for use. In other words, availability is the probability that a system is not failed or undergoing a repair action when it needs to be used. On the other hand, reliability represents the probability of the system to perform its required functions for a desired period of time without failure in specified environments with a desired confidence.

As stated above, the reliability and availability of EPS can be enhanced using available on-site WF. A case study is conducted using a NPP in El Dabaa site and a WF is erected near it. For calculations simplicity, only two WTs from the WF will be integrated to the EPS.

The reliability models of Ordinary Generation Units (OGUs) are originated from reliability data of failures and repair rates and the two state models covers all aspects of OGUs [42]. Figure 3.6 represents the Markov graph for single OGU. An OGU has a failure rate $\lambda$ and a repair rate $\mu$ based on the reliability data which have been collected through the field. When a generation unit fails, the state of the system will change by the rate of $\lambda$ from operating mode to failure mode. Similarly, when a generation unit is going to be repaired, the state of the system will change by the rate of $\mu$ from failure to operating mode. In this two state model OGUs are assumed to produce 0 MWs in the failure mode and to produce the rated power in the operating mode [42].

When a WT is going to be modeled for reliability studies, the two state model will not work properly. This is due to the behavior of WT compared with OGU. WT has a variable nonlinear output curve in different wind speeds shown typically in Figure 3.7. Several methods have been proposed to cover this difference, which mainly fall into two categories: the analytical and simulation methods. Simulation circumstances like Monte Carlo and the Auto Regressive and Moving Average are used to model the WFs for reliability assessment purposes. These methods need long term data of wind speed which these kinds of data may not be available for certain wind sites. The analytical methods are also used to model the WTs for many years. Using analytical methods reduce the computational efforts in comparison with the simulation methods. The multi-state model of output power has been used for reliability analysis of WF. The probability of occurrence for discrete output power states are considered by use of Probability Distribution Function (PDF) of wind speed data. The severity of the variation in the wind speed pattern will not be detected if the PDF based method is used for modeling the wind speed [42].

The Markov graph for single WT is shown in Figure 3.8. The arrows indicated by $Ø^s$ represent the probability of transition between output power states in a single WT which happen due to variation of wind speed [42].

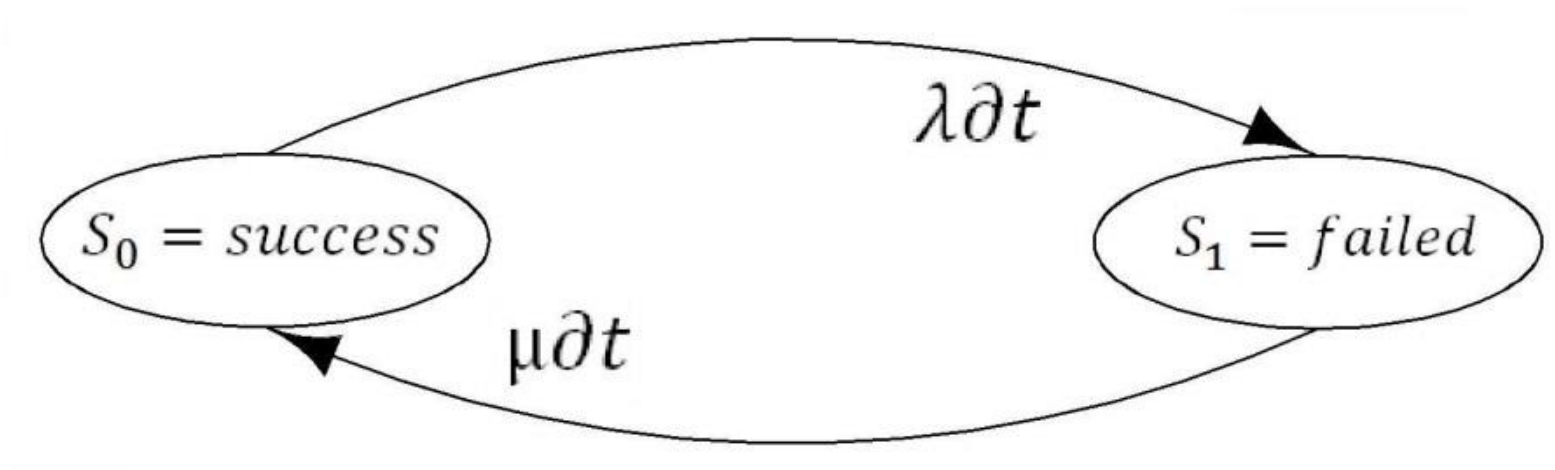

Figure 3.6: Markov graph for single OGU

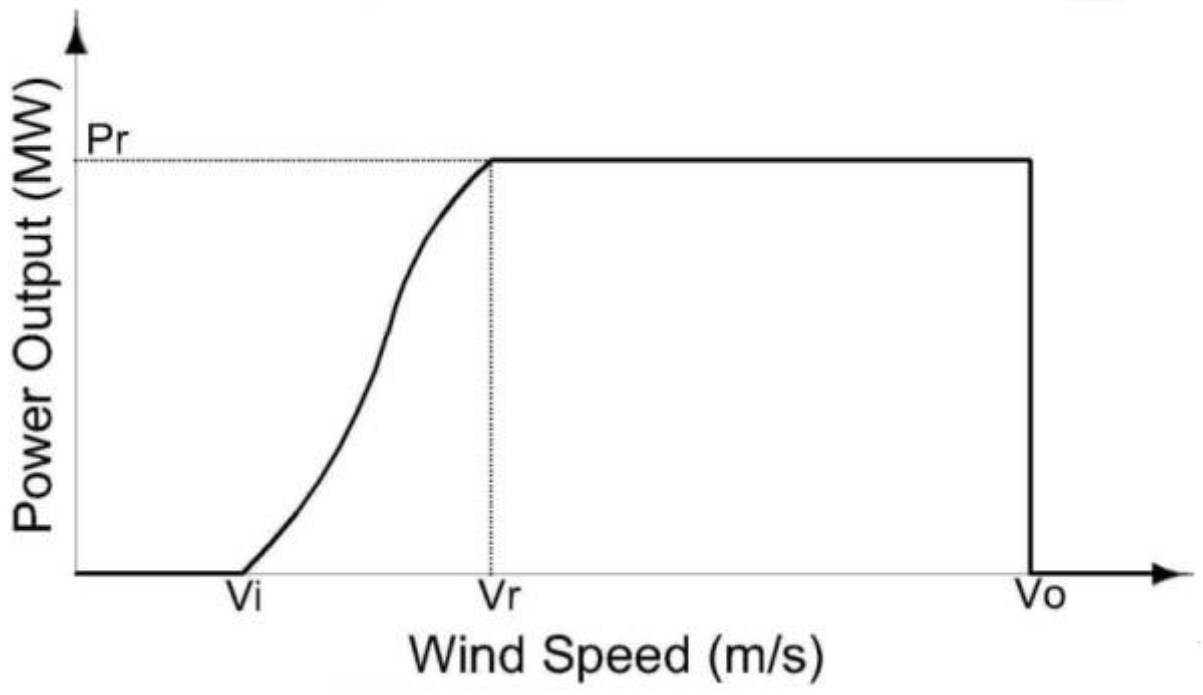

Figure 3.7: WT generator power curve

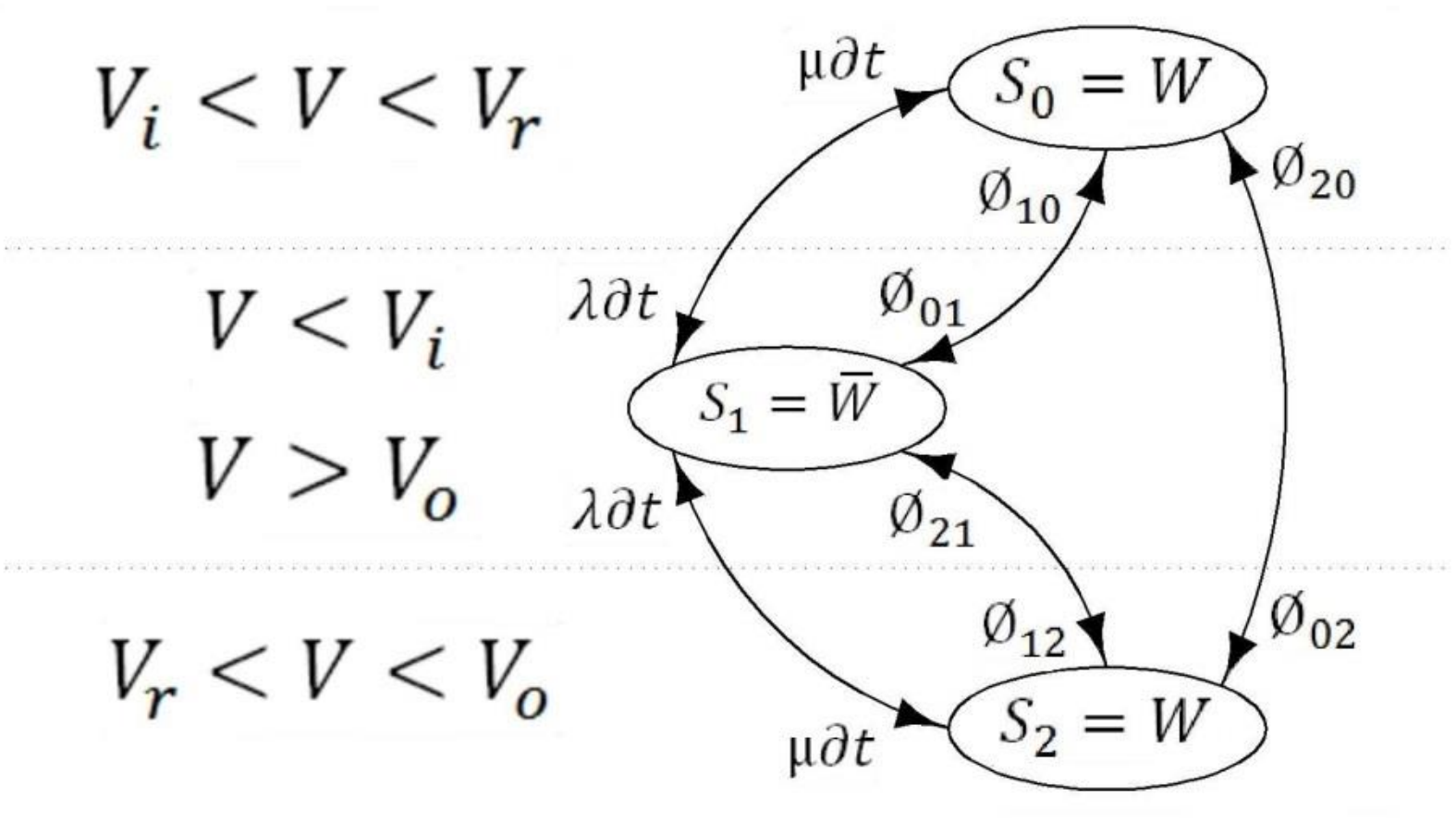

Figure 3.8: Markov graph for single WT

Any Markov graph can be represented by a system of linear first-order differential equations in the unknown state probabilities, representing the change with time of the probabilities of the system states. This is shown in Equation (3.11). The System instantaneous availability at time (t) is the sum of the probabilities of being in a success state at time (t), and it is given be Equation (3.12). The System instantaneous reliability at time (t) is the probability of the system being in success states continuously from t=0. To calculate System instantaneous reliability, exclude all the failed states from the transition matrix A. The new matrix A' contains the transition rates for transitions only among the success states (the 'reduced' system is virtually functioning continuously with no interruptions). The reduced system is represented by Equation (3.13). After that, the System instantaneous Reliability can be calculated using Equation (3.14).

$$\begin{bmatrix} \frac{dP_0(t)}{dt} \\ \frac{dP_1(t)}{dt} \\ \vdots \\ \frac{dP_N(t)}{dt} \end{bmatrix} = \begin{bmatrix} \alpha_{00} & \alpha_{10} & \cdots & \alpha_{N0} \\ \alpha_{01} & \alpha_{11} & \cdots & \alpha_{N1} \\ \vdots & \vdots & \ddots & \vdots \\ \alpha_{0N} & \alpha_{1N} & \cdots & \alpha_{NN} \end{bmatrix} . \begin{bmatrix} P_0(t) \\ P_1(t) \\ \vdots \\ P_N(t) \end{bmatrix} \quad \text{Or} \quad \frac{d\overline{P}}{dt} = \underline{A} \, . P(t) \tag{3.11}$$

Where:

$P_i(t)$: Probability of system to be in state (i) function in time;

$\underline{A}$: Transition rate matrix;

N: Number of states; and

$\alpha_{10}$: Transition rate from state (1) to state (0).

$$System\ instantaneous\ Availability = \sum_{i=S} P_i(t) = 1 - \sum_{j=F} P_j(t) \tag{3.12}$$

Where:

S: set of success states of the system

F: set of failure states of the system

$$\frac{d\overline{P^*}}{dt} = \underline{A'} \, . P^*(t) \tag{3.13}$$

$$System\ instantaneous\ Reliability = \sum_{i=S} P^*{}_i(t) \tag{3.14}$$

The Markov approach will be applied under the following basic assumptions:

1. Constant failure and repair rates are considered;
2. Only single repairman is permitted; and
3. Perfectly reliable switching is assumed.

#### 3.2.3.1 Calculation of reliability and availability for Emergency Diesel Generators (DGs) set

In NPP, the on-site AC power system typically consists of two DGs, either of which is sufficient to meet AC power load requirements for a design basis accident. This configuration has been designated by its success criterion as one-out-of two. Therefore, a system consisting of two identical DGs is considered here, the Markov graph which represents the system is given in Figure 3.9. Equation (3.15) represents the system in Figure 3.9, and is used to calculate the system instantaneous availability. States $S_0$, $S_1$, and $S_2$ are the success states and $S_3$ is the failed state. Equation (3.15) is solved using the ODE 45 Function in Matlab and after that system instantaneous availability is calculated using Equation (3.12). To calculate the reliability, as stated before, the row and column of failed state ($S_3$) in Equation (3.15) are deleted and the reduced system can be represented by Equation (3.16). Similarly, Equation (3.16) is solved using Matlab ODE 45 function and after that system instantaneous reliability is calculated using Equation (3.17). The values of $\lambda$, $\mu$, and $\lambda_{ccf}$ for DGs used in NPP are shown in Table 3.7. Figure 3.12.a shows the calculated availability versus time (in hours) for DGs set, and Figure 3.13.a shows the calculated reliability versus time (in hours) for DGs set.

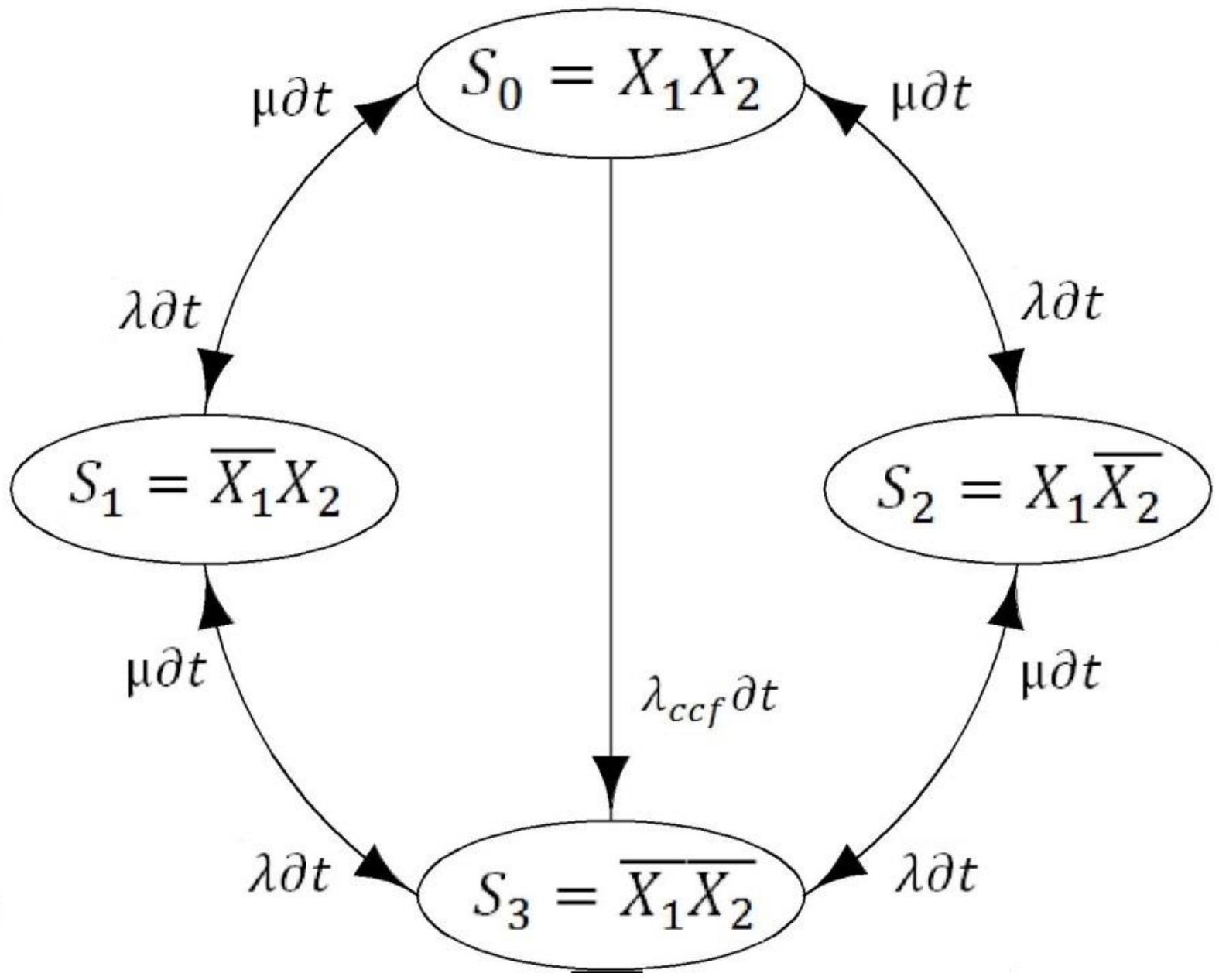

Figure 3.9: Markov graph for two identical DGs

$$\begin{bmatrix} \frac{dP_0(t)}{dt} \\ \frac{dP_1(t)}{dt} \\ \frac{dP_2(t)}{dt} \\ \frac{dP_3(t)}{dt} \end{bmatrix} = \begin{bmatrix} -(2\lambda + \lambda_{ccf}) & \mu & \mu & 0 \\ \lambda & -(\lambda + \mu) & 0 & \mu \\ \lambda & 0 & -(\lambda + \mu) & \mu \\ \lambda_{ccf} & \lambda & \lambda & -2\mu \end{bmatrix} \cdot \begin{bmatrix} P_0(t) \\ P_1(t) \\ P_2(t) \\ P_3(t) \end{bmatrix} \quad (3.15)$$

Where:

$P_0(t)$: Probability function in time of state $S_0$;

$\lambda$: Failure rate of one unit of DG;

$\mu$: Repair rate of one unit of DG; and

$\lambda_{ccf}$: Common cause failure rate of DGs set.

$$\begin{bmatrix} \frac{dP_0(t)}{dt} \\ \frac{dP_1(t)}{dt} \\ \frac{dP_2(t)}{dt} \end{bmatrix} = \begin{bmatrix} -(2\lambda + \lambda_{ccf}) & \mu & \mu \\ \lambda & -(\lambda + \mu) & 0 \\ \lambda & 0 & -(\lambda + \mu) \end{bmatrix} \cdot \begin{bmatrix} P_0(t) \\ P_1(t) \\ P_2(t) \end{bmatrix} \quad (3.16)$$

Table 3.7: Reliability and Availability data for DGs used in NPP [43]

| Symbol | Value |
|---|---|
| $\lambda$ | $5.2*10^{-3}$ /h |
| $\mu$ | 0.05 /h |
| $\lambda_{ccf}$ | $2.59*10^{-4}$ /h |

### 3.2.3.2 Calculation of reliability and availability for WTs set

The Markov graph which represents the system is given in Figure 3.10. The transition between states depend on WTs condition (operate or fail) and the wind speed. Transition between states based on WTs condition change is done through failure and repair rates. Transition between states based on change in wind speed region is done through $Ø_{ij}$, which is the probability of being in wind speed region of state (i) and the wind speed changes to be in the wind speed region of state (j). $Ø^s$ are calculated based on PDF of wind speeds in the area using Weibull distribution. Equations from (3.17) to (3.24) represent probability density function of the 2-parameter Weibull distribution of wind speed (V) and how to calculate different Ø<sup>s</sup>. Table 3.8 illustrates the conditions associated with each state of the states in Figure 3.10.

Matrix Equation (3.25) represents the system in Figure 3.10, and is used to calculate the instantaneous availability of WTs set. States from $S_0$ to $S_5$ are the success states and $S_6$ is the failed state. Equation (3.25) is solved using Matlab ODE 45 Function and after that system instantaneous availability is calculated using Equation (3.12). To calculate the reliability, as stated before, in Equation (3.25), the row and column of failed state ($S_6$) are deleted and the reduced system can be represented by Equation (3.26). Similarly, Equation (3.26) is solved using Matlab ODE 45 Function and after that system instantaneous reliability is calculated using Equation (3.14). Data for the used WTs and there site (El Dabaa) are shown in Table 3.9. Depending on these data, Figure 3.12.b shows the Availability versus time (in hours) for WTs set, and Figure 3.13.b shows the Reliability versus time (in hours) for WTs set.

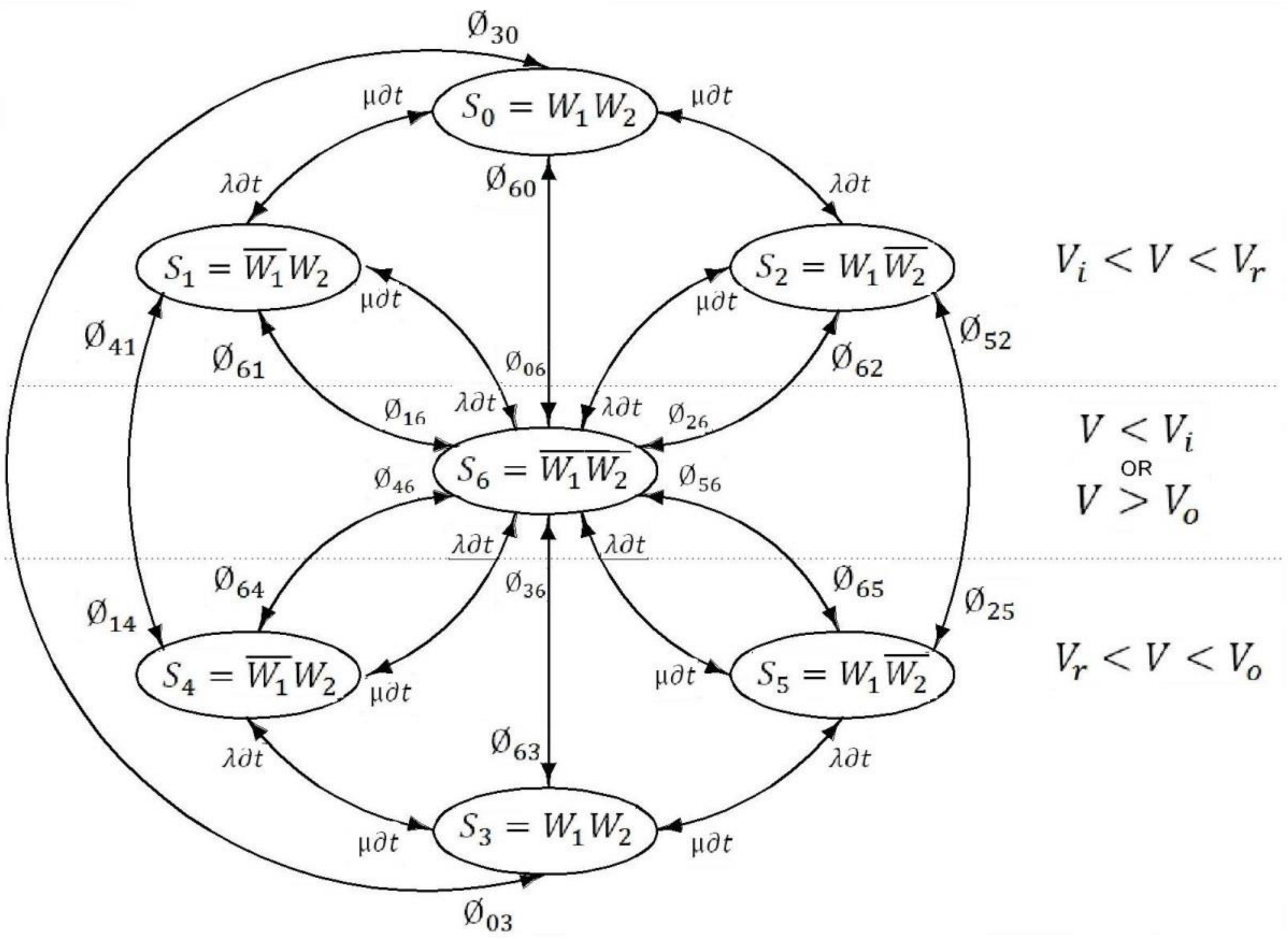

Figure 3.10: Markov graph for two identical WTs working in parallel

Table 3.8: Conditions associated with each state in Figure 3.10

| State | WTs conditions | Wind Speed |
|---|---|---|
| $S_0$ | Two healthy | $V_i$ <V< $V_r$ |
| $S_3$ | Two healthy | $V_r$ <V< $V_o$ |
| $S_1$, $S_2$ | One healthy | $V_i$ <V< $V_r$ |
| $S_4$, $S_5$ | One healthy | $V_r$ <V< $V_o$ |
| $S_6$ | Two dead | V< $V_i$ OR $V_r$ <V |

$$f(v) = \frac{k}{c}(\frac{v}{c})^{k-1}e^{(-(\frac{v}{c}))^k} \tag{3.17}$$

$$P_r(v_i < v < v_r) = \int_{v_i}^{v_r} f(v)\, dv \tag{3.18}$$

$$\emptyset_{06} = \emptyset_{16} = \emptyset_{26} = P_r(v < v_i) = 1 - \exp(-(\frac{v_i}{c})^k) \tag{3.19}$$

$$\emptyset_{60} = \emptyset_{61} = \emptyset_{62} = P_r(v > v_i) = \exp(-(\frac{v_i}{c})^k) \tag{3.20}$$

$$\emptyset_{30} = \emptyset_{41} = \emptyset_{52} = P_r(v < v_r) = 1 - \exp(-(\frac{v_r}{c})^k) \tag{3.21}$$

$$\emptyset_{03} = \emptyset_{14} = \emptyset_{25} = P_r(v > v_r) = \exp(-(\frac{v_r}{c})^k) \tag{3.22}$$

$$\emptyset_{63} = \emptyset_{64} = \emptyset_{65} = P_r(v < v_o) = 1 - \exp(-(\frac{v_o}{c})^k) \tag{3.23}$$

$$\emptyset_{36} = \emptyset_{46} = \emptyset_{56} = P_r(v > v_o) = \exp(-(\frac{v_o}{c})^k) \tag{3.24}$$

Where:

f(v):Probability density function of the 2-parameter Weibull distribution of wind speed (v);

K: Shape parameter of the 2-parameter Weibull distribution;

C: Scale parameter of the 2-parameter Weibull distribution;

$V_i$: Cut-in wind speed of WT;

$V_r$: Rated wind speed of WT;

$V_o$: Cut-out wind speed of WT;

$P_r$ ($V_i$<V<$V_r$ ): Probability of wind speed to be above (Vi) and below (Vr); and

$\emptyset_{06}$: Probability that wind speed is in the region of $S_0$ ($V_i$ <V< $V_r$) and decreases to be in the region of state $S_6$ (V< Vi).

$$\begin{bmatrix} \frac{dP_0(t)}{dt} \\ \frac{dP_1(t)}{dt} \\ \frac{dP_2(t)}{dt} \\ \frac{dP_3(t)}{dt} \\ \frac{dP_4(t)}{dt} \\ \frac{dP_5(t)}{dt} \\ \frac{dP_6(t)}{dt} \end{bmatrix} = \begin{bmatrix} A & \mu & \mu & \emptyset_{30} & 0 & 0 & \emptyset_{60} \\ \lambda & B & 0 & 0 & \emptyset_{41} & 0 & (\mu + \emptyset_{61}) \\ \lambda & 0 & C & 0 & 0 & \emptyset_{52} & (\mu + \emptyset_{62}) \\ \emptyset_{03} & 0 & 0 & D & \mu & \mu & \emptyset_{63} \\ 0 & \emptyset_{14} & 0 & \lambda & E & 0 & (\mu + \emptyset_{64}) \\ 0 & 0 & \emptyset_{25} & \lambda & 0 & F & (\mu + \emptyset_{65}) \\ \emptyset_{06} & (\lambda + \emptyset_{16}) & (\lambda + \emptyset_{26}) & \emptyset_{36} & (\lambda + \emptyset_{46}) & (\lambda + \emptyset_{56}) & G \end{bmatrix} . \begin{bmatrix} P_0(t) \\ P_1(t) \\ P_2(t) \\ P_3(t) \\ P_4(t) \\ P_5(t) \\ P_6(t) \end{bmatrix} \tag{3.25}$$

$$\begin{bmatrix} \frac{dP_0(t)}{dt} \\ \frac{dP_1(t)}{dt} \\ \frac{dP_2(t)}{dt} \\ \frac{dP_3(t)}{dt} \\ \frac{dP_4(t)}{dt} \\ \frac{dP_5(t)}{dt} \end{bmatrix} = \begin{bmatrix} A & \mu & \mu & \emptyset_{30} & 0 & 0 \\ \lambda & B & 0 & 0 & \emptyset_{41} & 0 \\ \lambda & 0 & C & 0 & 0 & \emptyset_{52} \\ \emptyset_{03} & 0 & 0 & D & \mu & \mu \\ 0 & \emptyset_{14} & 0 & \lambda & E & 0 \\ 0 & 0 & \emptyset_{25} & \lambda & 0 & F \end{bmatrix} . \begin{bmatrix} P_0(t) \\ P_1(t) \\ P_2(t) \\ P_3(t) \\ P_4(t) \\ P_5(t) \end{bmatrix} \tag{3.26}$$

Where:

$$A = -(\emptyset_{03} + 2\lambda + \emptyset_{06}) \tag{3.27}$$

$$B = -(\lambda + \mu + \emptyset_{16} + \emptyset_{14}) \tag{3.28}$$

$$C = -(\lambda + \mu + \emptyset_{26} + \emptyset_{25}) \tag{3.29}$$

$$D = -(\emptyset_{30} + 2\lambda + \emptyset_{36}) \tag{3.30}$$

$$E = -(\mu + \lambda + \emptyset_{46} + \emptyset_{41}) \tag{3.31}$$

$$F = -(\mu + \lambda + \emptyset_{52} + \emptyset_{56}) \tag{3.32}$$

$$G = -(\emptyset_{60} + 4\mu + \emptyset_{61} + \emptyset_{62} + \emptyset_{63} + \emptyset_{64} + \emptyset_{65}) \tag{3.33}$$

Table 3.9: Data for the used WTs and there site (El Dabaa)

| Parameter | Value |
|---|---|
| WT [44]: | |
| $\lambda$ | 0.00073266 /h |
| $\mu$ | 0.016423 /h |
| $V_i$ | 3 m/s |
| $V_r$ | 12 m/s |
| $V_o$ | 25 m/s |
| Site: El Dabaa [5]: | |
| K | 11.05 |
| C | 5.64 |

#### 3.2.3.3 Enhancement achieved in reliability and availability by adding WTs to EPSs of NPP (Wind – Diesel system)

To explain the benefit of integrating WTs to EPSs of NPP in enhancing the reliability and availability, two systems will be compared:

1. DGs working alone; and
2. DGs and WTs working in parallel as Wind - Diesel system.

For the second choice, the role of parallel system in Markov process will be used. For two components connected in parallel as shown in Figure 3.11, which is the case of Wind-Diesel system, the reliability and availability can be calculated as follows:

$$R_P = 1 - Q_A.Q_B = R_A + R_B - R_A.R_B \tag{3.34}$$

Or

$$Q_P = Q_A.Q_B \ \ \& \ \ R_P = 1 - Q_P \tag{3.35}$$

Where:

$R_p$: Reliability or availability of the parallel system;

$R_A$: Reliability or availability of the system A;

$Q_p$: Unreliability or unavailability of the parallel system; and

$Q_A$: Unreliability or unavailability of system A;

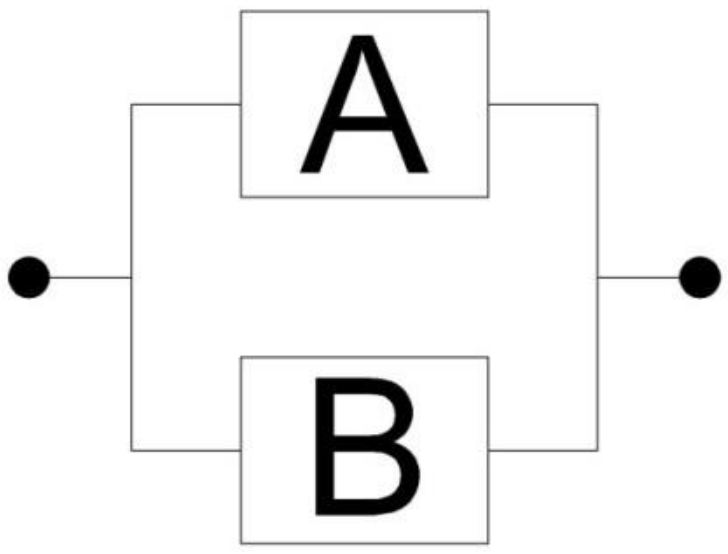

Figure 3.11: Two systems A and B connected in parallel

Applying equations (3.34) or (3.35) using Figures 3.12.a, 3.12.b, 3.13.a & 3.13.b, the availability and reliability of the Wind-Diesel system can be calculated. Figures 3.12.c & 3.13.c represent the availability and reliability of Wind-Diesel system respectively. Also, they compare this case with the case of DGs working alone. These figures illustrate the availability and reliability of each system regardless of its power generation level. DGs sets can generate at nearly 100% of installed capacity when they are on. However, generation from WTs (or WF) is fluctuating between 0% and 100% of installed capacity according to the weather and WTs conditions. Therefore, to be more accurate, minimum acceptable limit for WF generated power should be set and used in these calculations. Under this minimum limit, the WF will be considered unavailable and unreliable. This minimum limit depends on: NPP capacity, NPP house load during emergencies, WF capacity, etc.

As a case study, it is assumed that 1000 MW NPP is existed and 100 MW WF is adjacent to it. During emergency conditions, NPP requires about 2% of its installed capacity for supplying emergency house loads, and so this counts for 20 MW. Therefore, for WF to be reliable and available, they should supply these 20 MW which equals 20% of its installed capacity. So, the minimum limit is 20% (or 20 MW). Systems illustrated in Figure 3.9 and Figure 3.10 can be used in this case study. No changes for DGs system (Figure 3.9). In Figure 3.10, instead of having two WTs working in parallel, two parallel trains of WTs will be assumed. Each train consists of 50 MW, and should produce 20 MW as minimum limit to be available and reliable. From WT generator power curve (Figure 3.7), and data illustrated in Table 3.9, wind speed should be above 8.7 m/s at hub height for WF to be available and reliable. Applying the same procedure for systems given in Figures 3.11 and 3.12 with these simple changes results in Figure 3.14 and Figure 3.15 can be calculated. Figure 3.14 represents the enhancement in availability achieved by WF integration in the EPS system, and Figure 3.15 represents the enhancement in reliability. The system's reliability is only enhanced in the first six hours, which are vital during emergency conditions.

From Figures 3.12.c, 3.13.c, 3.14, and 3.15, it is clear that integrating WTs in the EPSs of NPP enhanced the system's availability and reliability considerably. This enhancement is very important especially during the first hours of operation. This integration of onsite WF to the NPP's EPSs benefits can be summarized as follows:

1. Increasing the Reliability and Availability of power source in the EPSs by adding another type of generation (WF). Generally, tens and hundreds of WTs have been distributed over an area, offer the advantages of wind diversity and distributed reliability [22].
2. Lower fuel consumption of DGs, so during LOOP accident, the availability time of power source in the EPS will be increased.
3. Increasing the onsite power source capacity allows supplying more loads than only safety related loads.

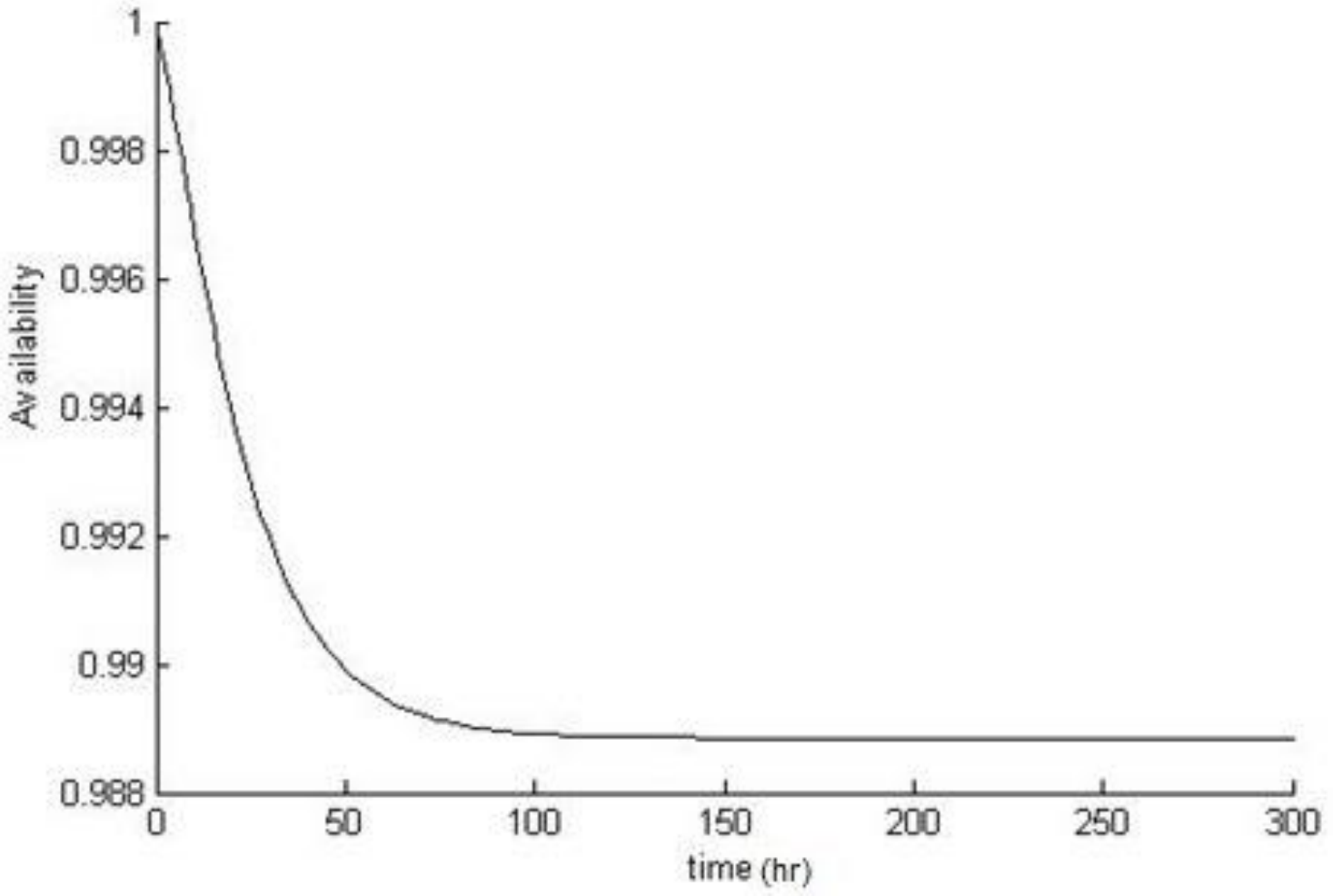

Figure 3.12.a: DGs set

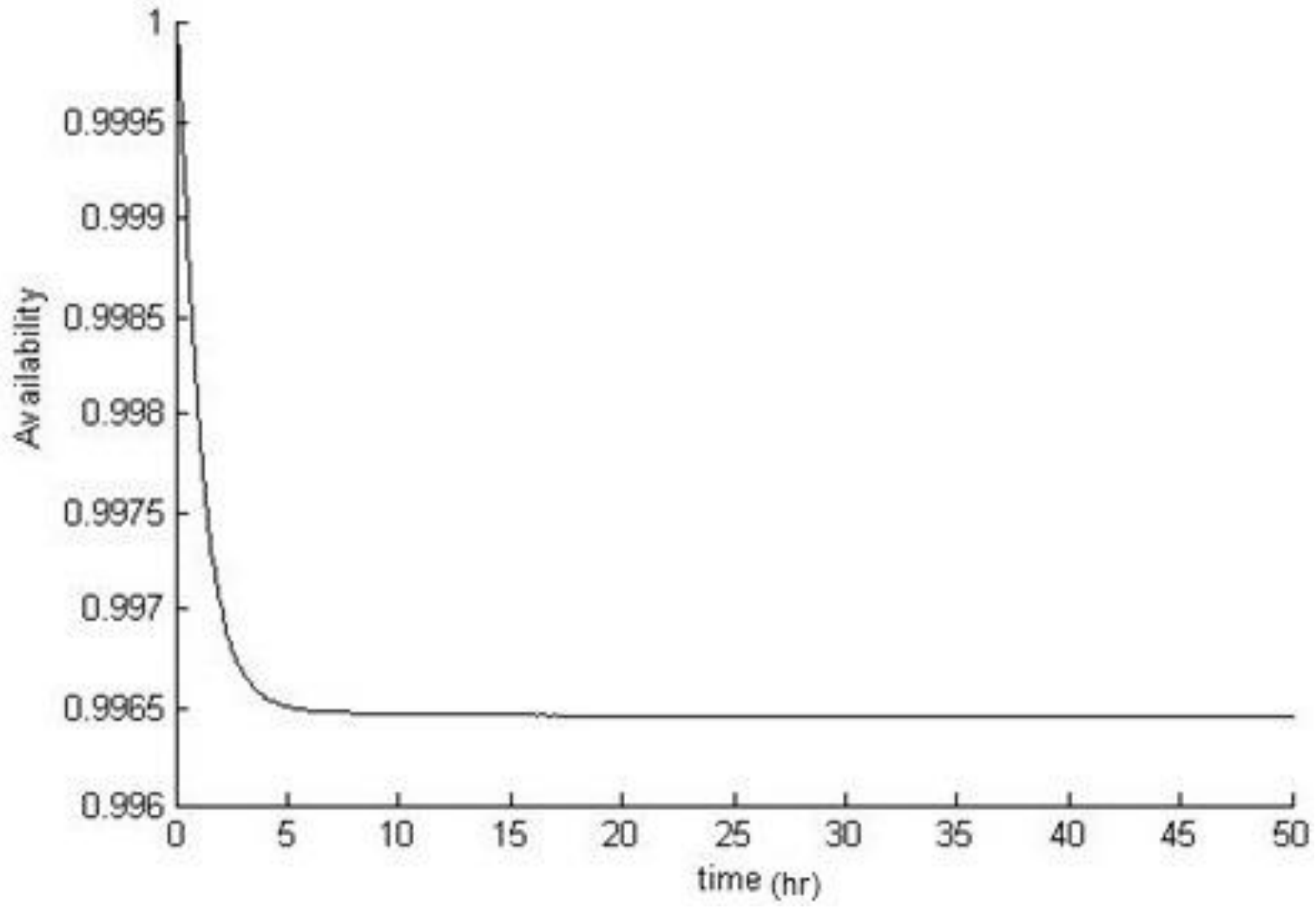

Figure 3.12.b: WTs set

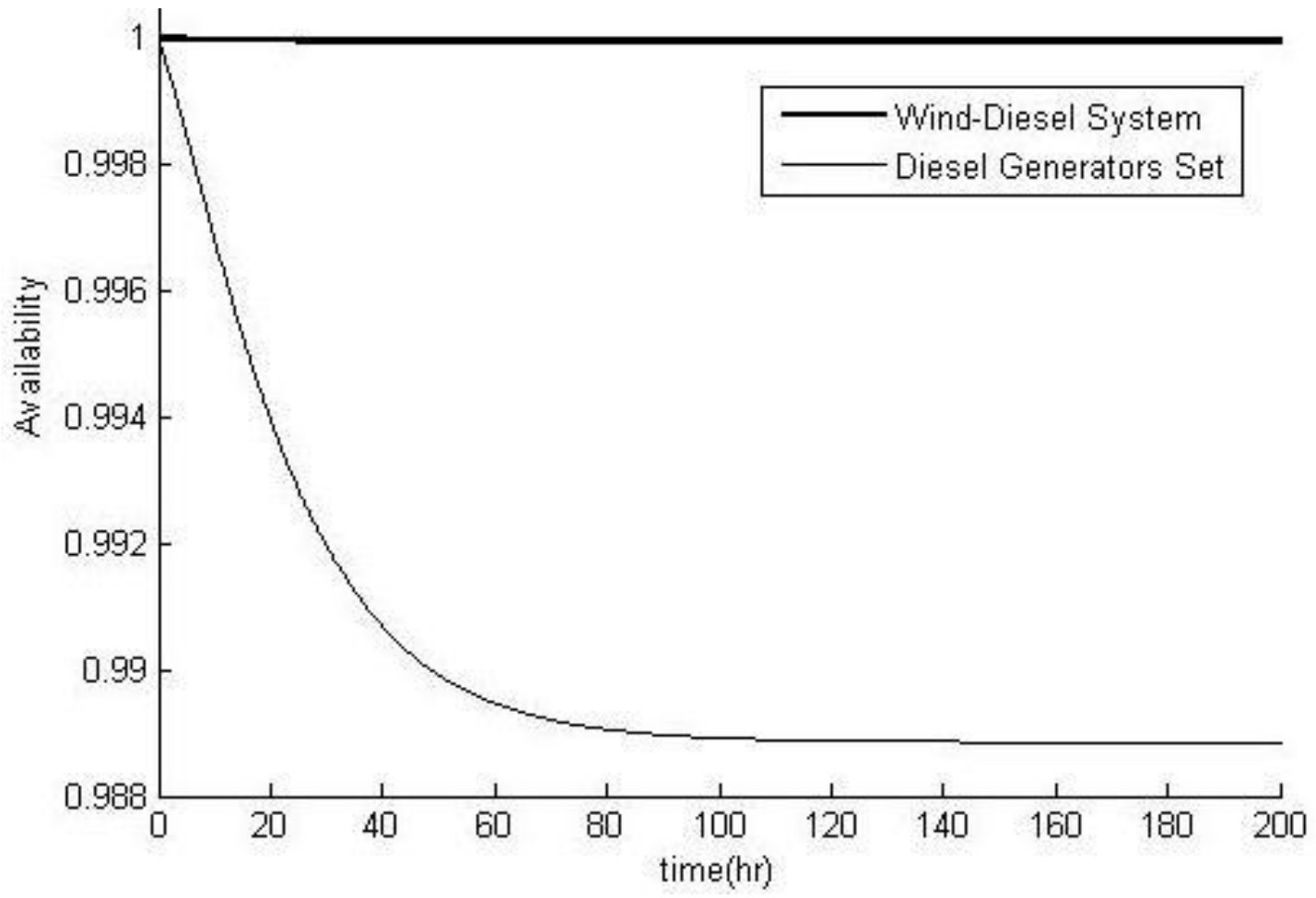

Figure 3.12.c: Wind-Diesel System versus DGs set

Figure 3.12: Availability versus time (hours)

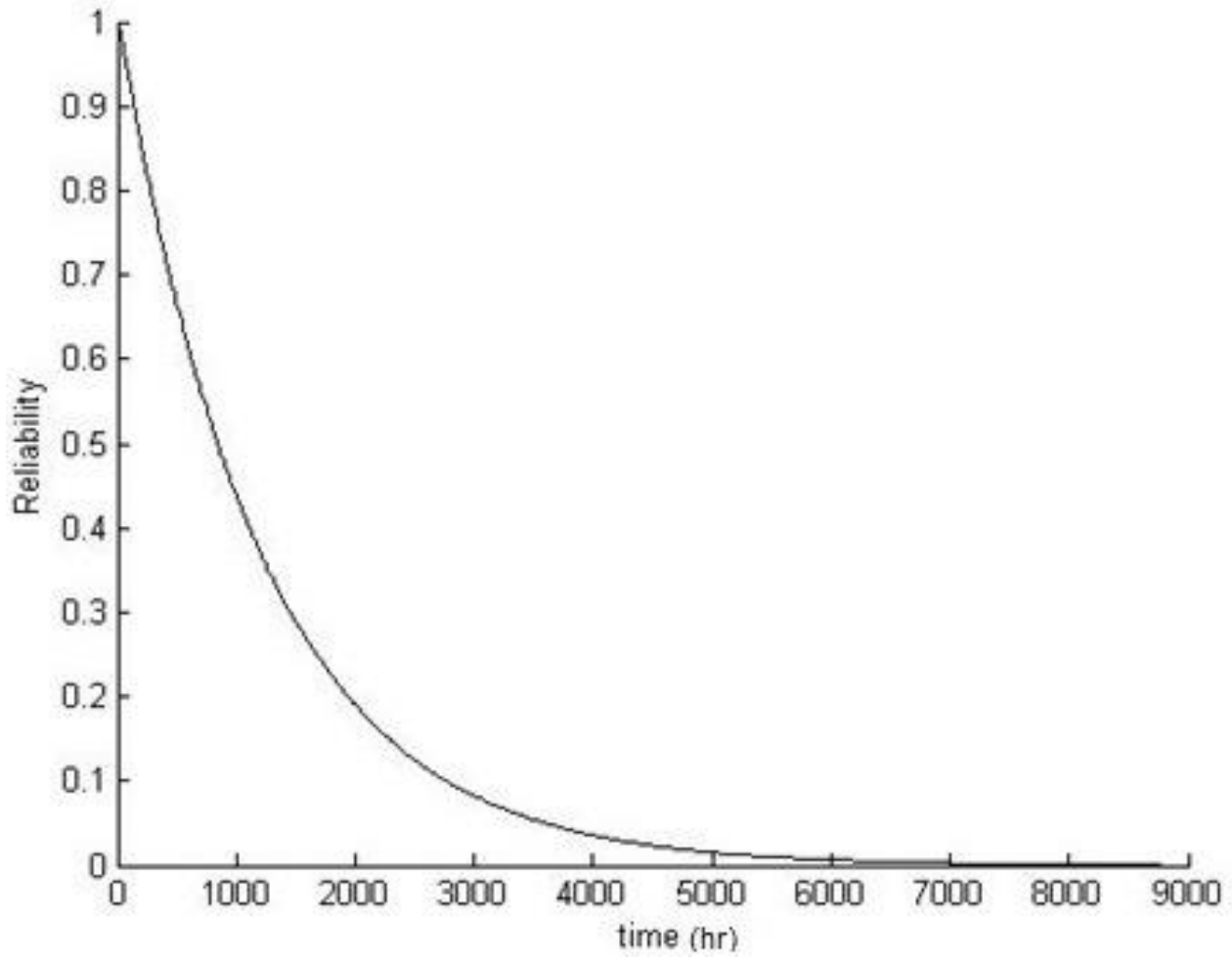

Figure 3.13.a: DGs set

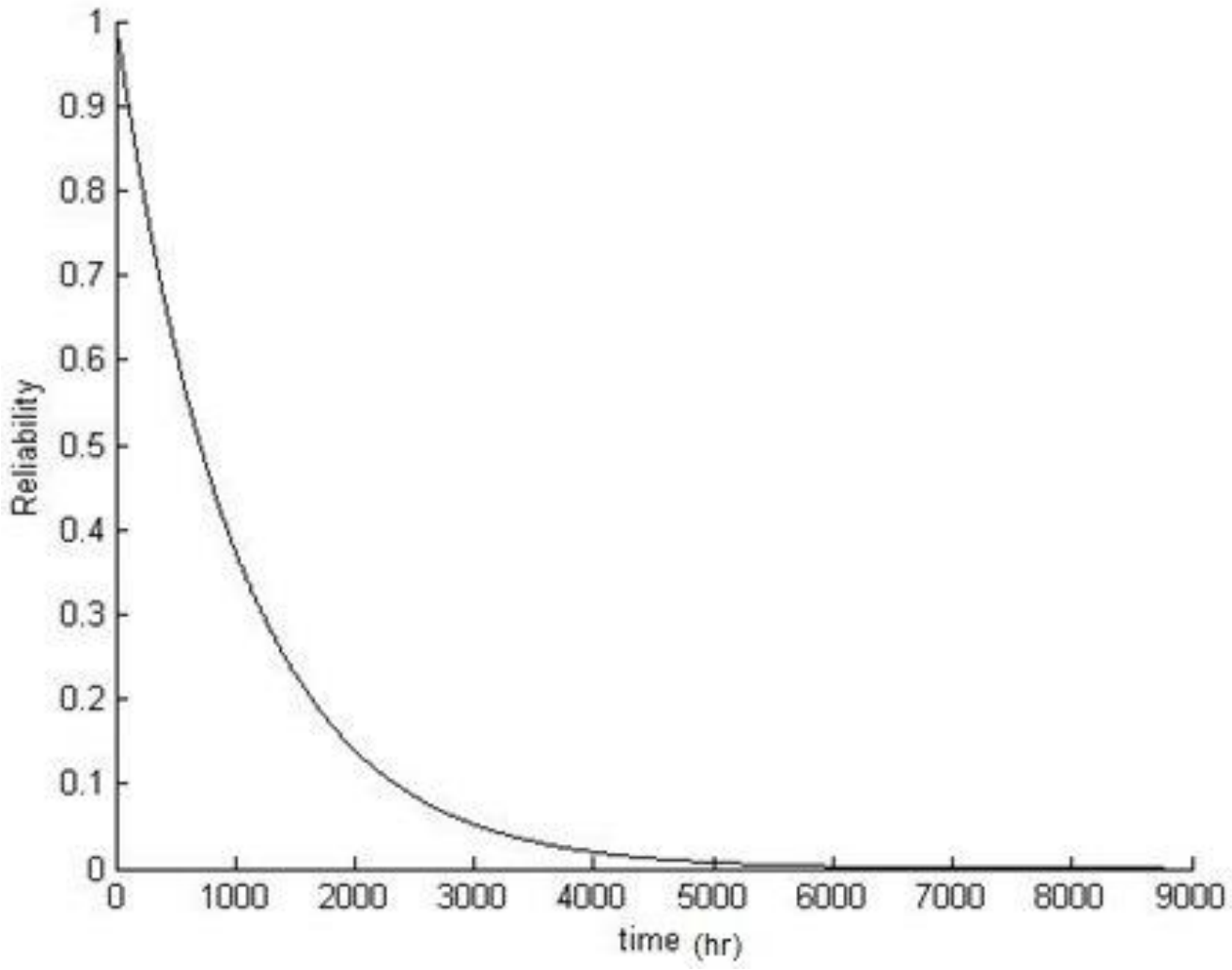

Figure 3.13.b: WTs set

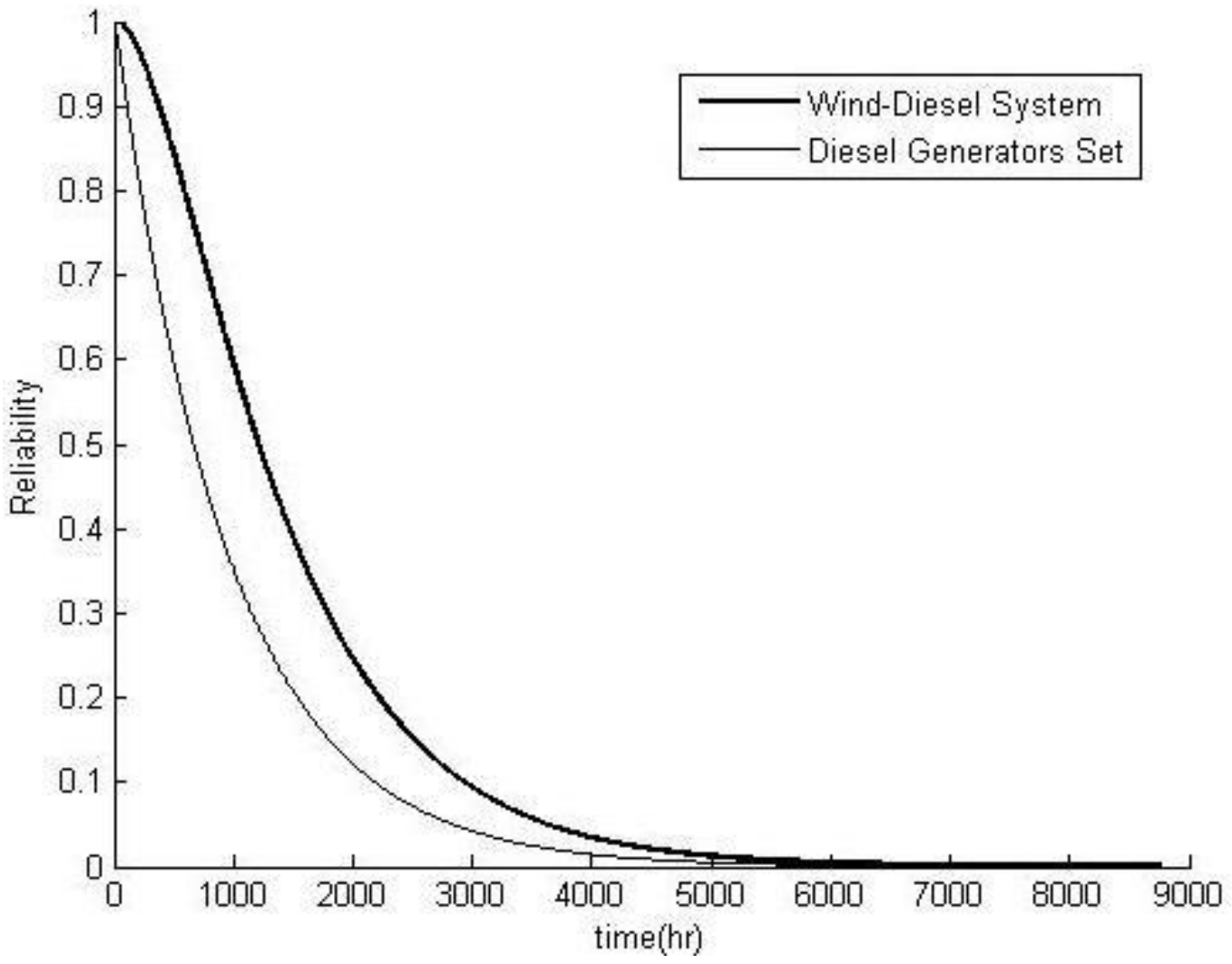

Figure 3.13.c: Wind-Diesel System versus DGs set

Figure 3.13: Reliability versus time (hours)

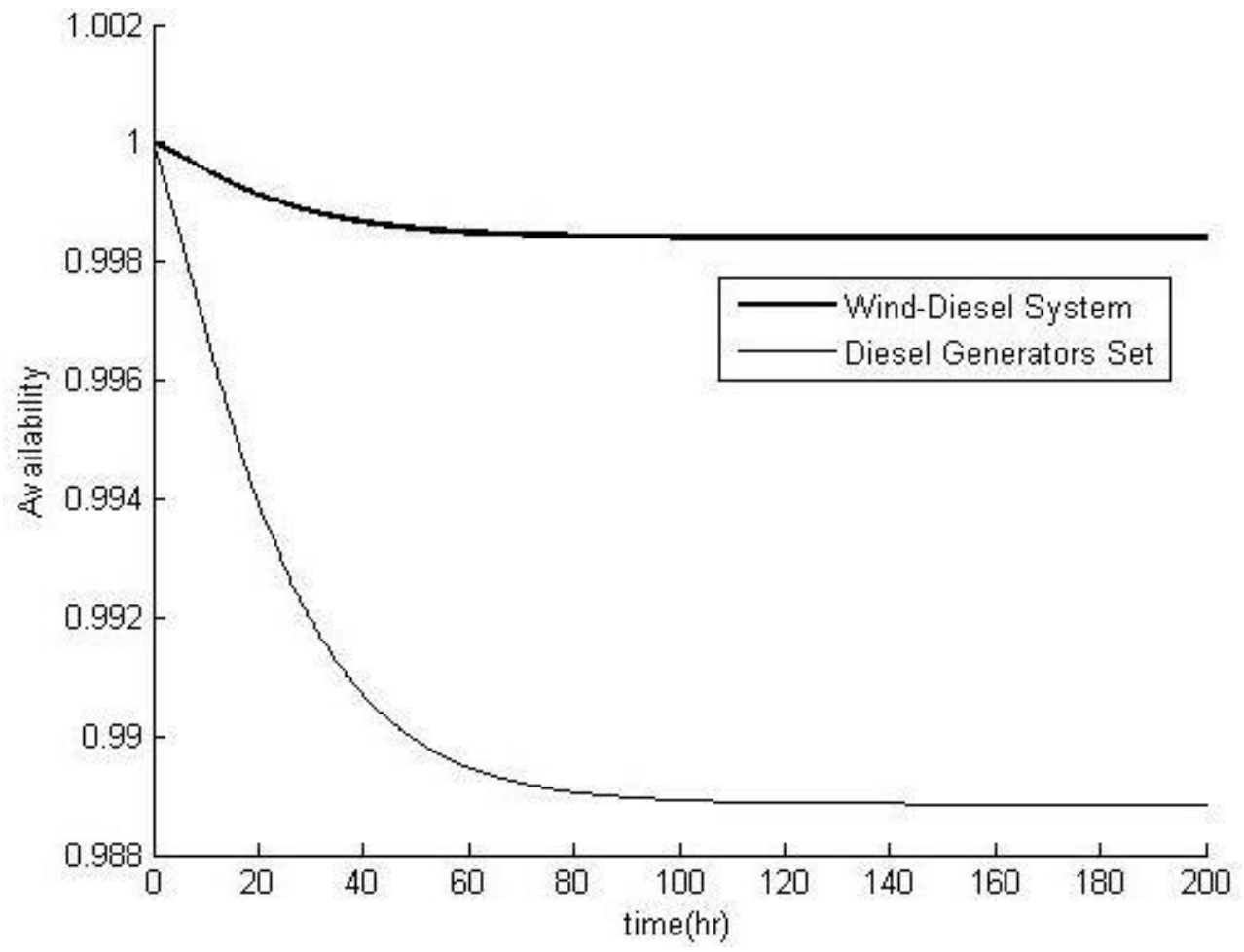

Figure 3.14: Availability versus time (hours) for Wind-Diesel System and DGs set, with minimum production of 20 MW from WF

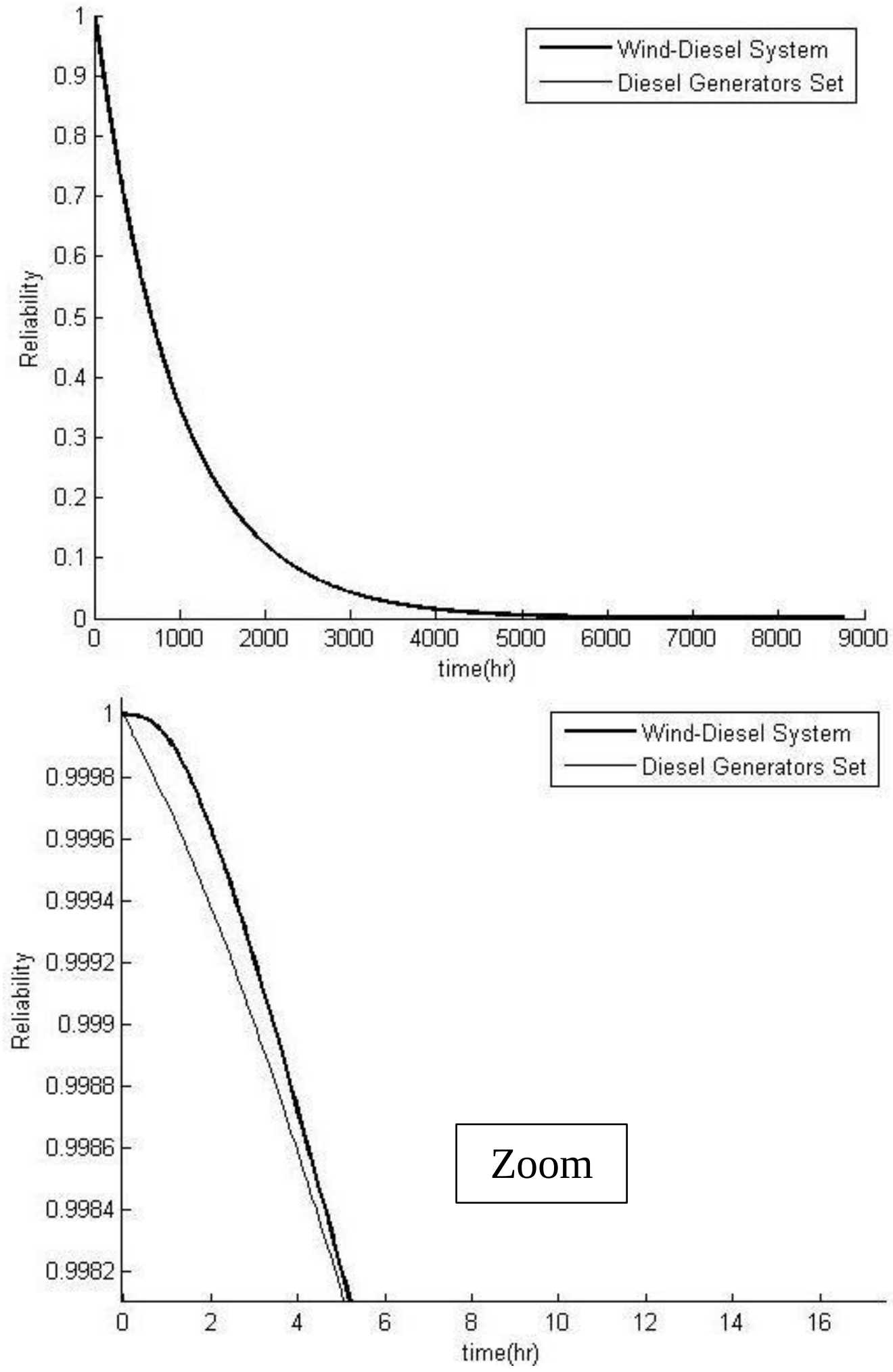

Figure 3.15: Reliability versus time (hours) for Wind-Diesel System and DGs set, with minimum production of 20 MW from WF

# Chapter 4

# Aspects Related to Adjacency Between WF and NPP

In this chapter the benefit of WFs geographical distribution is illustrated. A Case study is conducted to evaluate the WFs geographical distribution in Egypt and its impact on the capacity credit of wind energy in the grid.

## 4.1 Geographical distribution of WFs in the grid

The wind speed varies continuously as a function of time and height. Therefore, wind energy is characterized by large variations in production. Figure 4.1 illustrates an example of large-scale wind power production versus time in Denmark during January 2000 [18]. Just as consumer demands are smoothed by aggregation, so is the output from WF, and geographic dispersion dramatically reduces the wind fluctuations and there will be fewer instances of near zero or peak output [18, 22].

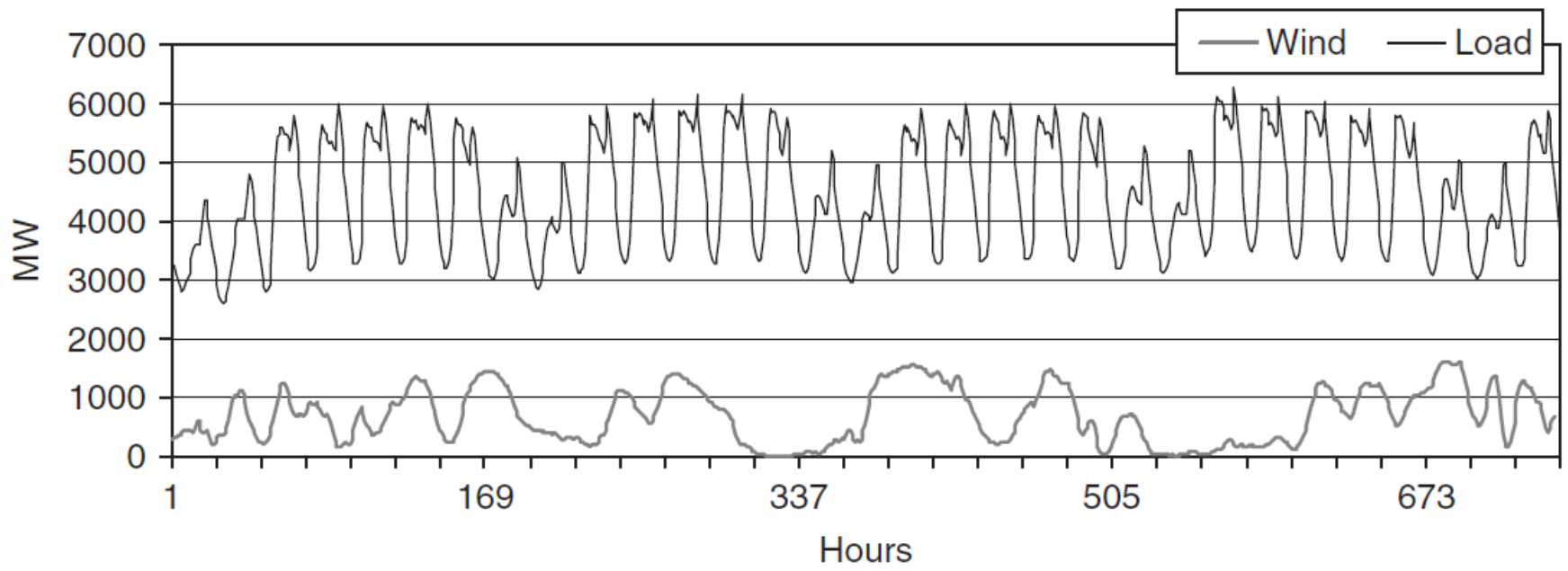

Figure 4.1: An example of large-scale wind power production in Denmark, January 2000 [18]

As wind energy is aggregated from multiple WTs in a WF, the short-term variations in wind resources is attenuated as a percentage of the overall power output [18, 16, 22]. When wind energy from multiple WFs is aggregated, minute to- hour variation in energy is attenuated and an hour-to-day variation in wind energy production and delivery to the grid is observed. Grid system operators manage such variations using spinning reserves [16]. With wind penetration of up to 20% of system total peak demand, additional spinning reserves will be required. It has been found that this results in additional cost of up to 10% of the wholesale value of wind energy [45]. So the smoothing effect of aggregation of wind power in the grid means that the operational penalties associated with running networks with wind may be lower than early estimates. Thus, the extra cost of running more thermal plant as spinning reserve is lowered [22]. A wide geographical distribution of WFs reduces the impact of the slower variations as changing weather patterns do not affect all WTs at the same time [18].

Figure 4.2 shows the principal of aggregated wind power production based on simulations by Pedro Rosas, where the time scale on graphics is seconds [18].

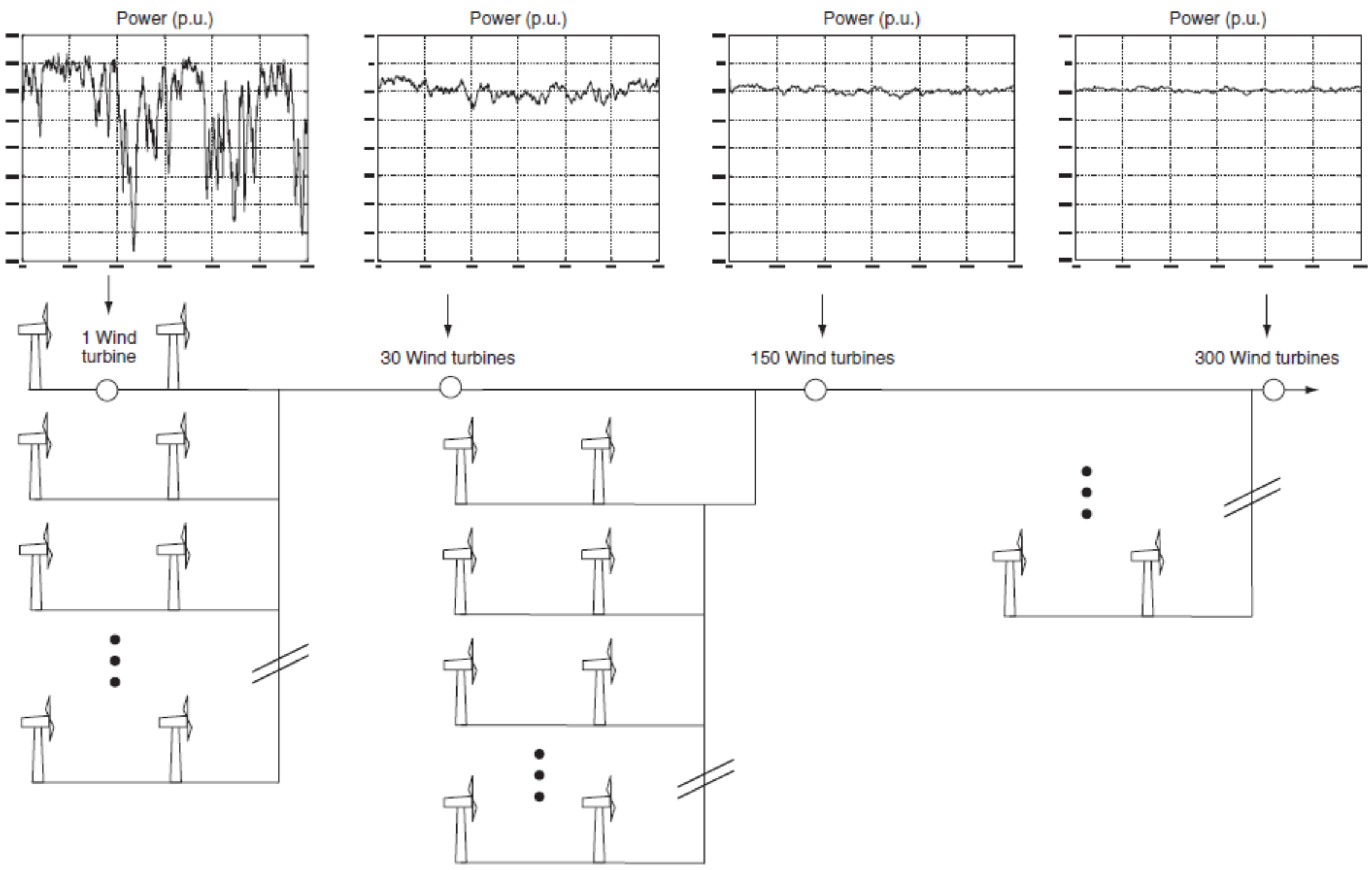

Figure 4.2: Impact of geographical distribution and additional WTs on aggregated power production [18]

### 4.1.1 Case study for the effect of WFs geographical distribution on the aggregation of wind power production in Egypt

To investigate the impact of the geographical distribution of WFs in Egypt, two cases are examined:

Case-A: Geographical distribution of WFs by their installation on both the shore of Suez Bay and near NPP sites (El Dabaa) as proposed through coupling

Case-B: Locating WFs as it is in the Egyptian Renewable Energy strategy, which is based on maximizing energy yield of the installed capacity. This means locating them only on the shore of Suez Bay.

According to the Egyptian plan for both WFs and NPPs, and the available wind data, two sites are chosen to conduct this study. They are El Dabaa (North Mediterranean coast), and the Zafarana (Suez Bay- Red Sea). Only mean hourly wind speed per month was made available to the authors in (m/s) at a height 24.5 m above the ground level from [15]. Therefore, 288 values for each site are used. For El Dabaa site, due to the limitation in data, data for EL Galala site (that is near to El Dabaa) is used. The geographical coordination of each site is illustrated in Table 4.1, and the wind speed data in (m/s) at a height 24.5 m *a.g.l.* for El Galala and Zafarana are shown in Table 4.2 and 4.3 respectively.

Table 4.1: Geographical coordination of the three sites [5]

| Site | Coordinates | | | | |
|---|---|---|---|---|---|
| | Latitude | | Longitude | | Elevation |
| | Deg. | Min. | Deg. | Min. | (m) |
| El Dabaa | 30 | 56 | 28 | 28 | 17 |
| El Galala | 31 | 1 | 28 | 11 | 59 |
| Zafarana | 29 | 06 | 32 | 36 | 25 |

Table 4.2: Monthly mean wind speeds (m/s), at a height 24.5 m *a.g.l.* for El Galala [15]

| Galala | Jan | Feb | Mar | Apr | May | June | July | Aug | Sep | Oct | Nov | Dec | Year |
|---|---|---|---|---|---|---|---|---|---|---|---|---|---|
| 0 | 7.1 | 6.3 | 5.2 | 5.5 | 4.7 | 4.4 | 4.6 | 4.1 | 4.8 | 4.4 | 6.3 | 5.9 | 5.3 |
| 1 | 7.5 | 6.2 | 5.4 | 5.4 | 4.5 | 4.6 | 4.4 | 4.2 | 5 | 4.7 | 6.5 | 5.7 | 5.3 |
| 2 | 7.4 | 6 | 5.3 | 5.5 | 4.6 | 4.4 | 4.6 | 4.3 | 5.1 | 4.9 | 6.3 | 6 | 5.4 |
| 3 | 7.6 | 6 | 5.2 | 5.8 | 4.9 | 4.5 | 4.6 | 4.5 | 5.2 | 4.8 | 6.6 | 6.1 | 5.5 |
| 4 | 7.7 | 6.5 | 5.7 | 6.1 | 4.6 | 4.7 | 4.9 | 4.5 | 4.8 | 4.6 | 6.3 | 5.8 | 5.5 |
| 5 | 7.5 | 6.3 | 5.9 | 6 | 4.3 | 5 | 5.1 | 4.6 | 4.6 | 4.8 | 6.4 | 5.8 | 5.5 |
| 6 | 7.5 | 6.3 | 5.7 | 4.8 | 4.7 | 5 | 5.1 | 4.7 | 4.5 | 4.8 | 5.9 | 5.9 | 5.4 |
| 7 | 7.3 | 6.4 | 5.6 | 5.3 | 4.5 | 5 | 5.1 | 4.5 | 4.5 | 4.8 | 5.7 | 5.8 | 5.4 |
| 8 | 7.1 | 6.2 | 5.6 | 5.9 | 5 | 5.5 | 5.5 | 5.3 | 4.8 | 5.2 | 5.6 | 5.7 | 5.6 |
| 9 | 7.1 | 6.5 | 6.2 | 5.9 | 5.2 | 5.8 | 5.8 | 5.6 | 5.3 | 5.3 | 6.3 | 5.9 | 5.9 |
| 10 | 7.6 | 6.8 | 6.3 | 6.1 | 5.2 | 6.2 | 5.9 | 5.8 | 5.3 | 5.4 | 6.3 | 6 | 6.1 |
| 11 | 8.5 | 7.3 | 6.2 | 6.5 | 5.6 | 6.3 | 6.2 | 6.2 | 5.5 | 5.5 | 6.4 | 6.4 | 6.4 |
| 12 | 8.7 | 7.6 | 6.3 | 6.8 | 6.2 | 6.6 | 6.6 | 6.6 | 6.1 | 5.9 | 6.7 | 6.4 | 6.7 |
| 13 | 8.8 | 7.7 | 6.7 | 7.2 | 6.7 | 7.2 | 6.8 | 7.1 | 6.6 | 6.4 | 7 | 6.4 | 7.1 |
| 14 | 8.5 | 7.7 | 7.1 | 7.8 | 7.2 | 7.2 | 7.3 | 7.5 | 7.1 | 6.5 | 7.2 | 6.3 | 7.3 |
| 15 | 8.4 | 7.2 | 7.3 | 8.1 | 7.4 | 7.3 | 7.5 | 7.6 | 7.4 | 6.5 | 6.7 | 6.1 | 7.3 |
| 16 | 7.9 | 7.4 | 7.3 | 8.3 | 7.6 | 7.5 | 7.6 | 7.9 | 7.5 | 6 | 6.2 | 5.8 | 7.2 |
| 17 | 7.4 | 6.7 | 7.1 | 8.2 | 7.5 | 7.2 | 7.4 | 7.6 | 7.2 | 5.4 | 6 | 5.4 | 6.9 |
| 18 | 6.4 | 6.1 | 6.8 | 7.3 | 7.2 | 7 | 7.1 | 7.1 | 6.7 | 5.3 | 6.1 | 5.6 | 6.6 |
| 19 | 6.2 | 5.4 | 6 | 6.5 | 6.4 | 6.6 | 6.3 | 6.4 | 5.8 | 4.9 | 6.3 | 5.4 | 6 |
| 20 | 6.6 | 5.5 | 5.9 | 5.9 | 5.5 | 5.8 | 5.4 | 5.6 | 5.5 | 4.5 | 5.9 | 5.6 | 5.6 |
| 21 | 6.9 | 5.8 | 5.4 | 6.1 | 5 | 5 | 4.6 | 5 | 5.5 | 4.6 | 6.1 | 5.5 | 5.5 |
| 22 | 6.8 | 5.9 | 5.2 | 6 | 4.7 | 4.9 | 4.4 | 4.5 | 5.1 | 4.3 | 6 | 5.7 | 5.3 |
| 23 | 6.8 | 6 | 5.3 | 6.1 | 4.3 | 4.6 | 4.5 | 4.3 | 4.7 | 4.3 | 6 | 5.7 | 5.2 |
| Mean | 7.5 | 6.5 | 6 | 6.4 | 5.5 | 5.8 | 5.7 | 5.6 | 5.6 | 5.2 | 6.3 | 5.9 | 6 |

Table 4.3: Monthly mean wind speeds (m/s), at a height 24.5 m *a.g.l.* for Zafarana [15]

| Zafarana | Jan | Feb | Mar | Apr | May | June | July | Aug | Sep | Oct | Nov | Dec | Year |
|---|---|---|---|---|---|---|---|---|---|---|---|---|---|
| 0 | 6.7 | 7.4 | 8.5 | 9.5 | 10.3 | 11.4 | 11 | 11.1 | 10.6 | 9.3 | 7.4 | 6.9 | 9.3 |
| 1 | 6.7 | 7.3 | 8.3 | 9.1 | 9.6 | 11 | 10.6 | 10.6 | 10 | 8.9 | 7.3 | 6.9 | 8.9 |
| 2 | 6.7 | 7.2 | 7.8 | 8.7 | 9.4 | 10.4 | 9.9 | 10.1 | 9.7 | 8.4 | 7.2 | 6.9 | 8.6 |
| 3 | 6.8 | 7 | 7.5 | 8.4 | 9.1 | 9.9 | 9.5 | 9.6 | 9.3 | 8.2 | 6.9 | 7 | 8.3 |
| 4 | 6.8 | 6.8 | 7.3 | 7.9 | 8.5 | 9.5 | 8.8 | 9.1 | 8.9 | 7.8 | 6.9 | 7.1 | 8 |
| 5 | 7 | 6.8 | 7.1 | 7.6 | 8.2 | 9 | 8.3 | 8.7 | 8.4 | 7.7 | 6.8 | 7.1 | 7.8 |
| 6 | 6.9 | 6.8 | 7 | 7.4 | 8 | 9 | 8 | 8.3 | 8.1 | 7.6 | 6.9 | 7.2 | 7.6 |
| 7 | 6.9 | 7 | 6.8 | 7.7 | 8.9 | 10 | 8.8 | 9.4 | 8.7 | 7.8 | 6.9 | 7.1 | 8.1 |
| 8 | 6.9 | 7.1 | 7.4 | 8.3 | 9.1 | 10.5 | 8.9 | 9.8 | 9.5 | 8.7 | 7.2 | 7.1 | 8.4 |
| 9 | 7.1 | 7.4 | 7.8 | 8.4 | 9.2 | 10.5 | 9.1 | 9.8 | 9.6 | 9 | 7.6 | 7 | 8.6 |
| 10 | 6.9 | 7.5 | 8 | 8.7 | 9.5 | 10.5 | 9.2 | 9.9 | 9.8 | 9.3 | 7.6 | 7 | 8.7 |
| 11 | 7.2 | 7.6 | 8.5 | 9 | 9.7 | 10.5 | 9.2 | 10 | 9.8 | 9.5 | 7.7 | 7.1 | 8.9 |
| 12 | 7.7 | 8 | 8.6 | 9.2 | 10 | 10.7 | 9.5 | 10.2 | 10 | 9.6 | 8.1 | 7.4 | 9.1 |
| 13 | 8.1 | 8.6 | 9 | 9.4 | 10.2 | 10.8 | 9.8 | 10.5 | 10.4 | 9.6 | 8.3 | 7.7 | 9.4 |
| 14 | 8.4 | 8.8 | 9 | 9.6 | 10.3 | 11 | 10 | 10.9 | 10.9 | 9.8 | 8.4 | 8.2 | 9.7 |
| 15 | 8.3 | 8.7 | 9.2 | 9.7 | 10.5 | 11.4 | 10.4 | 11.1 | 11.3 | 10 | 8.6 | 8.2 | 9.8 |
| 16 | 8.2 | 8.7 | 9.1 | 9.8 | 10.6 | 11.8 | 10.5 | 11.3 | 11.4 | 9.9 | 8.3 | 8.1 | 9.8 |
| 17 | 7.4 | 8 | 9 | 9.6 | 10.5 | 11.9 | 10.6 | 11.2 | 11 | 9.1 | 7.4 | 7.2 | 9.5 |
| 18 | 7 | 7.3 | 8.2 | 9 | 10.2 | 11.6 | 10.3 | 10.5 | 10 | 8.7 | 7.1 | 6.8 | 9 |
| 19 | 6.8 | 7.1 | 8.2 | 8.9 | 10.2 | 11.1 | 10 | 10.3 | 10.3 | 9.4 | 7.2 | 6.8 | 8.9 |
| 20 | 6.8 | 7.4 | 8.6 | 9.5 | 11 | 11.8 | 10.8 | 11.3 | 11.2 | 10 | 7.4 | 6.9 | 9.5 |
| 21 | 6.8 | 7.6 | 9 | 9.9 | 11.4 | 12.5 | 12 | 12 | 11.6 | 10.3 | 7.5 | 7 | 9.9 |
| 22 | 6.9 | 7.5 | 9 | 10 | 11.3 | 12.4 | 12 | 12.1 | 11.4 | 9.9 | 7.5 | 6.9 | 9.8 |
| 23 | 6.9 | 7.5 | 9 | 9.7 | 10.9 | 11.9 | 11.6 | 11.7 | 11 | 9.6 | 7.5 | 7 | 9.6 |
| Mean | 7.2 | 7.5 | 8.2 | 9 | 9.9 | 10.9 | 9.9 | 10.4 | 10.1 | 9.1 | 7.5 | 7.2 | 9 |

To calculate the WF generated power, the wind speed has to be measured at the WT hub. Therefore, the actual wind speed at the WT hub (80 m) is calculated as follows:

$$V_T = V.\left(\frac{H_T}{H}\right)^{\alpha} \tag{4.1}$$

Where:

V: Wind speed in m/s at H height in m,

$V_T$ : Wind speed at the WT hub height,

**α**: Correction factor.

The value of the correction factor is 0.1429 for the two sites, based on the location geography characteristics and wind direction and turbulence in the area of each site [16].

The output power of WT can be calculated using the power curve shown in Figure 3.7 and represented by Equation (4.2).

$$P(V_t) = \begin{cases} 0 & V_t < V_i \text{ or } V_t > V_o \\ (A + B \times V_t + C \times V_t^2) \times P_r & V_i \leq V_t < V_r \\ P_r & V_r \leq V_t < V_o \end{cases} \tag{4.2}$$

Where:

$$A = \frac{1}{(V_i - V_r)^2}\left\{V_i(V_i + V_r) - 4V_iV_r\left[\frac{V_i + V_r}{2V_r}\right]^3\right\} \tag{4.3}$$

$$B = \frac{1}{(V_i - V_r)^2}\left\{4(V_i + V_r)\left[\frac{V_i + V_r}{2V_r}\right]^3 - (3V_i + V_r)\right\} \tag{4.4}$$

$$C = \frac{1}{(V_i - V_r)^2}\left\{2 - 4\left[\frac{V_i + V_r}{2V_r}\right]^3\right\} \tag{4.5}$$

Where:

$P_r$: Rated power output of the WT

$V_t$: Wind speed at time (t)

$P(V_t)$: Power output of the WT at wind speed ($V_t$)

Assuming identical WTs, the total WF power is calculated as follows:

$$P_{WF}(V_t) = N_{WT} \times P_{WT}(V_t) \tag{4.6}$$

Where:

$P_{WF}(V_t)$: Electrical power generated by WF (MW)

$N_{WT}$: Number of WTs in the WF

The power generation in percentage of installed capacity ($\%P_{WF}(V_t)$) for each site will be:

$$\%P_{WF}(V_t) = \begin{cases} 0 & V_t < V_i \ or \ V_t > V_o \\ (A + B \times V_t + C \times V_t^2) & V_i \leq V_t < V_r \\ 1 & V_r \leq V_t < V_o \end{cases} \tag{4.7}$$

A 2 MW-WT type is assumed to be used. The installed WT capacity installed in Zafarana is about 500 MW WF. In addition to what is installed in Zafarana, it is assumed that a new 500 MW to be installed in two candidate sites, Zafarana or El Dabaa to have a total of 1000MW installed in the grid. The power generation for the two WFs can be calculated using the given wind data through Equations (4.1) to (4.7).

Two cases are evaluated:

Case-A: It examined the impact of installing new 500 MW in EL Dabaa to what already exists, 500 MW in Zafarana, which represents dispersed WFs in two different locations.

Case-B: A 1000 MW in Zafarana is assumed, which represents two WFs in the same region.

For each case under study (Case-A and Case B), three scenarios are discussed based on the used WT's characteristics. The 500 MW WTs already installed Zafarana site has cut-in, rated, and cut-out wind speeds which are 4, 10, and 23 respectively. The new 500 MW WTs to be installed in Zafarana and El Dabaa have the following characteristics:

Scenario-I: Cut-in, rated, and cut-out wind speeds of 4, 10, and 23 respectively.

Scenario-II: Cut-in, rated, and cut-out wind speeds of 4, 12, and 25 respectively.

Scenario-III: Cut-in, rated, and cut-out wind speeds of 4, 13, and 25 respectively.

Results in Figures 4.3 a to c show that, Case-A has a smoother annual generation in percentage of available capacity than that of Case-B. This is due to the effect of geographical distribution. In the meantime, it can be concluded that as the WT rated power and cut-out speeds increase, the output generated power becomes smoother and the impact of geographical distribution is better.

Similar to Figure 4.3, Figure 4.4 shows the same conclusions in the duration curves of wind power generation form, i.e. geographical spreading leads to a flattened duration curve for wind power production of Case-A than in Case-B. Furthermore, increasing the WT rated and cut-out speeds, reduces the overall generation level.

To illustrate clearly the smoothing effect, the variation in output power (in percentage of the installed capacity) from one reading to the next will be calculated using Equation (4.8) and the results are illustrated in Figure 4.5 a to c.

$$\Delta P_n = \begin{cases} P_n - P_{n-1} & for \quad 2 \le \mathrm{n} \le \mathrm{N}_r \\ P_1 - P_N & for \quad n = 1 \end{cases} \tag{4.8}$$

Where:

$\Delta P_n$: variation in output power (% installed capacity) from one reading to the next

$P_{n:}$ output power (% installed capacity) in according to reading number (n)

n: reading number;

$N_r$: total number of readings for each case (24*12=288)

From Figure 4.5, Table 4.4 shows that geographical distribution of WFs in Egypt will smoothen the aggregated wind energy production as shown in Case-A. Meanwhile, increasing the WT rated and cut-out speeds, reduces the output variations as well. Therefore, smaller output power variations means more efficient utilizing of the cross-section area of the overhead lines going to be used for transmitting the generated power. Thus, leading to economical cost saving.

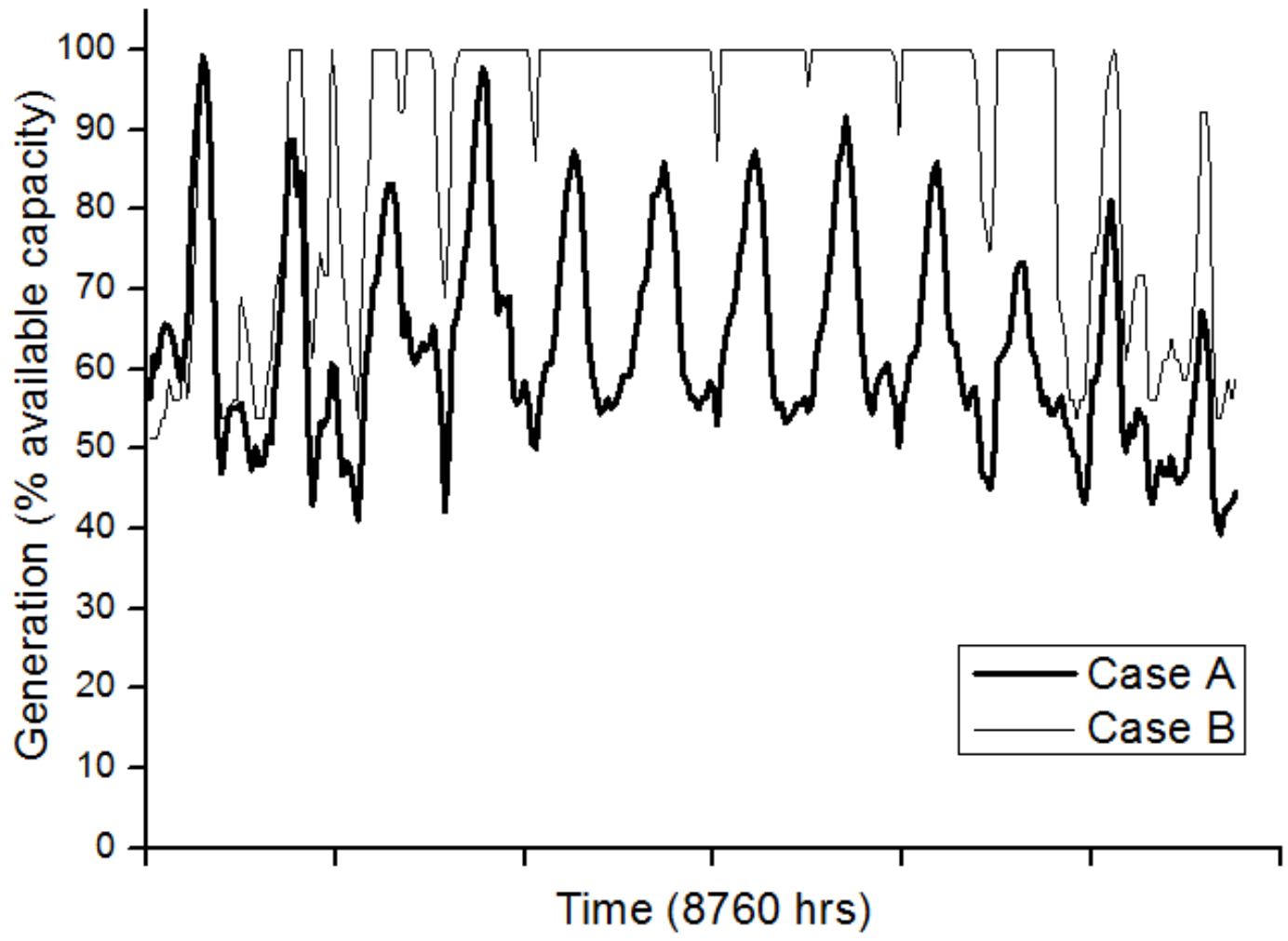

Figure 4.3 a: Scenario-I

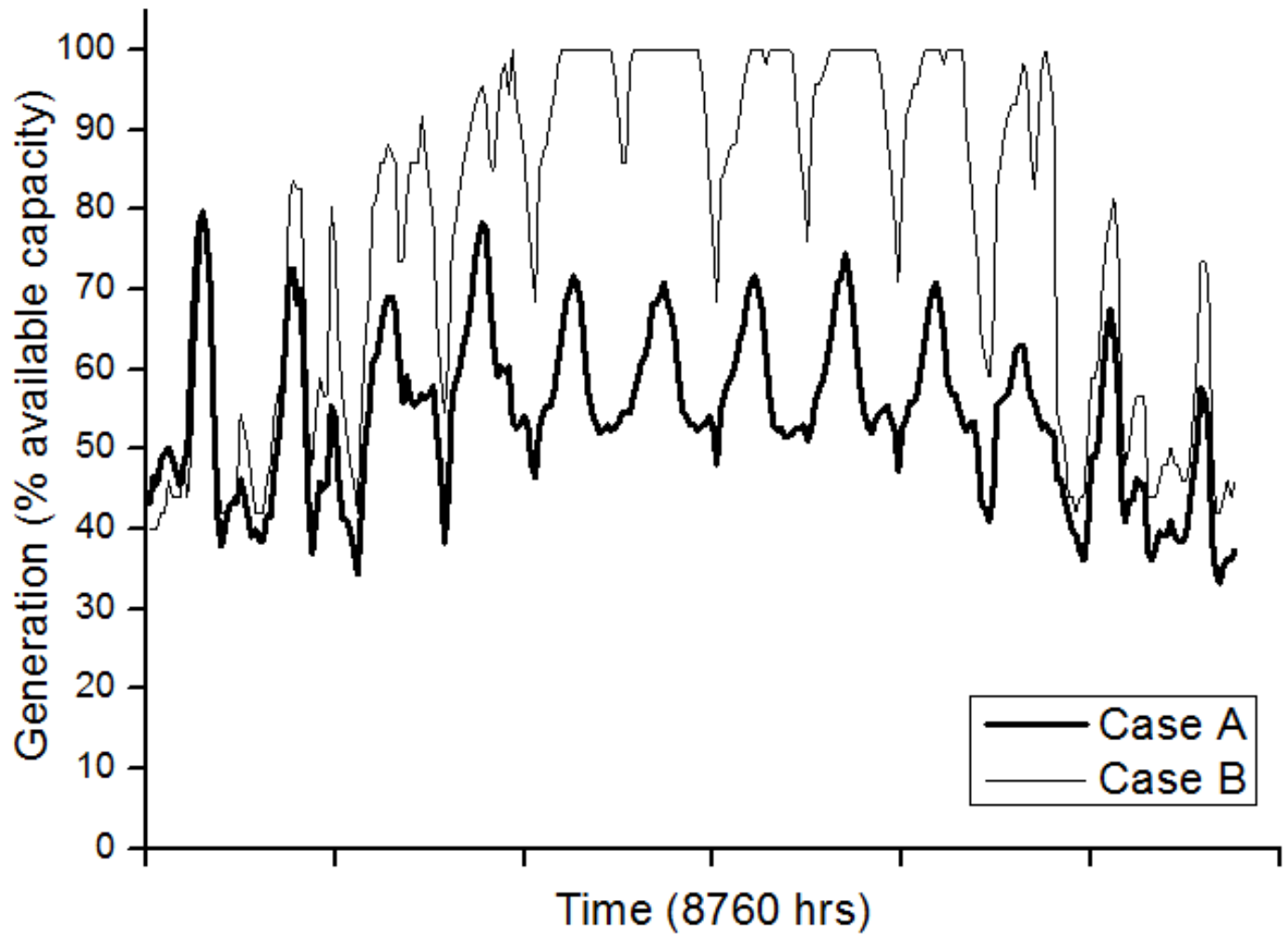

Figure 4.3 b: Scenario-II

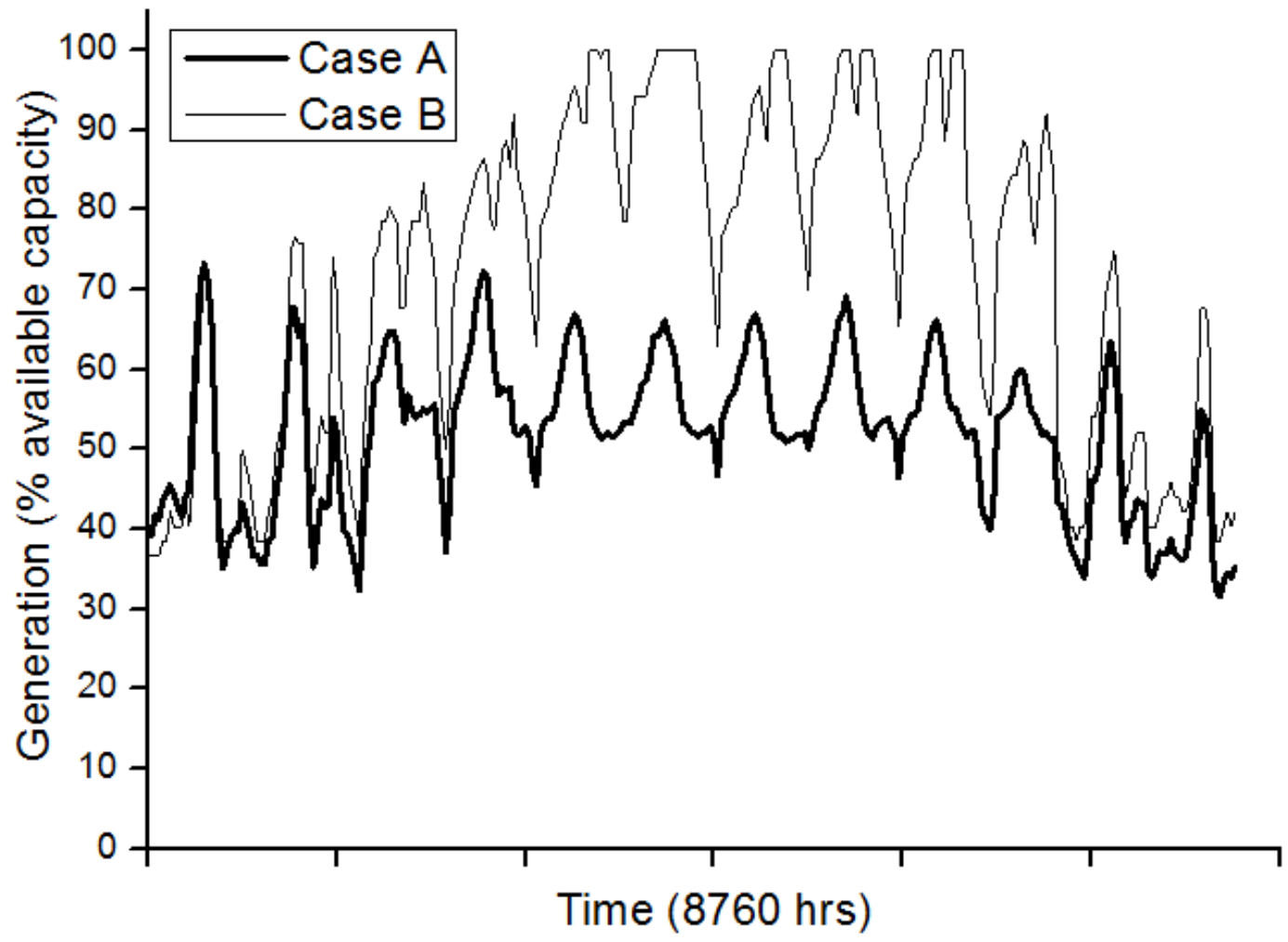

Figure 4.3 c: Scenario-III

Figure 4.3: Annual generation in percentage of available capacity

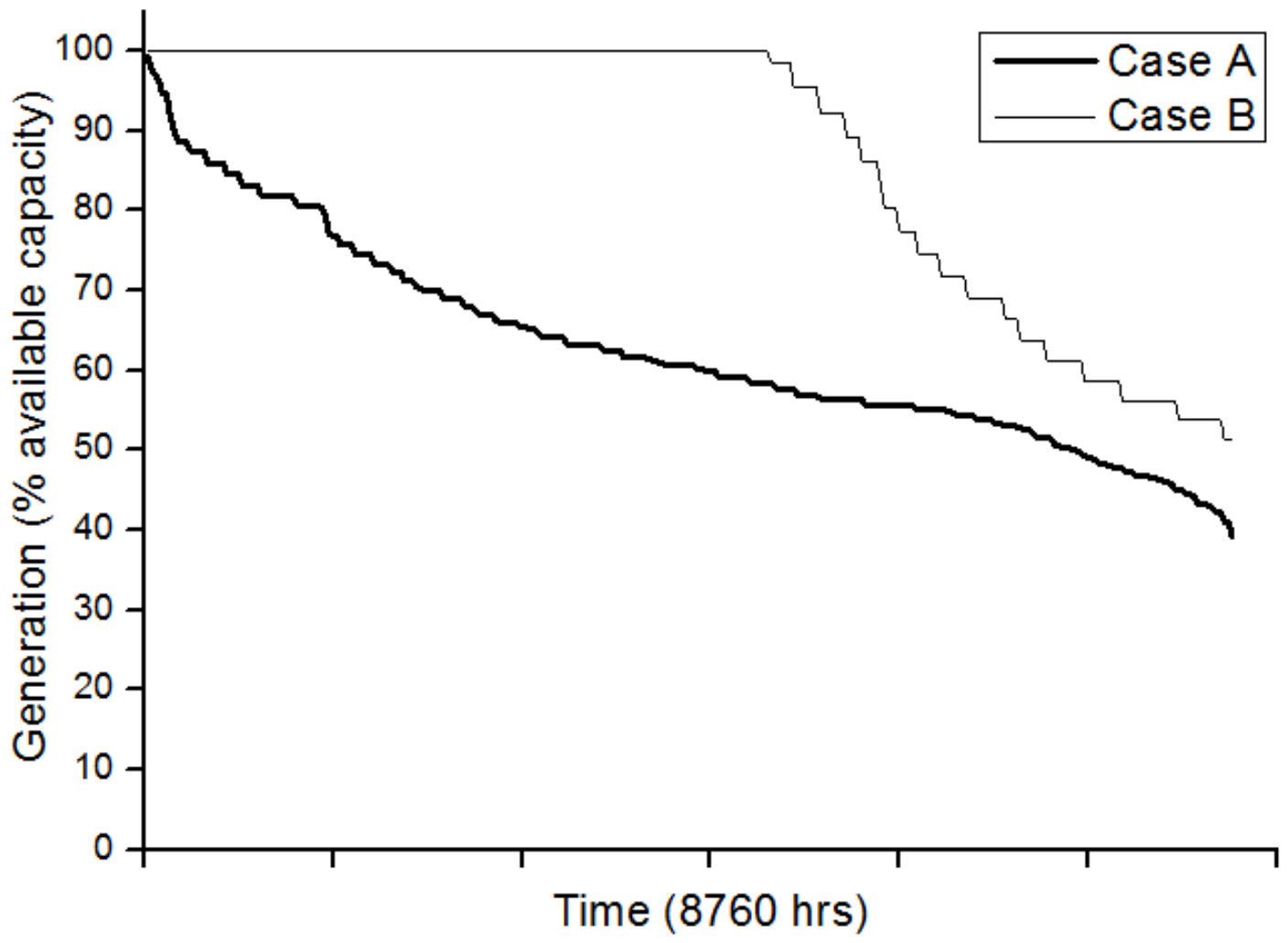

Figure 4.4 a: Scenario-I

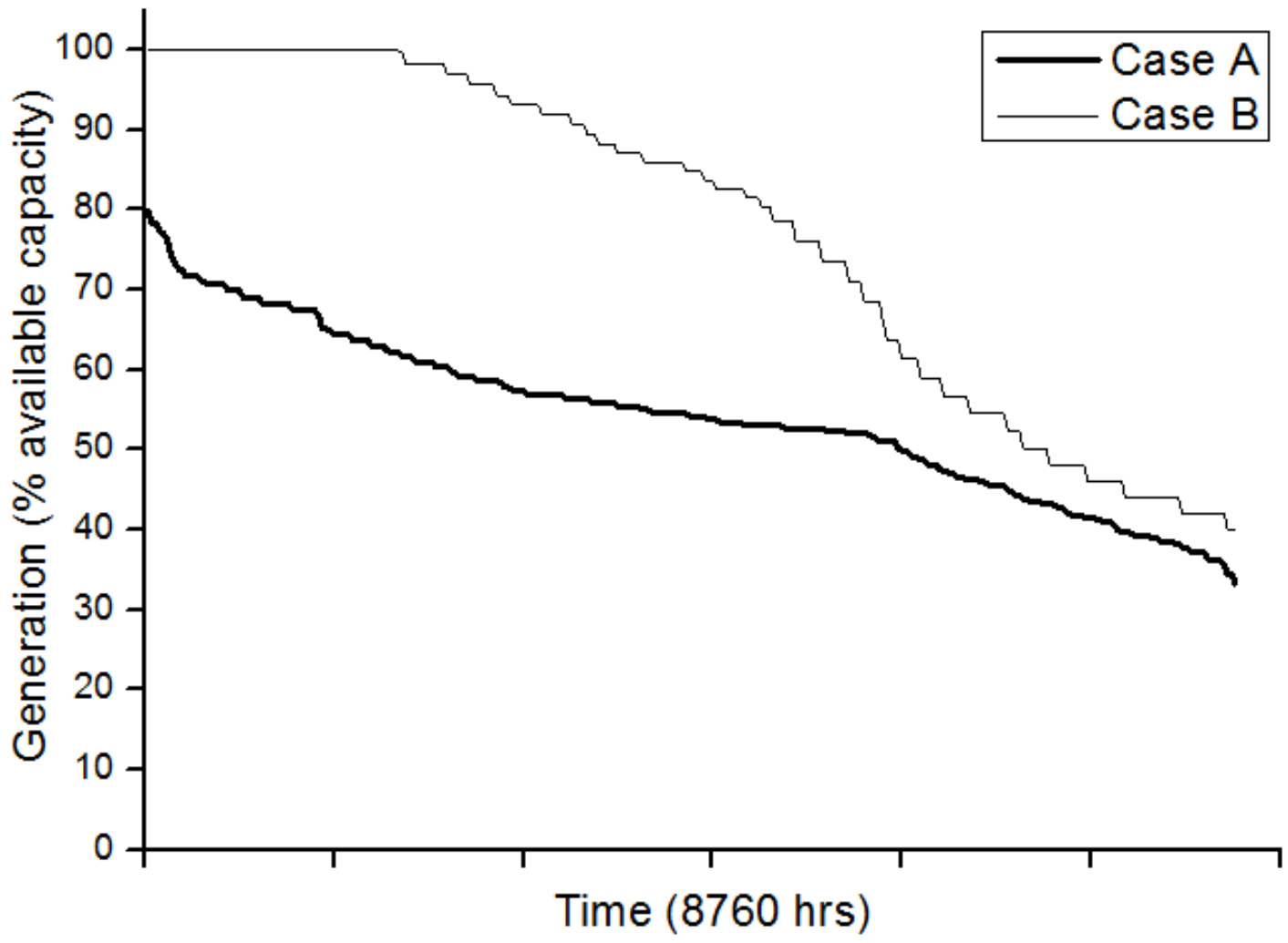

Figure 4.4 b: Scenario-II

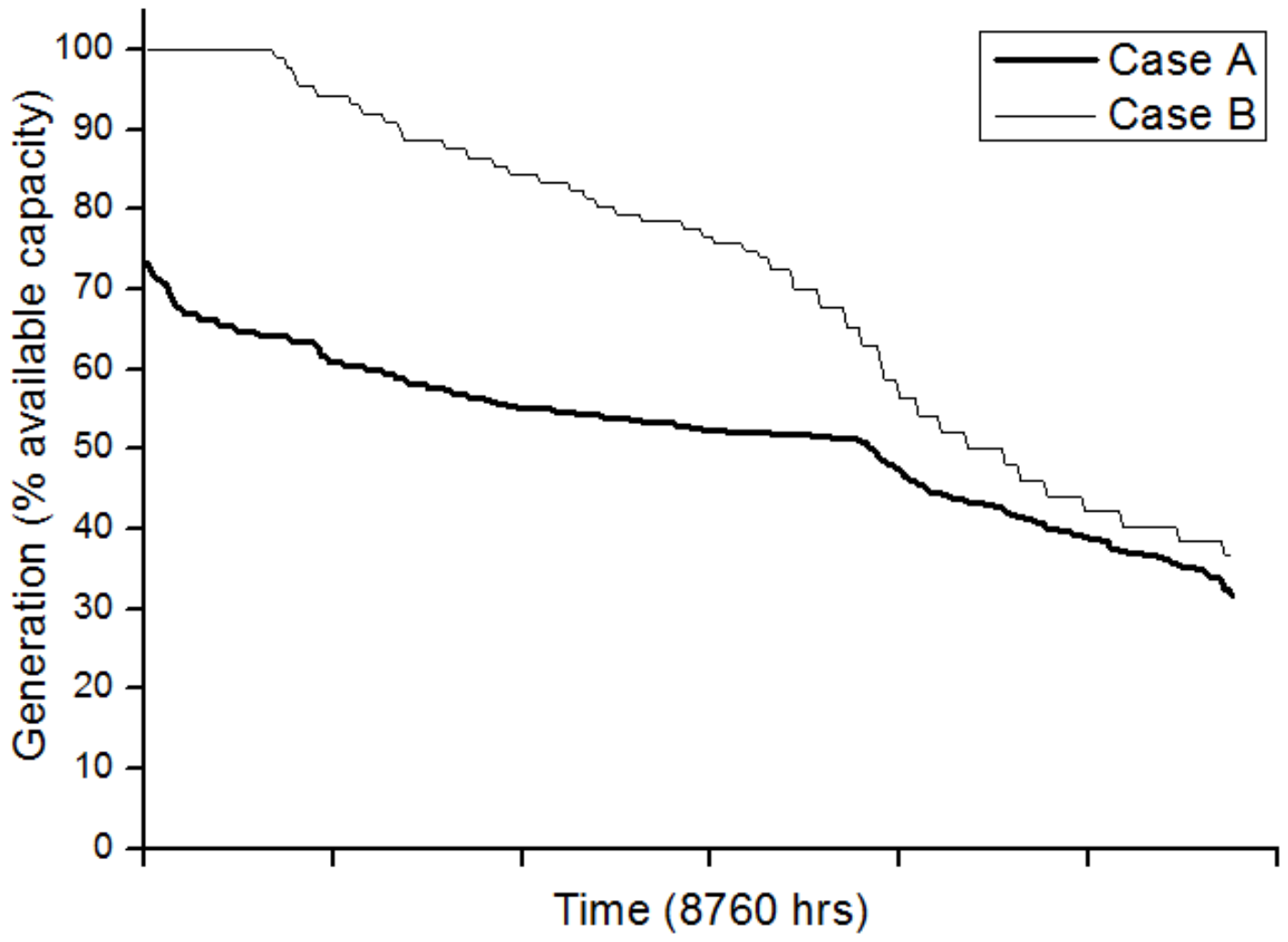

Figure 4.4 c: Scenario-III

Figure 4.4: Duration curves for the generation in percentage of available capacity cases

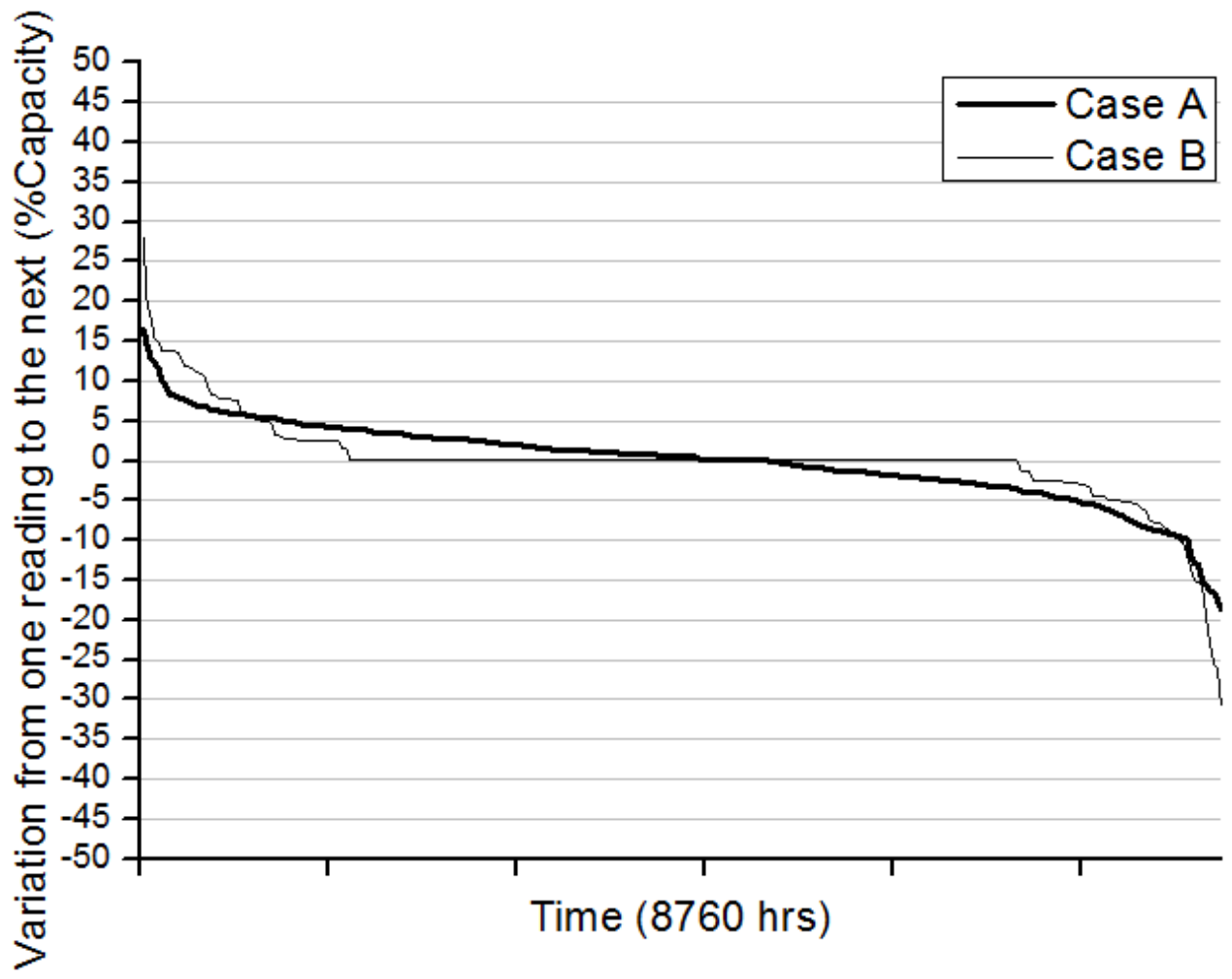

Figure 4.5 a: Scenario-I

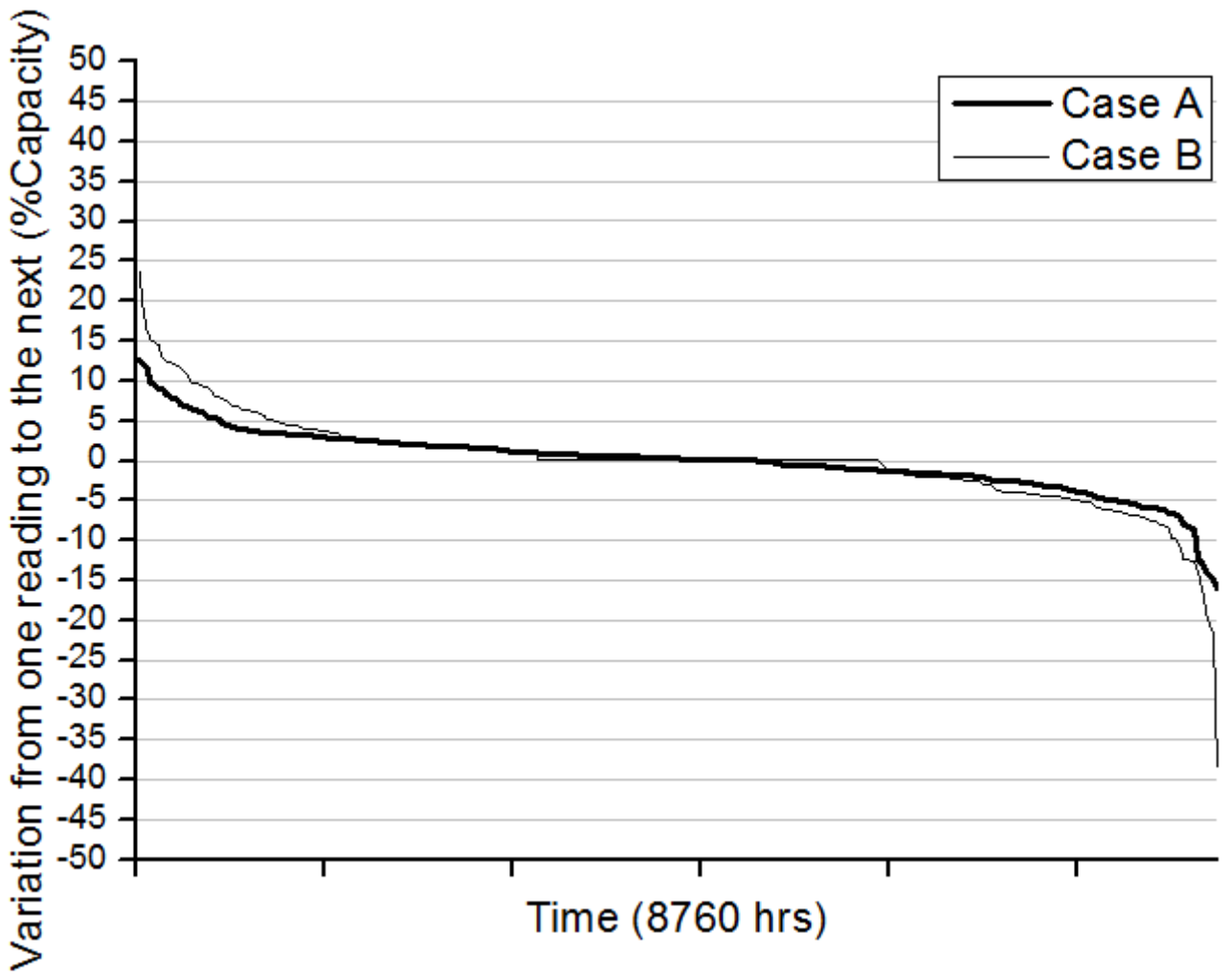

Figure 4.5 b: Scenario-II

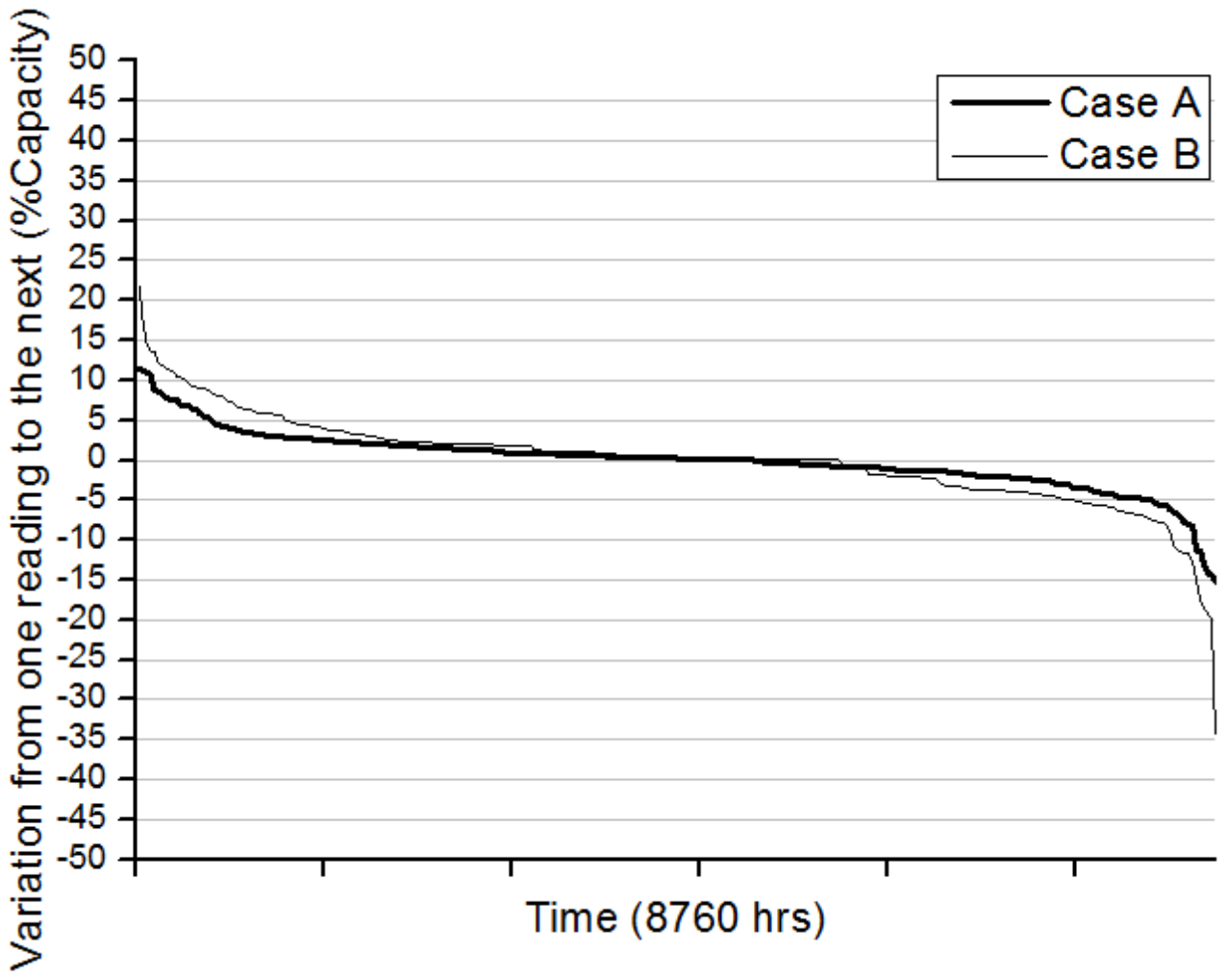

Figure 4.5 c: Scenario-III

Figure 4.5: Duration curves of monthly variations, as a percentage of the installed capacity

Table 4.4: Range of monthly variations, as a percentage of the installed capacity

| | Scenario-I | Scenario-II | Scenario-III |
|---|---|---|---|
| Case-A | 16 to -18 | 12 to -16 | 11 to -15 |
| Case-B | 28 to -31 | 24 to -37 | 22 to -34 |

## 4.2 Impact of geographical distribution on the capacity credit of wind energy

The capacity credit of a renewable plant is defined as the amount of conventional resources (mainly thermal) that could be replaced by renewable production without making the system less reliable [46].

With smaller grids, which are not large enough to benefit from geographical dispersion, the capacity credit may be smaller, or non-existent. A recent study of the Irish grid, for example, assumed that wind had no capacity credit, although it acknowledged that the evidence was conflicting and advocated further work to clarify the position. More recently, however, the network operator in Ireland has carried out an analysis of the impact of wind on the system and suggested that the capacity credit, with small amounts of wind, is about 30 per cent of the rated capacity of the WF, which is in line with most other European studies [22].

## 4.3 Wind energy Capacity Credit assessment (case study in Egypt)

It is hard to predict the behavior of wind; therefore, the evaluation of capacity credit provided by wind energy must be studied well. There are a number of methods to calculate the wind capacity values [46]:

*1. Probabilistic methods:* based on a system load duration curve. This method is best suited for system planners. The basic principles underlying probabilistic methods to assess the capacity credit of a wind plant are standard techniques normally used to evaluate the reliability of a power system. Probabilistic methods can be subdivided into two groups; Analytical and Simulation methods.

*2. The Effective Load Carrying Capability (ELCC):* is considered to be the preferred metric for evaluating the capacity value of a WF. The ELCC is typically calculated using a power system reliability model and the conventional ELCC calculation requires substantial reliability modeling. It is an iterative process and is computationally intensive. The Non-iterative method deals with minimal reliability modeling and is computationally less intensive than the conventional approach. Since calculation of ELCC involves considerable data and computational effort, some regions such as the Pennsylvania-New Jersey-Maryland (PJM) and Southwest Power Pool (SPP) use approximate methods to calculate wind capacity credit.

*3. Approximate methods:* are useful when the ELCC cannot be determined due to data or other limitations. In the PJM method, the capacity credit is calculated based on the wind generator's capacity factor during the hours from 3 PM to 7 PM, from June 1 through August 31. The capacity credit is a rolling three year average, with the most recent year's data replacing the oldest year data. On the otherhand, the New York Independent System Operator (NYISO) determines wind capacity credit using the wind generator's capacity factor between 2 PM and 6 PM from June through Aug and 4 PM through 8 PM from December through February. The California Public Utilities Commission (CPUC) uses a three year rolling average of the monthly average wind energy generation between 12 PM and 6 PM for the months of May through September.

Due to the limitation of data availability, the PJM method was implemented to calculate the capacity credit for EL-Zayt site- near to Zafarana. The data used is the grid peak hours in Egypt from 5 to 8 PM, during the period from May 1$^{st.}$ through August 31$^{st}$. Figure 4.6 shows the flow chart of calculating capacity credit using the PJM method. Here, the PJM method was applied on El-Zayt site using actual wind data of the years 2007, 2008, and 2009. Table 4.5 and Figure 4.7 show that the capacity credit of El-Zayt site is very high. This is due to the good correlation between wind speeds in this site and the grid peak times. Meanwhile, the PJM method was applied using different WT characteristics as discussed in Scenarios-I to III in geographical smoothing case study. Results show that as the rated and cut-out speeds increase, the capacity credit of the WF decreases as less power is generated from wind.

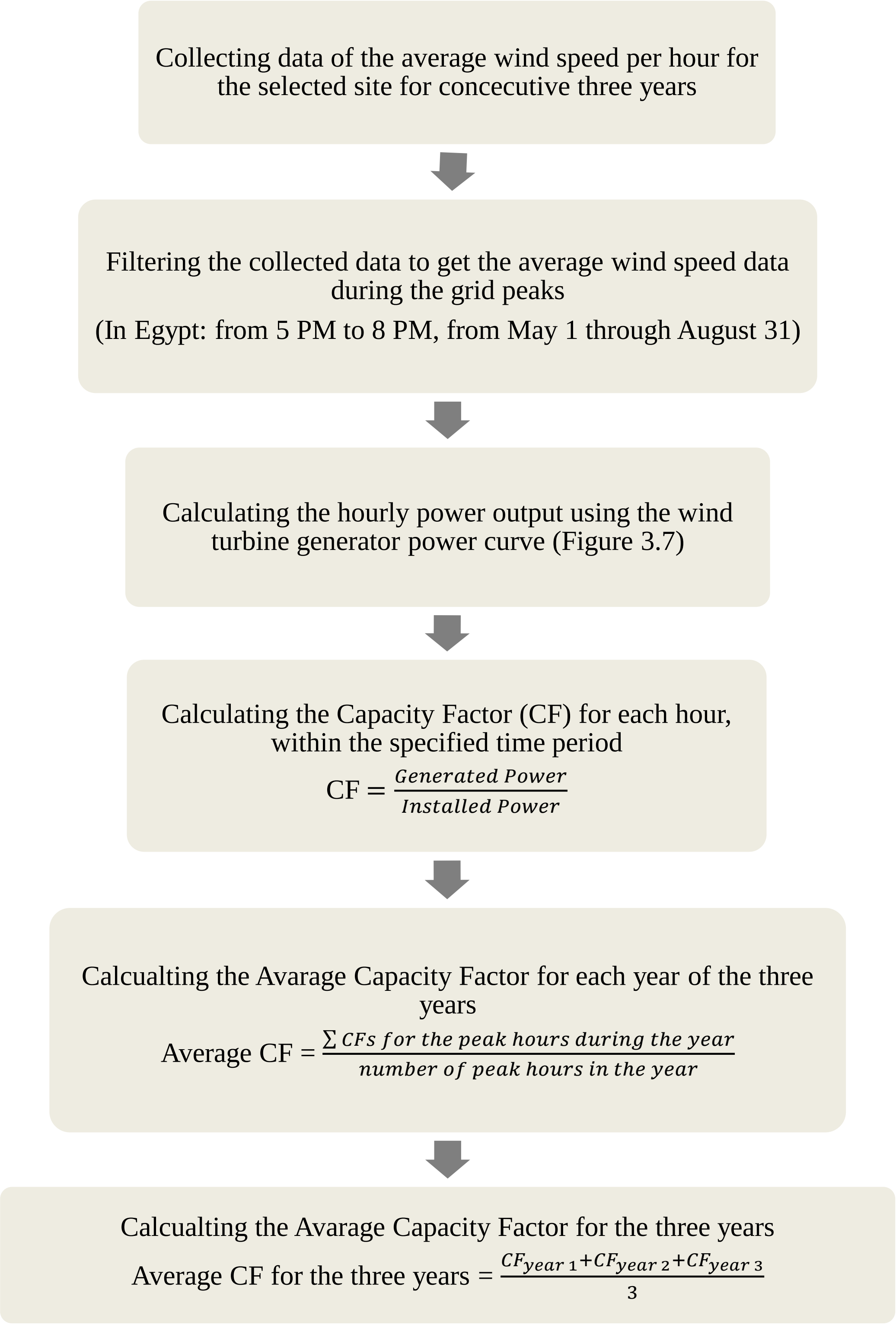

Figure 4.6: Flow Chart for Capacity Credit calculations using the PJM method

Table 4.5: Capacity Credit (%) for El Zayat Site for WTs used in various scenarios versus the times of peak loads

| Month, hour | Capacity Credit (%) | | |
|---|---|---|---|
| | Scenario (1) | Scenario (2) | Scenario (3) |
| May, 5 PM | 77.1 | 68.3 | 63.3 |
| May, 6 PM | 70.4 | 59.6 | 53.8 |
| May, 7 PM | 64.0 | 53.8 | 48.5 |
| May, 8 PM | 62.7 | 55.2 | 51.3 |
| June, 5 PM | 85.7 | 77.2 | 71.7 |
| June, 6 PM | 81.6 | 73.2 | 68.2 |
| June, 7 PM | 80.7 | 74.2 | 70.9 |
| June, 8 PM | 84.0 | 78.4 | 74.9 |
| July, 5 PM | 73.5 | 60.0 | 52.4 |
| July, 6 PM | 64.2 | 52.3 | 46.5 |
| July, 7 PM | 63.7 | 55.0 | 51.1 |
| July, 8 PM | 71.4 | 64.7 | 61.2 |
| Aug, 5 PM | 80.1 | 62.7 | 53.7 |
| Aug, 6 PM | 71.5 | 55.6 | 48.5 |
| Aug, 7 PM | 71.2 | 59.8 | 53.8 |
| Aug, 8 PM | 79.2 | 69.9 | 64.2 |
| Average | 73.8 | 63.7 | 58.4 |

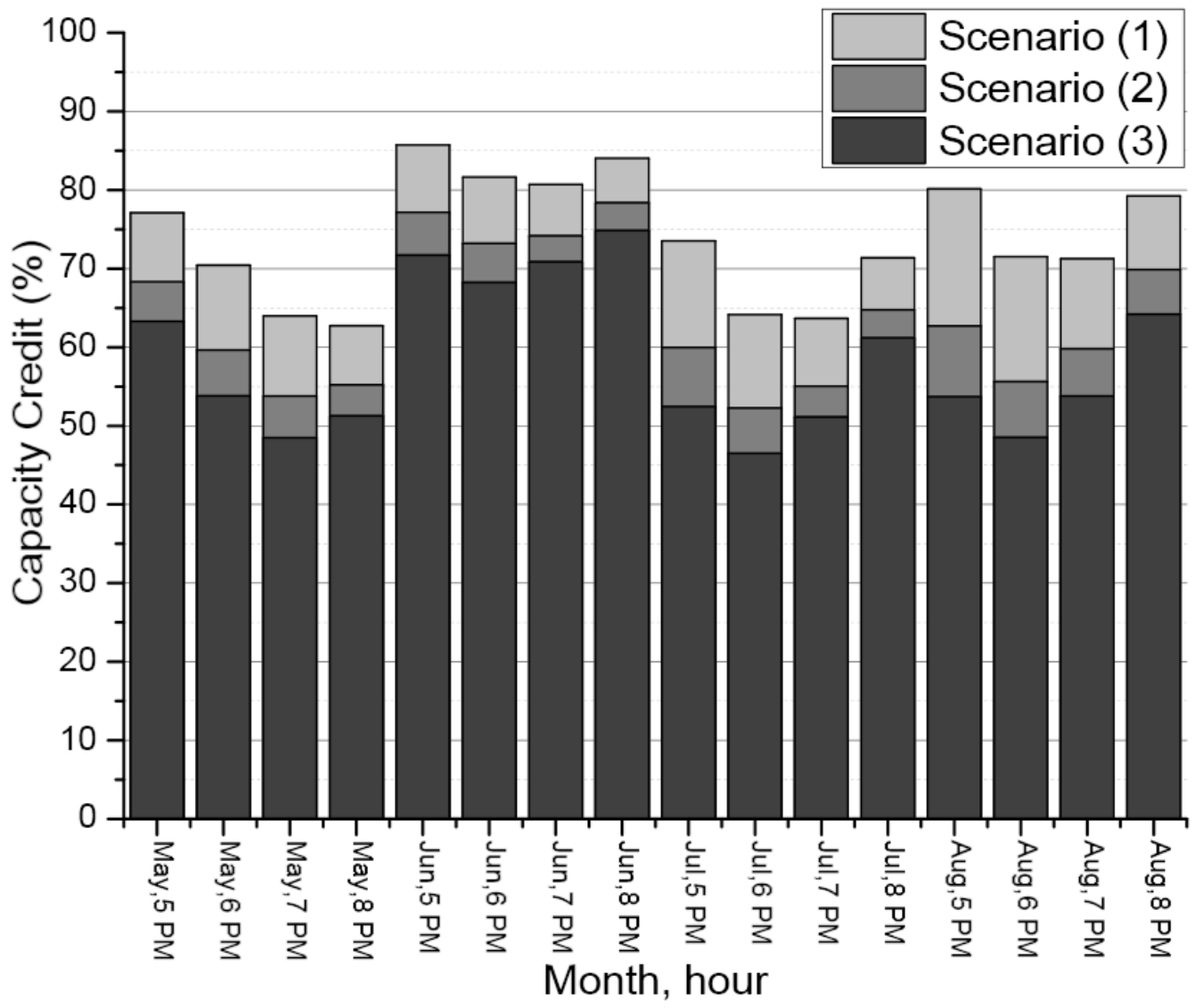

Figure 4.7: Capacity Credit (%) for El Zayat Site for WTs used in various scenarios versus the times of peak loads

## 4.4 Strategic Planning (case study in Egypt)

Thermal plants with the same value as the WFs' capacity credit have to be installed to satisfy the system's required reliability. In this section, it is proposed to set the timing and sizing of NPP to be installed in Egypt to cover the new WFs capacity credit. In practical applications, NPP will be running as base load, while, the same value of already existing load following thermal plants will be used to cover the wind fluctuations and balance the demand.

The previously mentioned cases (Case-A and B) and scenarios (Scenarios-I to III) are carried out to install a total of 3000 MW WFs. It has to be mentioned that around 500 MW is already installed at Zafarana, i.e. 2500 MW will be newly installed which will be distributed as follows:

- In Case-A, 1500 MW will be installed in Zafarana and another 1500 MW in El Dabaa region; and
- In Case-B, the whole 3000 MW will be in the Zafarana/Red Sea region.

Table 4.6 shows the coordination of the NPP timing and sizing with respect to WFs installations. For both cases A and B, as the rated and cut-out speeds increase, i.e. from Scenario-I to III, the output generated power decreases. This leads to reducing the capacity credit of the WF and in return the timing and sizing value of the NPP to be installed. Table 4.6 shows the values of NPP that cover the WF capacity credit for each scenario under study. The gray cells indicate that the timing and the rest of the NPP sizing can be delayed giving room for decision makers to evaluate the existing NPP installations. These results can play a significant role in the planning phase of selecting the WTs characteristics. Even though, Case-A (which represents the WF geographical distribution), generates less wind power, it gives other technical, economical benefits as well as provides more room for feedback from decision markers. That is shown by more gray cells than those reported by Case-B.

Table 4.6: NPP Timing and Sizing (MW) Selection Based on WFs MW Installations

| Year | | Till 2012 | 2016 / 2017 | 2017 / 2018 | 2018 / 2019 | 2019 / 2020 | 2020 / 2021 |
|---|---|---|---|---|---|---|---|
| Case-A | | | | | | | |
| WF | Z | 500 | 500 | 500 | | | |
| | D | | | | 500 | 500 | 500 |
| NPP-Scenario | I | | | | 2000 | | |
| | II | | | | 1000 | | |
| | III | | | | 1000 | | |
| Case-B | | | | | | | |
| WF | | 500 | 500 | 500 | 500 | 500 | 500 |
| NPP-Scenario | I | | | | 2000 | | 1000 |
| | II | | | | 2000 | | |
| | III | | | | 1000 | 1000 | |

Z: Zafarana, D: El Dabaa

## 4.5 Levelized Cost of Energy (LCOE)

The levelized cost of energy (LCOE) is defined as the constant price per unit of energy that causes the present value of the lifetime revenue equals the present value of the lifetime costs. In other words, it is the constant price per unit of energy that causes the investment to just break even: earn a present discounted value equal to zero. For any power plant, the levelized cost of energy can be calculated using Equations (4.9) to (4.17). Figure 4.8 shows the procedure of calculating the LCOE.

$$Total\ Investment\ (US\$) = 1000\ \times\ Capacity\ (MW)\ \times\ Capital\ Cost\ (\$/KW) \tag{4.9}$$

$$\text{Annual Capital Cost (US\$)} = \begin{cases} \dfrac{\text{Total Investment(US\$)}}{\text{Construction time (Year)}} & \text{during the construction period} \\ 0 & \text{during the life time} \end{cases} \tag{4.10}$$

$$Annual\ Energy\ Generation\ (MWh) = Capacity\ (MW) \times Capacity\ Factor\ \times\ 8760 \tag{4.11}$$

$$\begin{aligned} &Annual\ Fuel\ Costs (US\$) \\ &\qquad = Annual\ Energy\ Generation\ (MWh) \\ &\qquad \times\ Fuel\ Costs\ (US\$/MWh) \end{aligned} \tag{4.12}$$

$$\begin{aligned} &Annual\ \ Variable\ O\&M\ Costs (US\$) \\ &\qquad = Annual\ Energy\ Generation (MWh) \\ &\qquad \times\ Variable\ O\&M\ Costs (US\$/MWh) \end{aligned} \tag{4.13}$$

$$\begin{aligned} &Total\ annual\ cost\ (US\$) \\ &\qquad = Annual\ Capital\ Cost\ (US\$) \\ &\qquad +\ Annual\ Fixed\ O\ \&\ M\ Costs\ (US\$) \\ &\qquad + Annual\ Variable\ O\&M\ Costs\ (US\$) \\ &\qquad +\ Annual\ Fuel\ Costs\ (US\$) \end{aligned} \tag{4.14}$$

$$Total\ discounted\ annual\ cost\ (US\$) = \frac{Total\ annual\ cost\ (US\$)}{(1 + Discount\ Rate\ (\%))^{(t-1)}} \tag{4.15}$$

$$\begin{aligned} &Discounted\ annual\ energy\ generation\ (MWh) \\ &\qquad = \frac{Annual\ energy\ generation\ (MWh)}{(1 + Discount\ Rate\ (\%))^{(t-0.5)}} \end{aligned} \tag{4.16}$$

$$\begin{aligned} &Levelized\ Cost\ of\ Energy\ \left(\frac{US\$}{MWh}\right) \\ &\qquad = \frac{\sum_1^{\mathrm{T}} \text{Total discounted annual cost (US\$)}}{\sum_1^{\mathrm{T}} \text{Discounted annual enery generation (MWh)}} \end{aligned} \tag{4.17}$$

Where:

t = Year number

T = WF’s Life Time (Year) + WF’s Construction Time (Year)

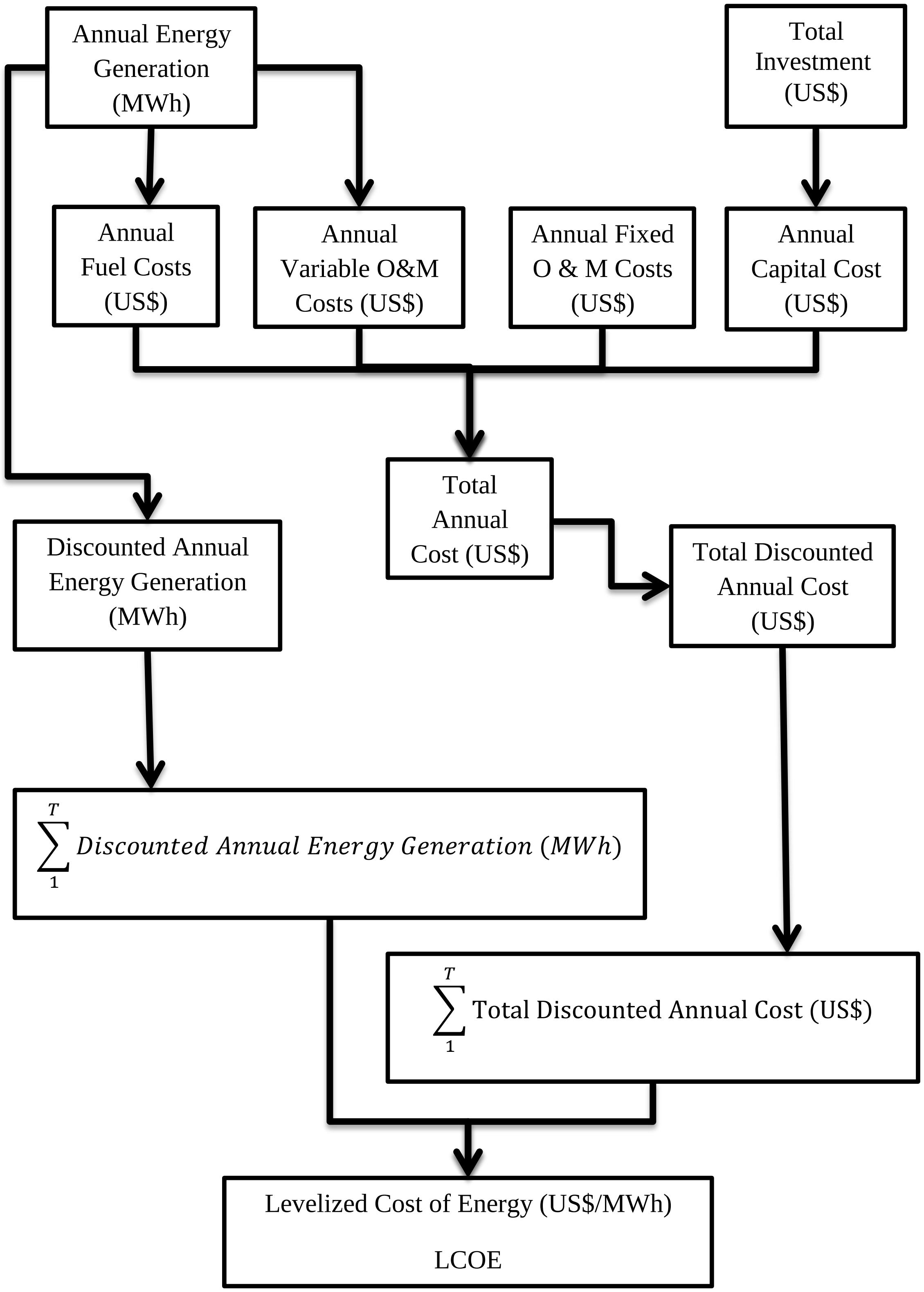

Figure 4.8: Flow chart for the procedure of calculating LCOE for any power plant

Based on the international experience of on shore WFs implementation, Table 4.7 is developed. It is based on data collected from twelve countries. Median case will be taken as a reference for calculations in this thesis. There are four main factors that affect the LCOE, which are: Capacity Factor, Discount Rate, Capital Cost, and Operation & Maintenance Cost. Figures (4.9) to (4.12) illustrate the impact of each factor on the levelized cost of energy when other factors are kept constant and equal to the values of median case. Discount Rate is assumed to be 8 % for the median case.

Table 4.7: Data for electricity generation costs from on shore WFs around the world [47]

| On shore WF | Capacity (MW) | Capacity Factor (%) | Capital Cost ($/KW)* | O & M Cost ( $/MWh)* |
|---|---|---|---|---|
| # of countries | 12 | 12 | 12 | 12 |
| count** | 13 | 13 | 13 | 13 |
| max | 150 | 41 | 3716.22 | 42.78 |
| min | 2 | 20.5 | 1845 | 8.63 |
| mean | 56 | 27.2 | 2422.64 | 23.79 |
| median | 45 | 25.7 | 2348.64 | 21.92 |
| delta | 148 | 20.5 | 1871.22 | 34.15 |
| Std. dev | 57 | 5.5 | 575.92 | 10.21 |

* All costs are expressed in US $ (2008 average values)

**Count refer to the number of data points or plants taken into account

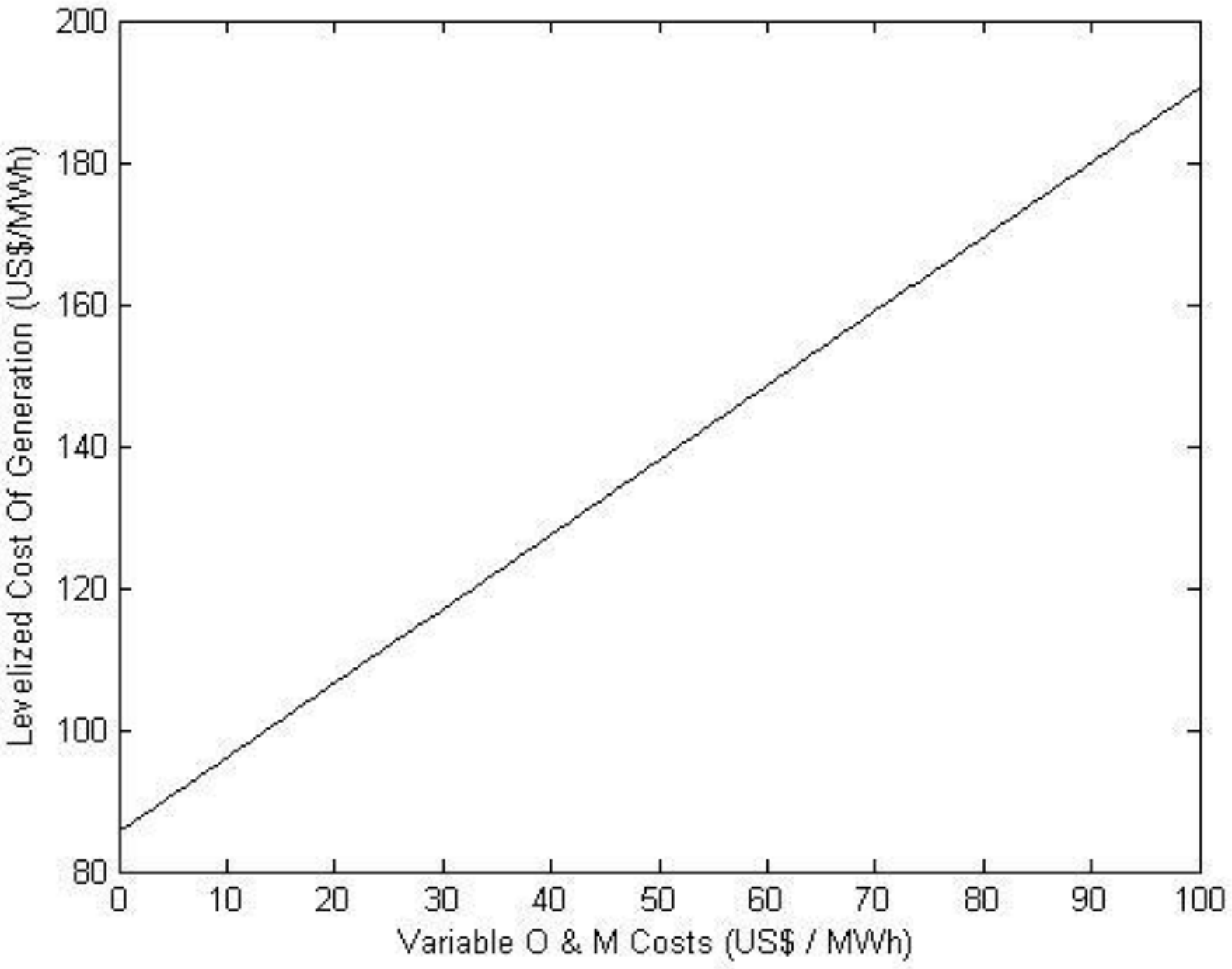

Figure 4.9: Relation between Variable O & M Costs and LCOE with other factors equals to that of the median case in Table 4.7

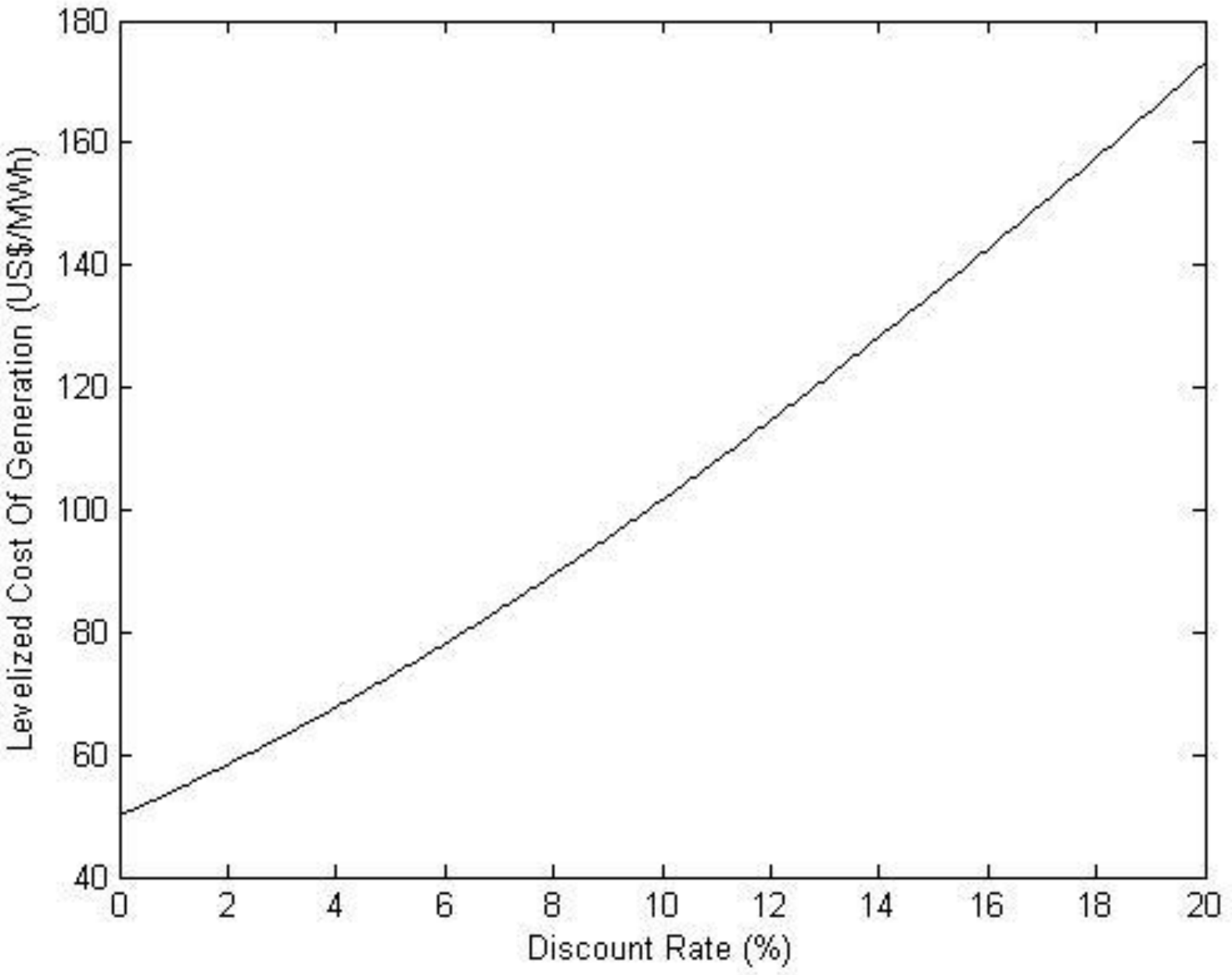

Figure 4.10: Relation between Discount Rate and LCOE with other factors equals to that of the median case in Table 4.7

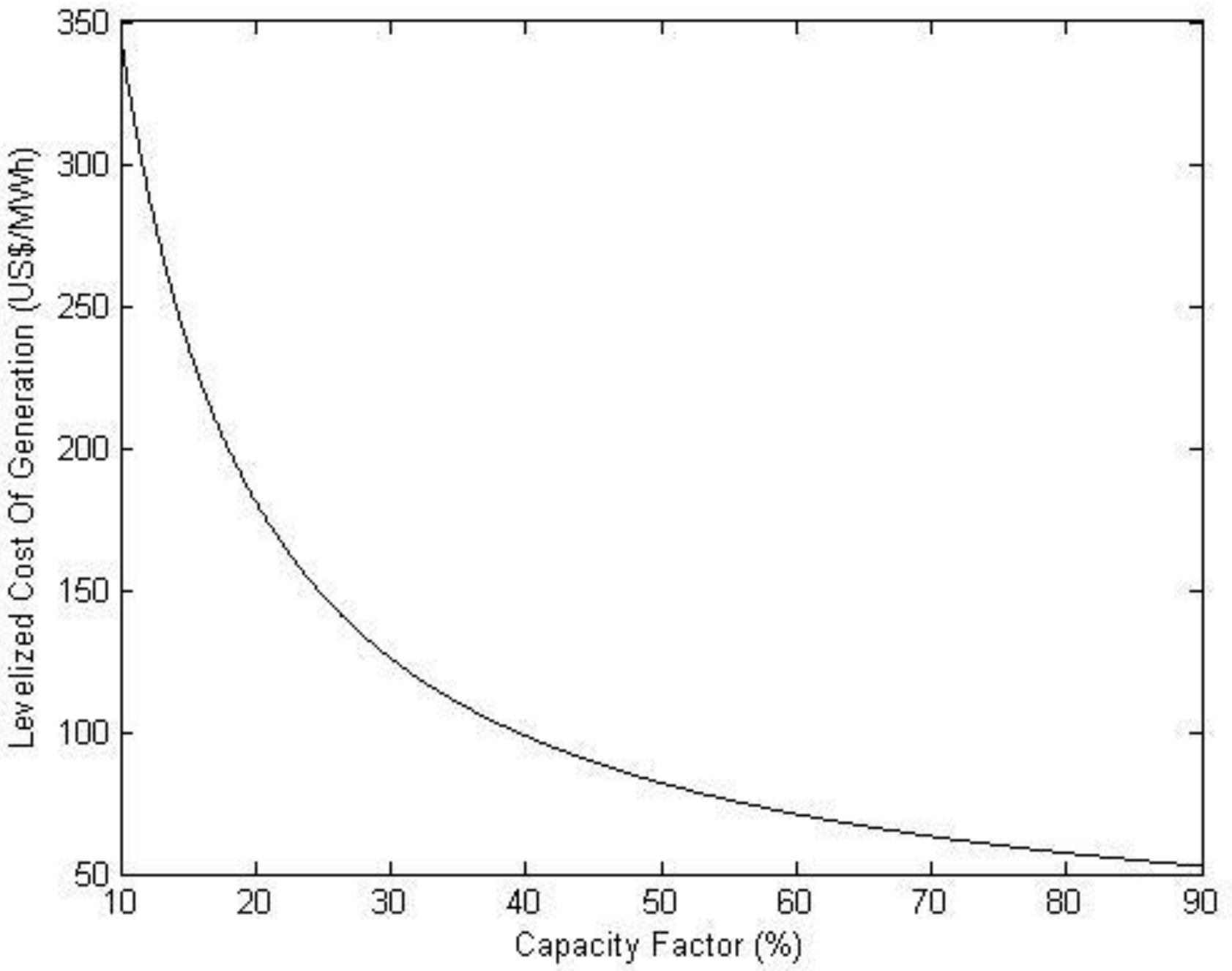

Figure 4.11: Relation between Capacity Factor and LCOE with other factors equals to that of the median case in Table 4.7

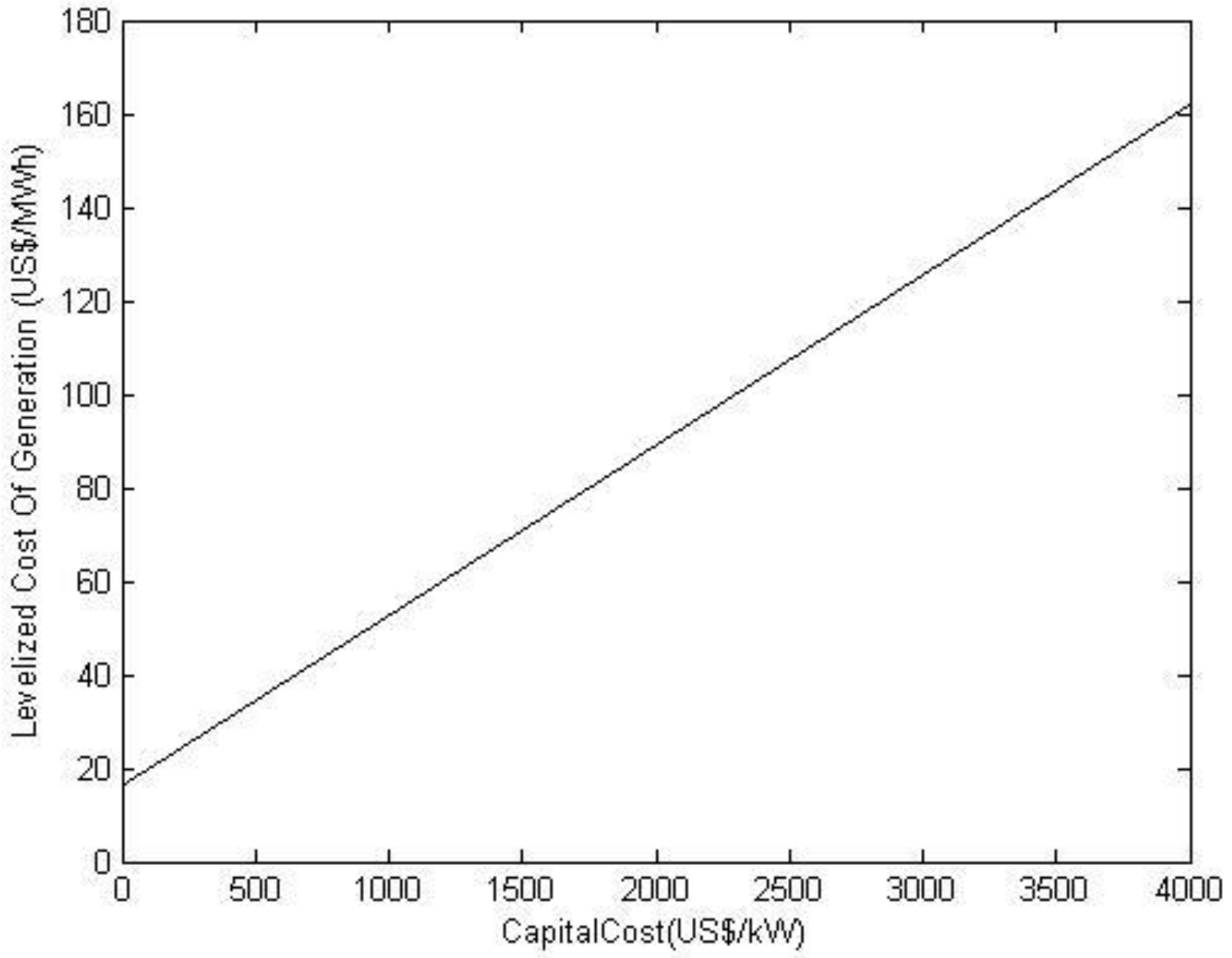

Figure 4.12: Relation between Capital Cost and LCOE with other factors equals to that of the median case in Table 4.7

### 4.5.1 Comparing LCOE for Cases A & B and Scenarios I, II, & III

LCOE will be calculated for different cases A & B, and different scenarios I, II, & III which are described in section 4.1.1. Life time and discount rate will be the same for all cases and scenarios. On the other hand, capacity factor and capital and operation & maintenance costs are site and/or WT specific, therefore values of these parameters are different for different cases and scenarios. As stated before, coupling of WFs with NPPs has some advantages and disadvantages. Some of these advantages and disadvantages are related to the economics of the WF, and they can be summarized in Table 4.8 with the projected increase/decrease in capital and operation & maintenance costs compared to median case. Other advantages and disadvantages are ignored as they do not have direct economical impact, but they have technical one which cannot be translated to economical impact. On the other hand, suitable site for coupling may have lower capacity factor, and so lower energy production for a certain installed capacity, compared to other potential sites in the country. This is clear in Figure 4.4, where case A which represents the case of coupling has lower energy production compared to case B. Therefore, to take a right decision which is better case A or B, the two cases have to be compared considering all key parameters which affect the energy production costs. This can be achieved by comparing LCOE for each case.

Applying Equations (4.9) to (4.17) to cases A & B, and considering the three scenarios, Figures 4.13, 4.14, and 4.15 can be obtained. Each figure represents the application of the two cases using one scenario. Any installations in Zafarana site, old or new, have the same costs of median case in Table 4.7. Installations in El Dabaa site are expected to have lower costs compared to Zafarana site (median case) due to coupling. LCOE for the two cases A & B are calculated considering different levels of reduction in capital cost and operation & maintenance cost for installations in El Dabaa site.

In Figures 4.13 to 4.15, LCOE for the two cases are drawn versus the capital cost reduction (in percentage of the median case) for installations in El Dabaa site. Also, the operation & maintenance cost reduction (in percentage of the median case) for installations in El Dabaa site is taken as a parameter. It is clear that with no reduction in both types of cost, LCOE for case B is cheaper. This is applicable for all scenarios. Considering Figure 4.13 (scenario I) LCOE for the two cases are equal when the capital cost reduction is about 55%. Above this value of capital cost reduction or with any reduction in the operation & maintenance cost, LCOE of case A will be cheaper. This is applicable also for Figures 4.14 and 4.15 with slightly different values of capital cost reduction required to achieve equality between the two cases. Therefore, for the coupling (case A) to be feasibly from the economical point of view, it has to achieve the required reduction in capital cost and/or operation and maintenance cost and so lower LCOE or at least to be equal to that of case B.

Table 4.8 presents some aspects associated with coupling and their projected increase/decrease in capital costs and/or operation & maintenance cost. It is based on the previous illustration of whole advantages and disadvantages of coupling. According to this table, coupling will decrease the operation & maintenance cost with about 10%, and the decrease in capital cost ranges from 45% to 65%. Although there is a great difference in the capacity fact between the two cases, expected reduction in costs is sufficient to make LCOE for case A cheaper that of case B or at least both will be equal. Therefore, the idea of coupling is economically feasible.

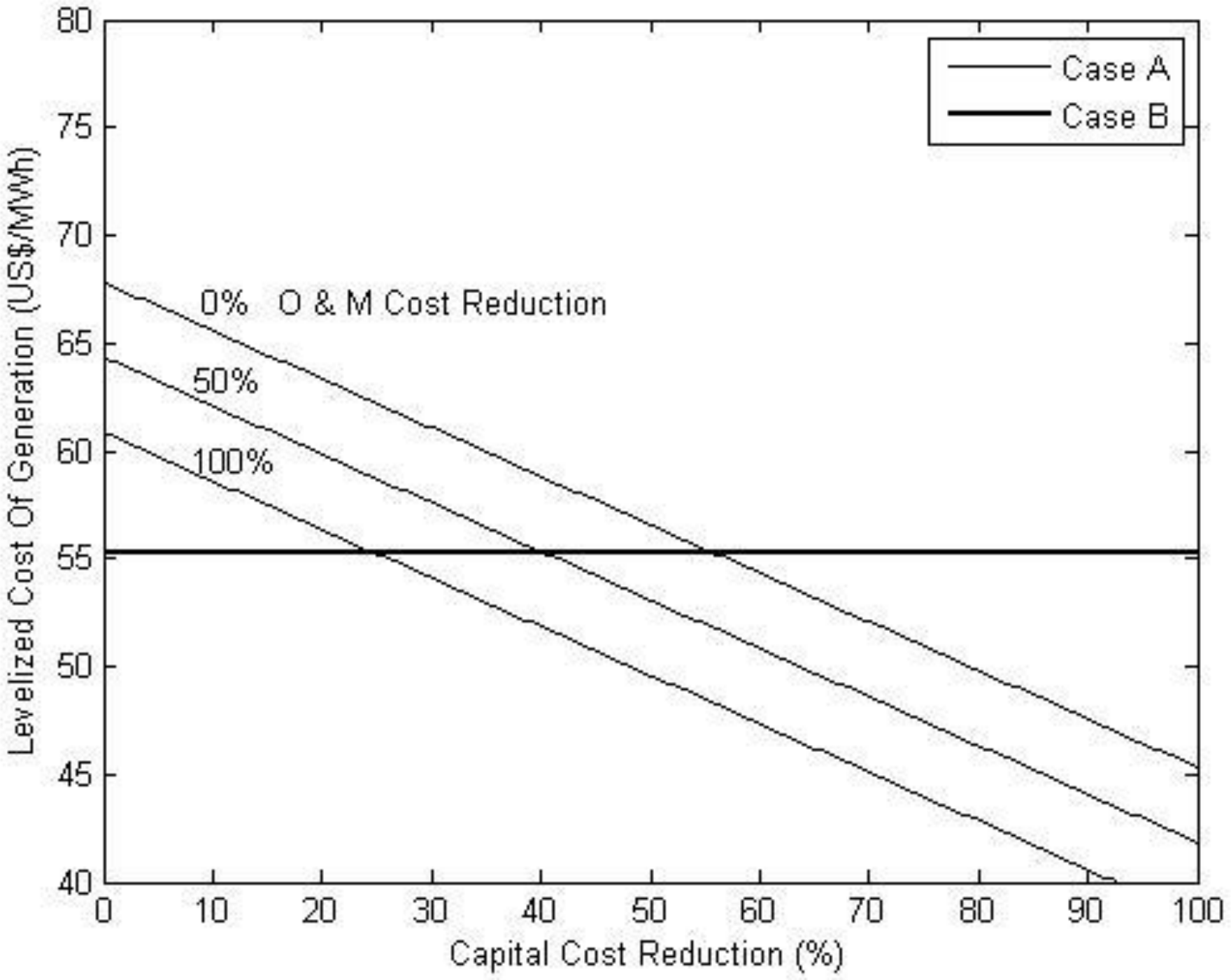

Figure 4.13: Scenario I: LCOE for cases A & B versus capital cost reduction (%) due to coupling, with operation & maintenance cost reduction (%) as a parameter

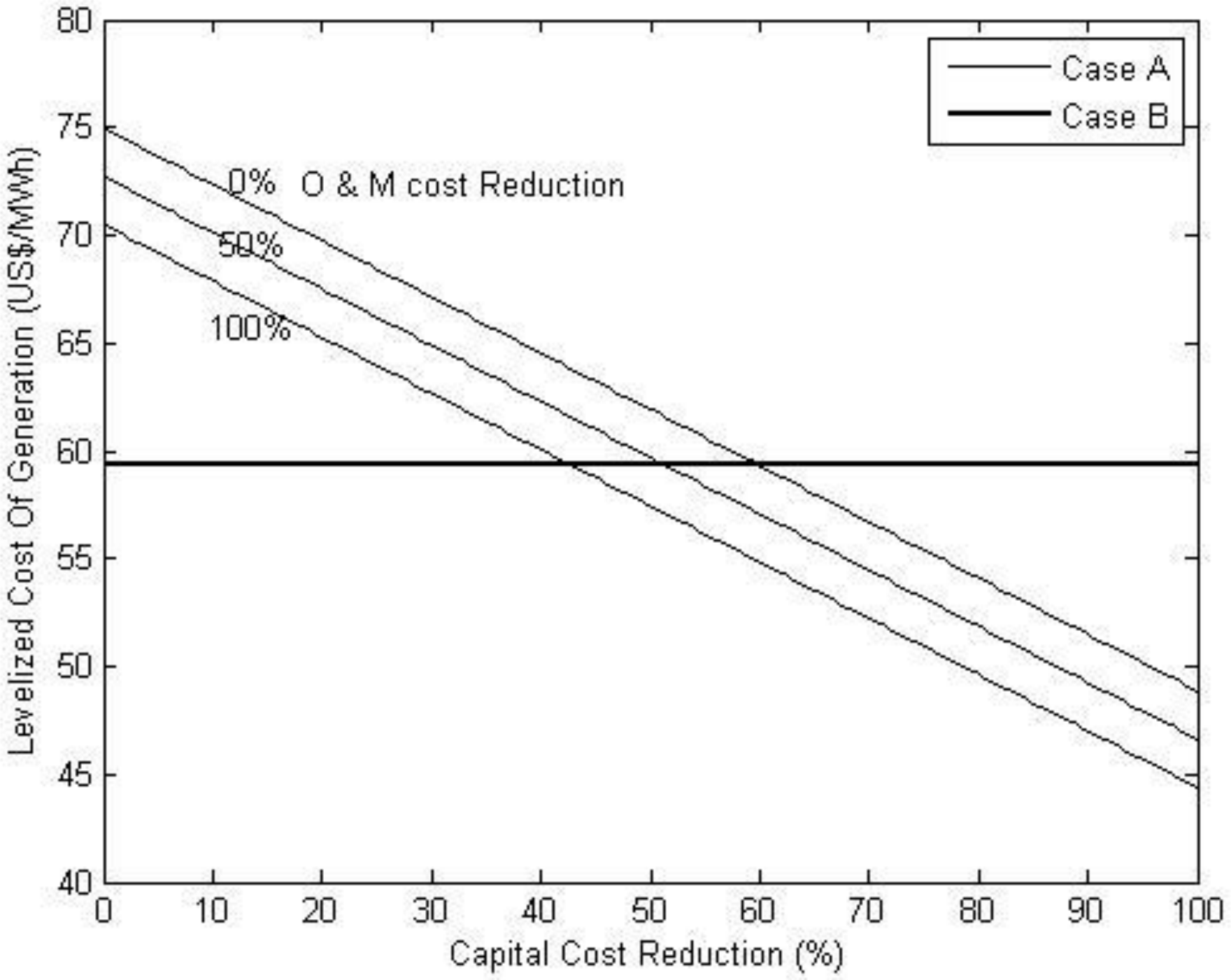

Figure 4.14: Scenario II: LCOE for cases A & B versus capital cost reduction (%) due to coupling, with operation & maintenance cost reduction (%) as a parameter

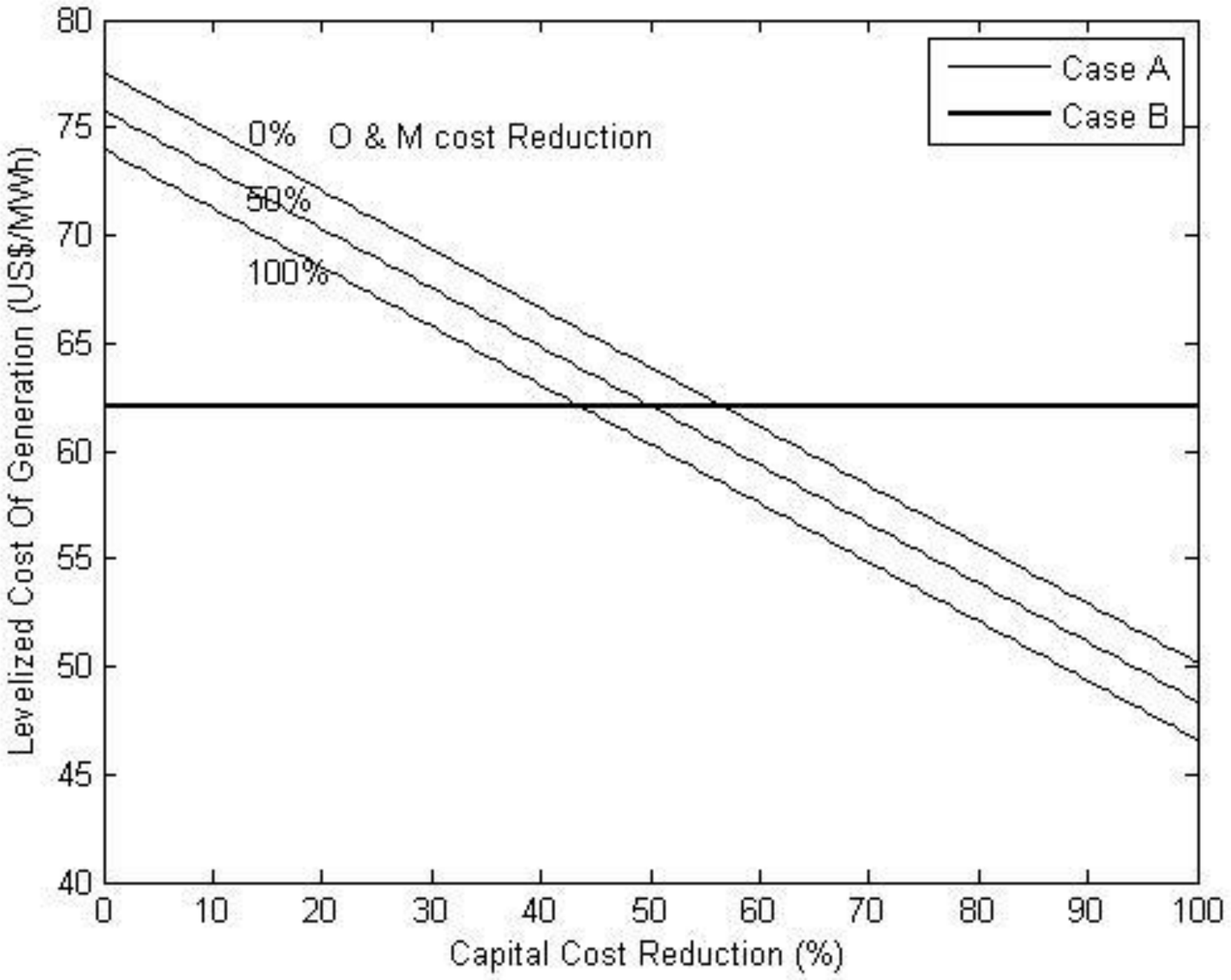

Figure 4.15: Scenario III: LCOE for cases A & B versus capital cost reduction (%) due to coupling, with operation & maintenance cost reduction (%) as a parameter

Table 4.8: Aspects related to coupling and their projected increase/decrease in costs of electricity generation

+: increase, -: decrease, C: Capital Cost, O&M: Operation and Maintenance Cost

| Aspect | Projected +/ - in cost | | |
|---|---|---|---|
| | +/- | C/ O&M | % |
| Availability of land with cheap prices | - | C | 5 |
| Existence of connection to grid | - | C | 0 - 10 |
| Existence of infrastructure required for WF | - | C | 5 |
| Smoothing effect of aggregated wind energy | - | O&M | 10 |
| Voltage variations at PCC | - | C | 5 |
| Reactive Power control capability | - | C | 30 - 40 |
| WF’s switch gear | + | C | 5 |
| Filters for WF’s produced harmonics | + | C | 5 |

# Chapter 5

# Conclusions and Future Work

In this thesis, different aspects related to the coupling of WFs with NPPs are studied. A brief literature survey on the previous research work related to this topic is introduced. This is followed by discussing the basic criteria for selecting the site that can accommodate both NPP and WF. This site selection criteria is then applied to Egypt. There are sites in Egypt that passes the initial site selection requirements for both NPP and WF. Next, the benefits of NPP and WF coupling are discussed. Also, some disadvantages of coupling and the proper solution for each disadvantage are illustrated.

After that, some coupling advantages are chosen for more illustrations and verifications with case studies. The benefits of connecting the WF and NPP to the same point of the grid are illustrated; one of them is the elimination of the voltage quality problems associated with the WF at the point of connection. The benefit of increasing reliability and availability of NPP EPSs by the on-site WF is also illustrated. A study using MARKOV process is conducted to quantify this increase in reliability and availability.

After that, case studies are conducted to illustrate the effect of WFs geographical distribution on the aggregated wind power production in the Egyptian grid along with evaluating the wind energy capacity credit assessment.

Other future work can be done based on this work. This future work can be summarized in the following points:

1. Studying the benefits of adding an energy storage facility to the system. It is proposed to study in details the benefits (technical and economical) of adding an energy storage facility, and choose the best technology of energy storage to be used in different operating conditions.

2. Conducting simulations to study the benefit of coupling on the WF's Low Voltage Ride Through capability if a fault occurred at PCC. This is done by comparing the results of different cases such as NPP and WF are in operation; NPP is shutdown, while WF is in operation; and WF is connected to weak grid.

3. Studying the coupling of offshore WFs with NPPs. It is proposed to study different aspects associated with this coupling including the usage of HVDC connection to WF.

# References

[1] Egyptian Electricity Holding Company, “Annual Report 2011/2012”

[2] Mohamed ElSobki and Peter Wooders, “Clean energy investment in developing countries: Wind power in Egypt”, International Institute for Sustainable Development (IISD), 2009

[3] Kate Rogers and Magdi Ragheb, “Symbiotic coupling of wind power and nuclear power generation”, International Nuclear and Renewable Energy Conference, Jordan, 2010

[4] Gregor Taljana, Michael Fowler, Claudio Canizares, and Gregor Verbic, “Hydrogen storage for mixed wind–nuclear power plants in the context of a hydrogen economy”, international journal of hydrogen energy 33, Elsevier, 2008

[5] Ahmed Shata, “Theoretical investigation and mathematical modeling of a wind energy system: Case study for Mediterranean and Red Sea”, MSc thesis, der Technischen Universität Berlin, 2008

[6] Mark Antkowiak, Richard Boardman, *et al.*, “Summary Report of the INL-JISEA Workshop on Nuclear Hybrid Energy Systems”, Technical Report NREL/TP-6A50-55650, July 2012

[7] Alistair I. Miller & Romney B. Duffey, “Co-Generation of Hydrogen from Nuclear
and Wind: the Effect on Costs of Realistic Variations in Wind Capacity and Power Prices”, Atomic Energy of Canada Limited, 2005

[8] Charles W. Forsberg, “Sustainability by combining nuclear, fossil, and renewable energy sources”, Progress in Nuclear Energy, Elsevier, 2008

[9] International Atomic Energy Agency, “Site evaluation for nuclear installations”, Safety requirements, NS-R-3, 2003

[10] Andrew Macintosh, “Siting nuclear power plants in Australia”, Australian institute for a just sustainable peaceful future, 2007

[11] L.R. Bishnoi and Prabir C. Basu, “Siting of Nuclear Installations”, Atomic Energy Regulatory Board, India, 2005

[12] European Commission, "European best practice guidelines for Wind Energy development", Online: http://ec.europa.eu/energy/res/sectors/doc/wind_energy/best_practice.pdf

[13] Auswind, "Best practice guidelines for implementation of wind energy projects in Australia", 2006.

[14] Egyptian Nuclear Power Plant Authority, "Egyptian nuclear power plant program, history and status", BULATOM Nuclear Conference, 2010.

[15] Niels G. Mortensen, Usama Said, and Jake Badger, "Wind atlas for Egypt", Risø National Laboratory, Egyptian New and Renewable Energy Authority, 2006.

[16] Pramod Jain, "Wind energy engineering", McGraw-Hill Companies, 2011

[17] Baharat Atomic Research Center, "Health physics & environment", BARC Highlights, Reactor Technology & Engineering, 2007

[18] Thomas Ackermann, "Wind power in power systems", John Wiley & Sons, 2005

[19] Deutsches Windenergie-Institut GmbH, Tech-wise A/S, DM Energy, "Wind turbine grid connection and interaction", 2001

[20] Jonathan D. Rose and Ian A. Hiskens, "Challenges of integrating large amounts of wind power", 1st Annual IEEE Systems Conference, Hawaii, USA, 2007

[21] Tony Burton, David Sharpe, Nick Jenkins, Ervin Bossanyi, "Wind energy handbook", John Wiley & Sons, Ltd, 2001

[22] Brendan Fox, Damian Flynn, *et al.*, "Wind power integration, connection and system operational aspects", The Institution of Engineering and Technology, 2007

[23] World Bank Group, "Environmental, health, and safety guidelines for wind energy", 2007

[24] Alexandra Gadawski and Greg Lynch, "The Real Truth About Wind Energy: A Literature Review on Wind Turbines in Ontario", Sierra Club Canada Interns, 2011

[25] Australian Wind Energy Association, "Wind farm safety in Australia", 2004

[26] M. Ragheb, "Safety of Wind Systems", University of Illinois, 2012

[27] EDS Consulting, "Land Use Planning for Wind Energy Systems in Manitoba",

2009

[28] Wind Energy , Defense & Civil Aviation Interests Working Group, “Wind Energy and Aviation Interests-Interim Guidelines”, Crown Copyright, 2002

[29] Remus Teodorescu, Pedro Rodriguez, Andrzej Adamczyk “Reactive power control for Wind Power Plant with STATCOM”, MSc Thesis, Institute of Energy Technology, Aalborg University, Denmark, 2010

[30] Gonzalo Abad, Jesu´s Lo´pez, *et al.*, “Doubly Fed Induction Machine- Modeling and Control for Wind energy Generation”, Institute of Electrical and Electronics Engineers, Inc., John Wiley & Sons, Inc, 2011

[31] Carson W. Taylor, “Power system voltage stability”, Electrical Power Research Institute, 1994

[32] L. H. Hansen, L. Helle, *et al.*, “Conceptual survey of Generators and Power Electronics for Wind Turbines”, Risø National Laboratory, 2001

[33] Manfred Stiebler, “Wind energy systems for electric power generation”, Springer-Verlag Berlin Heidelberg, 2008

[34] IEC Standard 61400-21, “Wind turbine generator systems – Part 21: Measurement and assessment of power quality characteristics of grid connected wind turbines”, Reference number CEI/IEC 61400-21:2001, 2001

[35] Cezary Szafron, "Wind Energy Conversion Systems Grid Connection", MSc thesis, Wroclaw University of Technology, Poland, 2011

[36] Miller AI, Duffey RB, “Integrating large-scale co-generation of hydrogen and electricity from wind and nuclear sources (NUWIND)”, IEEE EIC Climate Change Technology, 2006

[37] Gregor Taljan, “The use of hydrogen in electric power systems”, PhD thesis, University of Ljubljana, 2009

[38] International Atomic Energy Agency, “Electric grid reliability and interface with nuclear power plants”, NG-T-3.8, 2012.

[39] International Atomic Energy Agency, “Interfacing Nuclear Power Plants with the Electric Grid: the Need for Reliability amid Complexity”, Paper, online: http://www.iaea.org/About/Policy/GC/GC53/GC53InfDocuments/English/gc53inf-3-att5_en.pdf

[40] International Atomic Energy Agency, “Design of Emergency Power Systems for Nuclear Power Plants”, NS-G-1.8, 2004

[41] S.A. Eide, C.D. Gentillon, T.E. Wierman, D.M. Rasmuson, “Reevaluation of Station Blackout Risk at Nuclear Power Plants”, Idaho National Laboratory, NUREG/CR-6890, Vol. 2, 2005

[42] Mazaher Haji Bashi, Akbar Ebrahimi, “Markovian approach applied to reliability modeling of a wind farm”, Turkish Journal of Electrical Engineering & Computer Sciences, 2012

[43] D. J. Campbell, R. E. Battle, “Reliability of Emergency AC Power Systems at Nuclear Power Plants”, NUREG/CR-2989, Jul 1983

[44] Yannick Degeilh, “Wind farm diversification and its impact on power system reliability”, MSc thesis, Texas A&M University, USA, 2009

[45] Smith, J. C., Milligan, M. R., DeMeo, E. A., and Parsons, B. “Utility Wind Integration and Operating Impact State of the Art,” IEEE Transactions on Power Systems, 2007

[46] Sunanda Mishra, “Wind Power Capacity Credit Evaluation Using Analytical Methods”, MSc thesis, University of Saskatchewan, Canada, 2010

[47] International Energy Agency, and Nuclear Energy Agency “Projected Costs of Generating Electricity, 2010 Edition”, 2010.

جامعة عين شمس
كلية الهندسة
قسم هندسة القوي والآلآت الكهربية

# اقتران مزرعة للرياح مع محطة للطاقة النووية

## رسالة

مقدمة للحصول علي درجة الماجستير في الهندسة الكهربية
تخصص هندسة القوي و الآلات الكهربية

مقدمه من

## محمد كريم عبد الرحمن ابراهيم العشيري

بكالوريوس الهندسة الكهربية
قسم هندسة القوي والآلآت الكهربية
جامعة عين شمس، ٢٠٠٨

تحت إشراف

ا.د/ محمد عبد الرحيم بدر
د/ وليد علي سيف الاسلام الختام
د/ مصطفي صالح القليل

جامعة عين شمس
كلية الهندسة
قسم هندسة القوي والآلآت الكهربية

# اقتران مزرعة للرياح مع محطة للطاقة النووية

رسالة للحصول علي درجة الماجستير مقدمه من

## محمد كريم عبد الرحمن ابراهيم العشيري

## لجنة الحكم

| الاسم | التوقيع |
|---|---|
| أ.د. عصام الدين محمد أبو الذهب<br>جامعة القاهرة – كلية الهندسة – قسم القوي و الآلات الكهربية | ................ |
| أ.د. المعتز يوسف عبد العزيز محمد<br>جامعة عين شمس – كلية الهندسة – قسم القوي و الآلات الكهربية | ................ |
| أ.د. محمد عبد الرحيم بدر<br>جامعة المستقبل – عميد كلية الهندسة و التكنولوجيا | ................ |
| أ.د. وليد علي سيف الاسلام الختام<br>جامعة عين شمس – كلية الهندسة – قسم القوي و الآلات الكهربية | ................ |

التاريخ: / / ٢٠١٤

جامعة عين شمس
كلية الهندسة
قسم هندسة القوي والآلآت الكهربية

# رسالة ماجستير

اسم الطالب : محمد كريم عبد الرحمن ابراهيم العشيري

عنوان الرساله : اقتران مزرعة للرياح مع محطة للطاقة النووية

اسم الدرجه : ماجستير العلوم في الهندسة الكهربائية (هندسة القوي و الألات الكهربية)

## لجنة الاشراف

أ.د/ محمد عبد الرحيم بدر — عميد كلية الهندسة و التكنولوجيا جامعة المستقبل

د/ وليد علي سيف الاسلام الختام — قسم هندسة القوي و الألات الكهربية كلية الهندسه- جامعة عين شمس

د/ مصطفي صالح القليل — مركز البحوث النووية- هيئة الطاقة الذرية

## الدراسات العليا

ختم الاجازه

أجيزت الرساله بتاريخ / / ٢٠١٤

موافقة مجلس الكلية / / ٢٠١٤

موافقة مجلس الجامعه / / ٢٠١٤

# تعريف بمقدم الرسالة

الاسم: محمد كريم عبد الرحمن ابراهيم

تاريخ الميلاد: ٢٥ / ٦ / ١٩٨٦

محل الميلاد: المملكة العربية السعودية

الدرجة الجامعية: بكالوريوس الهندسة الكهربية، تخصص قوي و آلات كربية

الجهة المانحة: جامعة عين شمس

تاريخ المنح: ٢٠٠٨

الوظيفة الحالية: معيد بهيئة الطاقة الذرية المصرية

# الملخص

أصبحت قضية التغير المناخي واحدة من أكبر التحديات التي تواجه الدول والحكومات والشركات و المواطنين في العالم . و تتطلب مواجهة تهديدات التغير المناخي زيادة في حصة الطاقة المتجددة من مجموع توليد الطاقة. أيضا هناك جهود هائلة لتقليل الاعتماد على مصادر الطاقة من الوقود الأحفوري الذي يفتح المجال لزيادة استخدام الموارد البديلة مثل الطاقة النووية. لهذا تخطط العديد من البلدان (مثل مصر) لتلبية الطلبات المتزايدة علي الكهرباء عن طريق زيادة المساهمات لكل من الطاقات المتجددة ( بخاصة طاقة الرياح ) والطاقة النووية في توليد الكهرباء.

في مرحلة التخطيط لتحديد مواقع كلا من مزارع الرياح و محطات الطاقة النووية ، توجد العديد من التحديات. أحد الجوانب الهامة التي تؤخذ بعين الاعتبار أثناء تحديد موقع محطة الطاقة النووية هو وجود مياه للتبريد في الموقع، وتأخذ هذه المياه من البحار في معظم الحالات. لذلك معظم محطات الطاقة النووية تقع على سواحل البحار. من ناحية أخرى ، خلال تحديد موقع مزرعة الرياح ، الجانب الرئيسي المؤثر هو وجود سرعات رياح جيدة في الموقع في معظم أوقات العام. العديد من المناطق الساحلية في مختلف أنحاء العالم توفي بهذا المطلب لبناء مزرعة رياح. اقتران كلا من مزرعة رياح و محطة للطاقة النووية في نفس الموقع أو المتاخمة بينهم لها فوائد كثيرة، كما أن لها عقبات أيضا. في هذه الأطروحة ، على أساس الخبرة العالمية في هذا المجال، تم دراسة و تقييم فوائد و معوقات هذا الاقتران / المتاخمة. وكذلك للتحقق من مفهوم الاقتران / المتاخمة تم القيام ببعض دراسات الحالة المختلفة علي الشبكة الكهربية في مصر.

# محتويات فصول الرسالة

هذه الأطروحة تدور حول دراسة وتقييم فوائد و معوقات اقتران مزارع الرياح مع محطات الطاقة النووية، وتنقسم هذه الرسالة إلى خمسة فصول نظمت على النحو التالي:

الفصل الأول: يحتوي علي تعريف بفكرة اقتران مزارع الرياح مع محطات الطاقة النووية. كما أنه ناقش الدافع من وراء هذه الفكرة، وعرض مخطط الأطروحة.

الفصل الثاني: يعرض هذا الفصل الدراسة الاسترجاعية للموضوع. ثم يوضح خصائص الموقع الملائم لبناء مزرعة رياح و محطة للطاقة النووية. كما يقيم هذا الفصل المواقع المختلفة في مصر التي يمكن أن تستوعب هذا اقتران. و أخيرا ، تم شرح مزايا وعيوب اقتران مزارع الرياح مع محطات الطاقة النووية.

الفصل الثالث: يوضح فوائد ربط مزرعة الرياح و محطة الطاقة النووية بنفس النقطة في الشبكة الكهربية. كما يعرض نتائج دراستي حالة للتحقق من فائدتين رئيسيتين لهذا الاقتران. تم تقييم تأثير ارتفاع مستوي الماس الكهربي عند نقطة الربط لمزرعة الرياح علي جودة الجهد الكهربي عند هذه النقطة. كما تم حساب الزيادة في الاتاحية و العول لنظم الطاقة في حالات الطواريء لمحطات الطاقة النووية عن طريق دمج مزرعة الرياح المتواجدة في الموقع.

الفصل الرابع: يناقش بالتفصيل الفائدة من التوزيع الجغرافي لمزارع الرياح في الشبكة الكهربية . تم اجراء دراسة حالة لتوضيح تأثير التوزيع الجغرافي لمزارع الرياح على اجمالي الطاقة المولدة من هذه المزارع و التغيرات اللحظية لها. بعد ذلك، تم تقييم القدرة الائتمانية لطاقة الرياح في مصر. و أيضا ، تم وضع خطة استراتيجية لهذا الاقتران في مصر ، أخذين في الاعتبار التنسيق بين مزارع الرياح و محطات الطاقة النووية المخطط بناؤها في المستقبل. أخيرا ، تم إجراء دراسة حالة للتحقق من القدرة التنافسية لتكلفة توليد الطاقة من مزرعة رياح مجاورة لمحطة طاقة نووية في مصر.

الفصل الخامس: يختم هذه الأطروحة باستخراج الاستنتاجات ووضع نقاط البحث المستقبلية التي يمكن القيام بها بناءا على هذا العمل.